\documentclass[a4paper,fleqn]{cas-dc}

\usepackage[authoryear,longnamesfirst]{natbib}
\usepackage{subcaption}
\usepackage{graphicx}
\usepackage{caption}
\usepackage{bm}
\usepackage{xcolor}
\usepackage{pifont}
\makeatletter
\renewcommand{\@maketitle}{%
  \begin{center}\LARGE\@title\vskip1.5em\large\@author\vskip1em\@affiliation\vskip0.5em\@email\vskip0.5em\@cortext\vskip2em\end{center}%
}
\makeatother
\def\tsc#1{\csdef{#1}{\textsc{\lowercase{#1}}\xspace}}
\tsc{WGM}
\tsc{QE}

\begin{document}
\let\WriteBookmarks\relax
\def\floatpagepagefraction{1}
\def\textpagefraction{.001}



\title [mode = title]{Data-Driven Flow Initialization Framework for CFD Acceleration of Underwater Vehicle in Vertical-Plane Oblique Motion}  



%

\author[1,3,4]{Tianli Hu}






\affiliation[1]{organization={China Ship Scientific Research Center},
            city={Wuxi},
            postcode={214082}, 
            country={PR China}}
\author[1,3,4]{Chengsheng Wu}
\author[1,3]{Jun Ding}
\author[1,3,4]{Xing Wang}
\author[2]{Yu Yang}
\cormark[1]
\ead{yuyangyy@sjtu.edu.cn}




\cortext[1]{Corresponding author}

\author[1,3,4]{Jianchun Wang}

\affiliation[2]{organization={Marine Numerical Experimental Center},
            organization={State Key Laboratory of Ocean Engineering}, 
            organization={School of Ocean and Civil Engineering},
            organization={Shanghai Jiao Tong University},
            city={Shanghai},
            postcode={200240}, 
            country={PR China}}

\affiliation[3]{organization={Taihu Laboratory of Deepsea Technological Science},
            city={Wuxi},
            postcode={214082}, 
            country={PR China}}
\affiliation[4]{organization={National Key Laboratory of Hydrodynamics},
            city={Wuxi},
            postcode={214082}, 
            country={PR China}}





\begin{abstract}
Accurate prediction of flow fields around underwater vehicles undergoing vertical-plane oblique motions is critical for hydrodynamic analysis, but it often requires computationally expensive CFD simulations. This study proposes a Data-Driven Flow Initialization (DDFI) framework that accelerates CFD simulation by integrating deep neural network (DNN) to predict full-domain flow  fields. Using the suboff hull under various inlet velocities and angles of attack as an example,  a DNN is trained to predict velocity, pressure, and turbulent quantities based on mesh geometry, operating conditions, and hybrid vectors. The DNN can provide reasonably accurate predictions with a relative error about 3.3\%. To enhance numerical accuracy while maintaining physical consistency, the DNN-predicted flow fields are utilized as initial solutions for the CFD solver, achieving up to 3.5-fold and 2.0-fold speedup at residual thresholds of  $5\times10^{-6}$ and  $5\times10^{-8}$, respectively. This method maintains physical consistency by refining neural network outputs via traditional CFD solvers, balancing computational efficiency and accuracy. Notably, reducing the size of training set does not exert an essential impact on acceleration performance. Besides, this method exhibits cross-mesh generalization capability. 
In general, this proposed hybrid approach offers a new pathway for high-fidelity and efficient full-domain flow field predictions around complex underwater vehicles.
\end{abstract}





\maketitle

\section{Introduction}\label{}
Underwater vehicles  serve as essential equipment for ocean exploration, with significant applications in both civilian and military domains \citep{LUO2021109050,ardeshiri2022efficient}. The accurate and efficient prediction of the flow field around submarine geometries is of importance to understand the hydrodynamic behaviors and optimize the parameters of submarine model such as drag coefficient under different flow conditions \citep{LIU2021109361,LI202165}.
The hydrodynamic performance, including resistance, maneuvering stability, and acoustic signature, is fundamentally determined by the flow characteristics such as  velocity distribution, pressure fields, and turbulence properties \citep{panda2021review}.  The suboff generic submarine hull has been widely adopted as a standard benchmark configuration for validating computational fluid dynamics (CFD) methods in underwater applications, owing to its well-documented geometry and extensive experimental database \citep{POSA2018116,SARRAF2022112849}. Understanding the steady-state flow physics around such configuration is fundamental for performance evaluation and operational analysis \citep{MENG2019106528,QU2021109866,ZHOU2022113107}.

Traditional approaches to obtaining detailed flow field information rely heavily on high-fidelity CFD simulations \citep{SEZEN2018258,QIU2020107285,ROCCA2022103360}. Although modern CFD solver like Reynolds-Averaged Navier-Stokes (RANS) solvers can provide a high degree of accuracy, they come with significant computational costs \citep{CHU2024118250}. For instance, a single simulation for a complex geometry like suboff under a specific operating condition may require hours to days of computational time on high-performance clusters, even for steady-state solutions \citep{zbMATH07735933}. This high computational cost is primarily due to the need for a fine spatial discretization to adequately resolve boundary layers, wake regions, and vortical structures, combined with the iterative process of solving the coupled, non-linear governing equations until convergence is achieved \citep{blazek2015computational}.

Conducting high-precision, large-scale, and multiscale 3D numerical simulations of flow fields around underwater vehicles remains a computationally intensive task  \citep{LU2020102343}. In recent years, Machine learning (ML) \citep{liong2022data,li2021uncertainty,2024Ultra,YANG2025379,jiang2024development} has emerged as an effective approach for modeling complex systems by uncovering patterns from data without the knowledge of underlying physics. Among the various ML techniques, deep neural networks (DNNs) \citep{jordan2015,YANG2023113470} have gained prominence for their ability to approximate highly nonlinear relationships. DNNs are biologically inspired computational models composed of interconnected layers of artificial neurons. During training, the weights of neural networks are iteratively optimized through backpropagation to minimize prediction errors.  Due to their universal approximation capability, DNNs demonstrate good proficiency in  tasks requiring high-dimensional function approximation, making them particularly suitable for complex dynamics modeling \citep{samek2021explaining,YANG2023336}. By implementing a deep learning model as a surrogate, the computational cost  associated with conventional fluid dynamics simulations
 can be effectively alleviated \citep{XIE2024104074}. For example, the DNN has been successfully used to predict ship resistance and wake field \citep{kim2006wake,grabowska2015ship}. Besides, a deep neural network-based reduced-order model  has been applied for the rapid prediction of the steady-state velocity field \citep{Peng2021}. Note that, for ML predictions of flow field, most existing studies focus on the wake flow field prediction or near-wall pre-sampled flow field prediction. While, the focus of this work will lie on the prediction of the full flow field, i.e., the entire computational domain.

\begin{figure*}[!h]
\centering
    \begin{subfigure}[b]{0.95\textwidth}
    \centering
    \includegraphics[height=4cm]{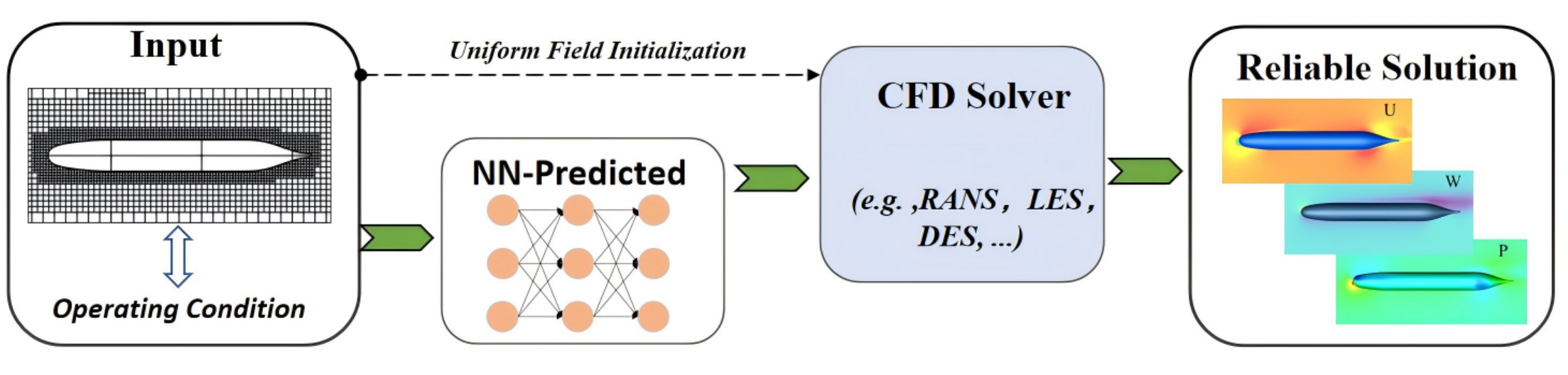} 
    \end{subfigure}
\caption{\label{Fig:suboff_DDFI} Schematic diagram of the DDFI computational workflow.}
\end{figure*}

It should be noted  that, while machine learning model can function as an efficient surrogate model, providing rapid flow field predictions, this often comes at the expense of solution accuracy and physical consistency. For scientific and engineering applications that demand high-fidelity results, traditional CFD solvers should continue to play a central role in the computational workflow. At present, it is essential to mitigate the risk of misapplying neural network predictions and to ensure the reliability of results  \citep{zhou2025neural}.
For tasks requiring high precision, a hybrid strategy can be employed where neural network-based predictions are further refined or corrected using traditional CFD solvers, which can establish a balance between computational efficiency and numerical accuracy. \citet{zhou2025neural} proposed a super-fidelity method by using a neural network to map low-fidelity solutions to high-fidelity initial guesses, which can  accelerate the convergence for problems such as two-dimensional laminar flow around elliptical cylinders and turbulent flow over airfoils or wings. However, constructing an ML-based operator integrated with a CFD solver to predict a full, high-resolution flow field around a complex three-dimensional body like the suboff submarine remains a significant challenge. Besides, compared with using low-fidelity flow fields as inputs of neural network \citep{zhou2025neural}, we try to use the mesh topological information and operating parameters as inputs in this study, so that this scheme can be directly integrated into traditional CFD workflows.

In this study, we propose a Data-Driven Flow Initialization (DDFI) framework to accelerate the acquisition of full-domain flow fields using CFD solver. Taking the suboff model as an example for steady flow problem, we construct a  DNN model to predict flow field parameters—including velocity, pressure, and turbulent quantities, for given mesh geometry, operating conditions, and hybrid vectors. The DNN provides a reasonably accurate initial prediction. To achieve higher accuracy while ensuring consistency with the governing physical equations, the DNN-predicted flow field is used as the initial solution for the CFD solver, which significantly accelerates the convergence of the numerical simulation. Besides, we investigate the impact of the training dataset size on both the  DNN’s predictive performance and the CFD solver’s convergence efficiency. Furthermore, the proposed neural network-accelerated CFD framework is applicable to higher-resolution meshes. The proposed strategy offers a new pathway for accelerating flow field predictions around intricate underwater vehicle structures.

\section{Methodology}
\label{method}

For steady-state problems, any physical quantity can be described by the following general transport equation, which is expressed as:
\begin{equation}\label{transport}
\centering
\rho\nabla\cdot(\mathbf{V})\phi-\nabla\cdot(\zeta\nabla\phi)+Q=0,
\end{equation}
where $\rho$ is the fluid density, $\mathbf{V}=(U,V,W)$ is the velocity vector, $\nabla$ is the generalized diffusion coefficient, $\phi$ is the generalized physical quantity in the flow field and $Q$ is the generalized source term. For the momentum equation, $\phi$ is replaced by $\mathbf{V}$, and $Q$ is replaced by $\nabla p+Q$. For specific operating conditions, the physical quantity $\phi$ and 
velocity vector $\mathbf{V}$ satisfying the left-hand side (LHS) of Eq.~(\ref{transport}) being equal to zero are unknown. A classical approach adopts a uniform field as the initial condition and iteratively drives the LHS to approach zero, but this process is typically highly time-consuming, especially when the uniform field deviates significantly from the actual flow field. A natural idea is to provide an initial condition that is sufficiently accurate to reduce the number of iterations and thereby speed up the solution process. However, acquiring such a sufficiently accurate initial condition is often extremely challenging. Fortunately, the advancement of data-driven machine learning techniques has opened up a feasible path for this vision. 

Based on this, this study proposes the DDFI strategy, whose core is to output a sufficiently accurate initial condition through a DNN during flow field initialization, thereby improving the iterative convergence speed of CFD solvers, as shown in Fig.~\ref{Fig:suboff_DDFI}. Notably, the inputs of the neural network in this study are mesh topological information and operating parameters. Compared with using low-fidelity flow fields as inputs \citep{zhou2025neural}, the proposed scheme  achieves a closer integration with the traditional CFD workflows, offering stronger engineering practicality but also posing greater challenges. To this end, the input features of the neural network have been carefully designed to ensure the reliability and engineering applicability of the prediction results.

\subsection{Data pre-processing}
\label{predata}
\begin{table*}[!h]
    \centering
    \normalsize
    \renewcommand{\arraystretch}{1.3}
    \caption{Classification, description, and definition of input feature $\mathbf{F}$ for the DNN model.}
    \label{tab:feature_definitions}
    \begin{tabular*}{\textwidth}{@{\extracolsep{\fill}} l l l l @{}}
        \toprule
        \bfseries Category & \bfseries Feature & \bfseries Description & \bfseries Definition \\
        \midrule
        \multirow{6}{*}{\makecell{Mesh\\geometric}} 
        & $\hat{x}$ & Hull relative $x$-coordinate & $\hat{x}=x-x_{0}$ \\
        & $\hat{y}$ & Hull relative $y$-coordinate & $\hat{y}=y-y_{0}$ \\
        &$\hat{z}$& Hull relative $z$-coordinate &  $\hat{z}=z-z_{0}$ \\
        & $\hat{\lambda}$ & Minimum wall distance &  $\hat{\lambda}=\lambda/\lambda_{max}$ \\
        & $\hat{\gamma}$ & Cell volume & $\hat{\gamma}$=$\gamma$/$\gamma_{max}$\\ 
        & $\hat{\kappa}$ & Hull distance & $\hat{\kappa}=\kappa/\kappa_{max}$\\ 
        & $\eta$ & Flag & cell: 0,face: 1 \\
        \midrule
        \multirow{3}{*}{\makecell{Operating\\condition}}
        & $\hat{u}$ & Inlet total velocity & $\hat{u}=|\mathbf{V}|/|\mathbf{V}_{max}|$ \\
        & $\theta_{\text{rad}}$ & Inlet velocity angle & $\theta_{\text{rad}}=\alpha\frac{\pi}{180^{\circ}}$  \\
        & $\mathbf{u}_{inf}$ & Unit inlet velocity  & $(\cos\theta, 0, \sin\theta)$  \\
         \midrule
        \multirow{2}{*}{\makecell{Hybrid\\vector}}
        & $\bm{\omega}_{w}$ & Wall hybrid vector \ & $\bm{\omega}_{w}=\mathbf{n}_{w}\times \mathbf{u}_{inf}$ \\
        & $\bm{\omega}_{h}$ & Hull hybrid vector & $\bm{\omega}_{h}=\mathbf{n}_{h}\times \mathbf{u}_{inf}$\\
       \bottomrule
    \end{tabular*}
\end{table*}

Overall, the input features $\mathbf{F}$ of the DNN proposed in this study can be categorized into three parts, including mesh geometric and topological information, operating condition information, and mesh-operating condition hybrid information. Without loss of generality, the following illustration is based on the ascent/descent motions of the suboff model investigated in this study (see Section~\ref{sec3} for details). For the $i$-th mesh component centroid in the mesh, its input features can be  expressed as:

\begin{figure*}[!h]
\centering
    \begin{subfigure}[b]{0.9\textwidth}
    \centering
    \includegraphics[height=6.5cm]{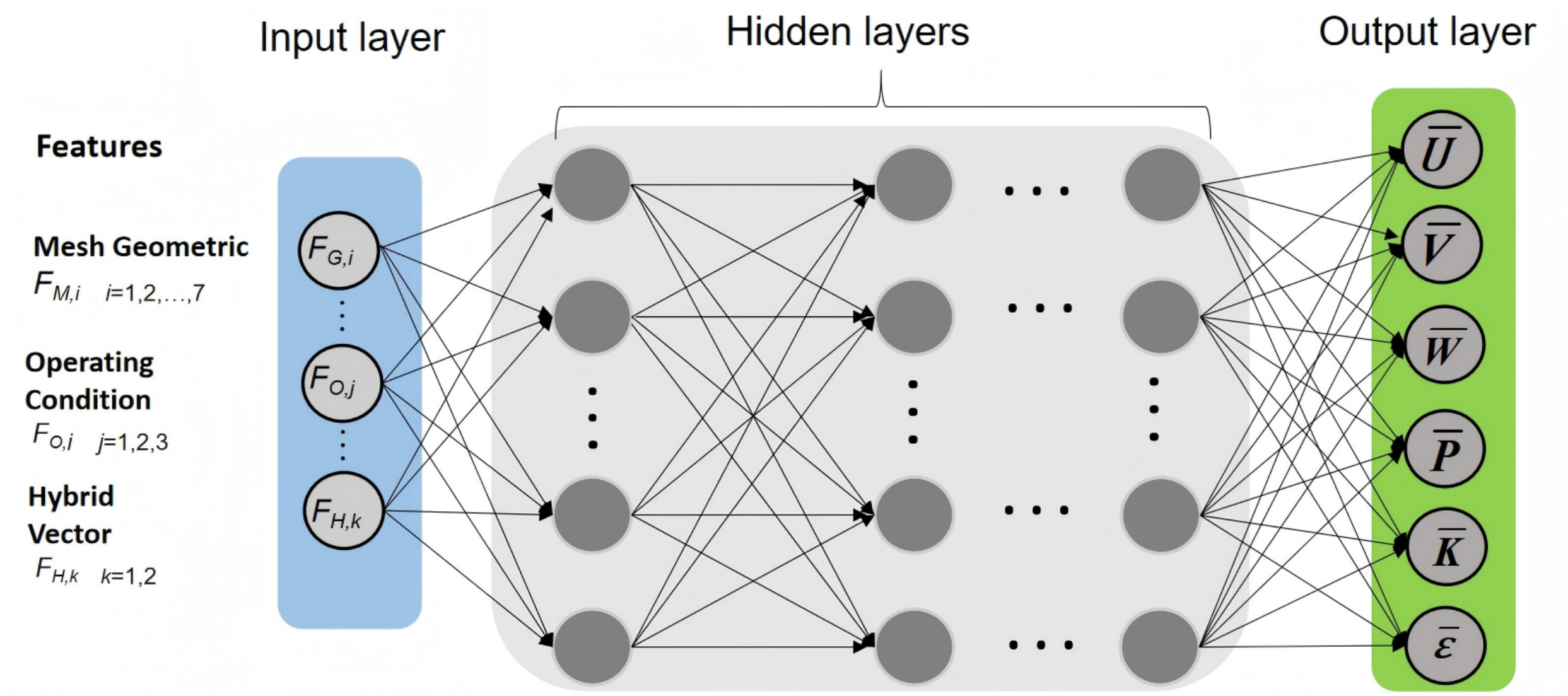}
    \end{subfigure}
\caption{\label{Fig:DNN} Schematic diagram of the DNN model architecture.}
\end{figure*}

\begin{equation}\label{input}
\begin{gathered}
\mathbf{F}^{(i)} = \underset{
     \underbrace{\hspace{8.5em}}_{G^{(i)}} \; 
    \underbrace{\hspace{5.5em}}_{O^{(i)}} \; 
    \underbrace{\hspace{3.5em}}_{H^{(i)}}\quad
}{\left[
        \hat{x},\; \hat{y},\; \hat{z},\; \hat{\lambda},\; \hat{\gamma},\; \hat{\kappa},\; \eta,\;
        \hat{u},\; \theta_{\mathrm{rad}},\; \mathbf{u}_{\mathrm{inf}},\;
        \bm{\omega}_{w},\;\bm{\omega}_{h}
    \right]^{(i)}}, \\
i \in \Omega
\end{gathered}
\end{equation}
herein, $G^{(i)}$, $O^{(i)}$ and $H^{(i)}$ correspond to the three categories mentioned above. Specifically, $\hat{x}$, $\hat{y}$ and $\hat{z}$ are the relative coordinates with the origin at the bow of suboff ($x_{0},y_{0},z_{0})$, $\hat{\lambda}$ and  $\hat{\gamma}$ denote the normalized minimum  wall distance and cell volume, respectively, $\hat{\kappa}$ denotes the normalized distance from the bow of suboff, $\eta$ denotes the flag symbol, $\hat{u}$ and  $\theta_{\text{rad}}$ denote the normalized inlet velocity scalar $|\mathbf{V}|$ and the inlet angle $\alpha$ in radians, respectively, $\mathbf{u}_{\text{inf}}$ refers to the vector of the projection of the unit inlet velocity onto each coordinate direction.  

Furthermore, to more accurately characterize the coupling interaction between operating conditions and the mesh, two hybrid features, denoted as $\bm{\omega}_{w}$ and $\bm{\omega}_{h}$, are introduced in this study. Specifically, $\bm{\omega}_{w}$ represents the cross product of the face normal vector of the nearest wall and $\mathbf{u}_{\text{inf}}$, which is used to characterize the flow field characteristics in the near-wall region. $\bm{\omega}_{h}$ refers to the cross product of the vector from the mesh component centroid to the bow of the suboff and $\mathbf{u}_{\text{inf}}$, which serves to describe the spatial orientation of the corresponding mesh cell in the global flow field. Detailed feature definitions are provided in Table.\ref{tab:feature_definitions}.

Besides, $i\in\Omega$ denotes the set of all input mesh components, consisting  of the cell centroid of the mesh $\Omega_{c}$ and the face centroid on the suboff wall $\Omega_{f}$, namely: 
\begin{equation}\label{set}
\Omega=\Omega_{c}\cup\Omega_{f}.
\end{equation}

\subsection{Neural Network Architecture Construction}

For a typical deep feedforward neural network \citep{jordan2015}, the computation of its $p$-th layer can be written as:
\begin{equation}\label{NN}
a^{[p]}=g^{[p]}(z^{[p]}),
\end{equation}
where $g^{[p]}$  denotes the activation function and $z^{[p]}$ denotes the affine transformation, which is expressed as:
\begin{equation}\label{affine}
z^{[p]}=W^{[p]}a^{[p-1]}+b^{[p]},
\end{equation}
where $W^{[p]} \in \mathbb{R}^{m^{[p]}\times m^{[p-1]}}$ denotes the weight matrix, $b^{[p]}$ denotes the bias vector, and $m^{[p]}$ denotes the number of neurons in the $p$-th layer.

\begin{table}[htbp]
    \centering
    \renewcommand{\arraystretch}{1.3}
    \caption{The architecture and initialization method of the DNN model}
    \label{tab:sampling_dataset_params}
    \begin{tabular*}{\linewidth}{@{\extracolsep{\fill}} c c@{}}
        \toprule
        \multirow{1}{*}{\bfseries } & \bfseries DNN model 
\\
        \midrule
        Input neuron count   &17  \\
        Hidden layer count    &5  \\
        Output neuron count   & 6    \\
        Architecture   & $($17,80,80,80,80,80,6$)$  \\
        Activation function   & LeakyReLU;Linear(last layer)     \\
        \multirow{3}{*}{Weight initialization}   &  Xavier uniform\\
                                   &\quad  gain=0.7(input$\&$hidden layers)  \\
                                   &\quad  gain=default(output layers)  \\               
        \bottomrule
    \end{tabular*}
    \label{Architecture}
\end{table}

The universal approximation theorem  \citep{hornik1989multilayer,cybenko1989approximation} has verified that the feedforward neural network equipped with nonlinear activation functions and hidden layers has the potential to act as a universal flow field approximator, provided that a sufficient number of neurons are deployed. Based on this conclusion, the deep neural network architecture illustrated in Fig.~\ref{Fig:DNN} is established in this paper. Detailed architecture and initialization strategy of the DNN are presented in Table.~\ref{Architecture}. Note that, input features consist of 9 scalars and 3 vectors, and the consistently zero term in $\bm{u}_{inf}$  are neglected, resulting in a total of 17 features (i.e.,9+2+3+3=17). Numerical results demonstrate that, with the input features designed in this paper, the adoption of the aforementioned lightweight DNN can achieve desirable prediction accuracy.

\begin{table}[htbp]
    \centering
    \renewcommand{\arraystretch}{1.3}
    \caption{Dimensionless normalization  of the output data of the DNN model}
    \label{tab:sampling_dataset_params}
    \begin{tabular*}{\linewidth}{@{\extracolsep{\fill}} c c c@{}}
        \toprule
        \multirow{1}{*}{\bfseries Symbol} & \bfseries Description & \bfseries  Definition
\\
        \midrule
        $\tilde{U}$    &$x$-velocity   & $\tilde{U}=U/|\mathbf{V}|$\\
        $\tilde{V}$    &$y$-velocity  & $\tilde{V}=V/|\mathbf{V}|$ \\
        $\tilde{W}$   & $z$-velocity   & $\tilde{W}=W/|\mathbf{V}|$  \\
         $\tilde{P}$   & Pressure   &  $\tilde{P}=P/\rho|\mathbf{V}|$  \\
        $\tilde{K}$   & Turbulent kinetic energy   &   $\tilde{K}=K/|\mathbf{V}|^{2}$   \\
          $\tilde{\epsilon}$   & Dissipation rate  &   $\tilde{\epsilon}=\epsilon L/|\mathbf{V}|^{3}$  \\ 
        \bottomrule
    \end{tabular*}
    \label{stand}
\end{table}

In addition, two-step processing is required for the generalized  physical quantity $\phi$ given by the CFD solver to obtain the output value $\bar{\phi}$ of the DNN:

1). Normalize $\phi$ to the dimensionless quantity $\tilde{\phi}$ using the inlet velocity $\mathbf{V}$, density $\rho$ and characteristic length $L$, as shown in Table~\ref{stand}.

2). Convert $\tilde{\phi}$ to $\bar{\phi}$ by standardizing it to the range of 0 to 1 using the following formula: 

\begin{equation}\label{affine}
\bar{\phi}^{(i)}=(\tilde{\phi}^{(i)}-\tilde{\phi}_{min}^{(i)})/(\tilde{\phi}_{max}^{(i)}-\tilde{\phi}_{min}^{(i)}).\quad \quad \quad i \in \Omega
\end{equation}
The loss function consists of cell loss $l_{c}$ and face loss $l_{f}$, which are written as: 
\begin{equation}\label{loss_c}
l_{c}=\frac{1}{|\Omega_{c}|}\Sigma_{\Omega_{c}}(\bar{\phi}_{pred}-\bar{\phi}_{true})^{2},
\end{equation}

\begin{equation}\label{loss_f}
l_{w}=\frac{1}{|\Omega_{f}|}\Sigma_{\Omega_{f}}(\bar{\phi}_{pred}-\bar{\phi}_{true})^{2}.
\end{equation}
Where $\bar{\phi}_{pred}$ denote the neural network(NN)-predicted results, $\bar{\phi}_{true}$ denote the CFD results. Then, the total error can be expressed as:
\begin{equation}\label{loss_total}
Loss=c_{1}l_{c}+c_{2}l_{w},
\end{equation}
where without loss of generality, we set $c_{1}=c_{2}=1$.

\section{Numerical Results}
\label{sec3}

\begin{figure*}[htbp]
\centering
    \begin{subfigure}[b]{0.9\textwidth}
    \centering
    \includegraphics[height=4cm]{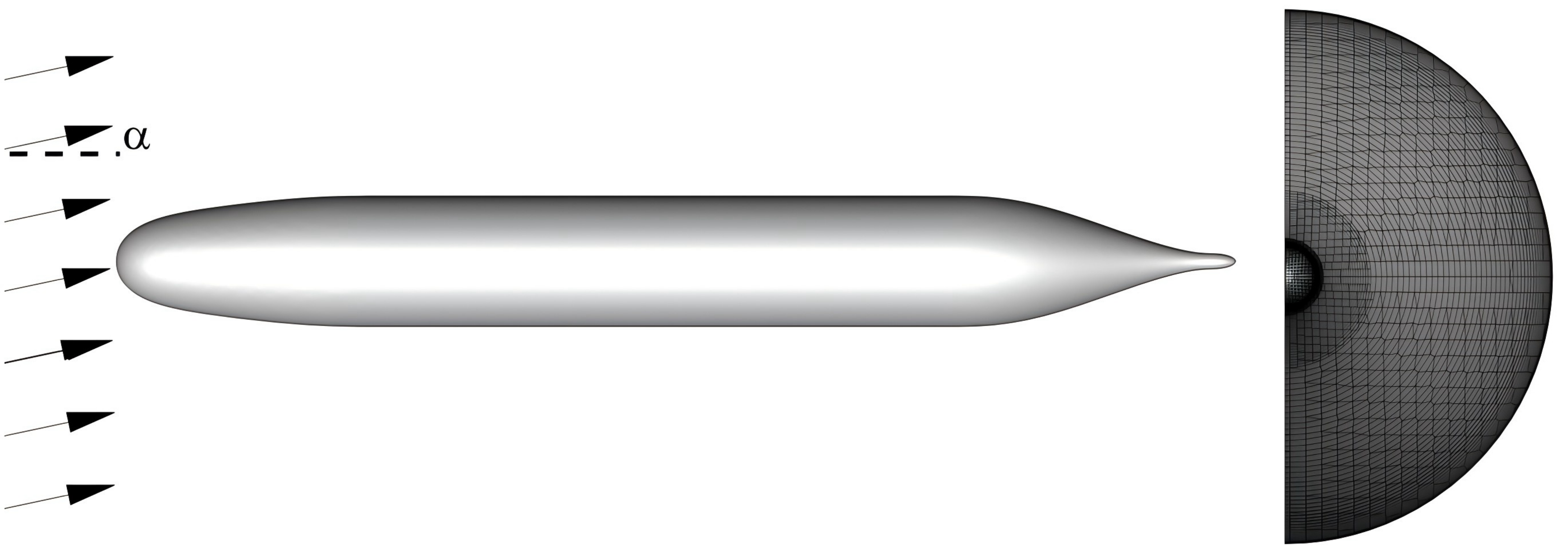}
    \end{subfigure}
\caption{\label{Fig:suboff1} Schematic diagram of the suboff model hull, inlet flow angle of attack and wall mesh distribution.}
\end{figure*}

\begin{figure*}[htbp]
\centering
    \begin{subfigure}[b]{0.85\textwidth}
    \centering
    \includegraphics[height=8cm]{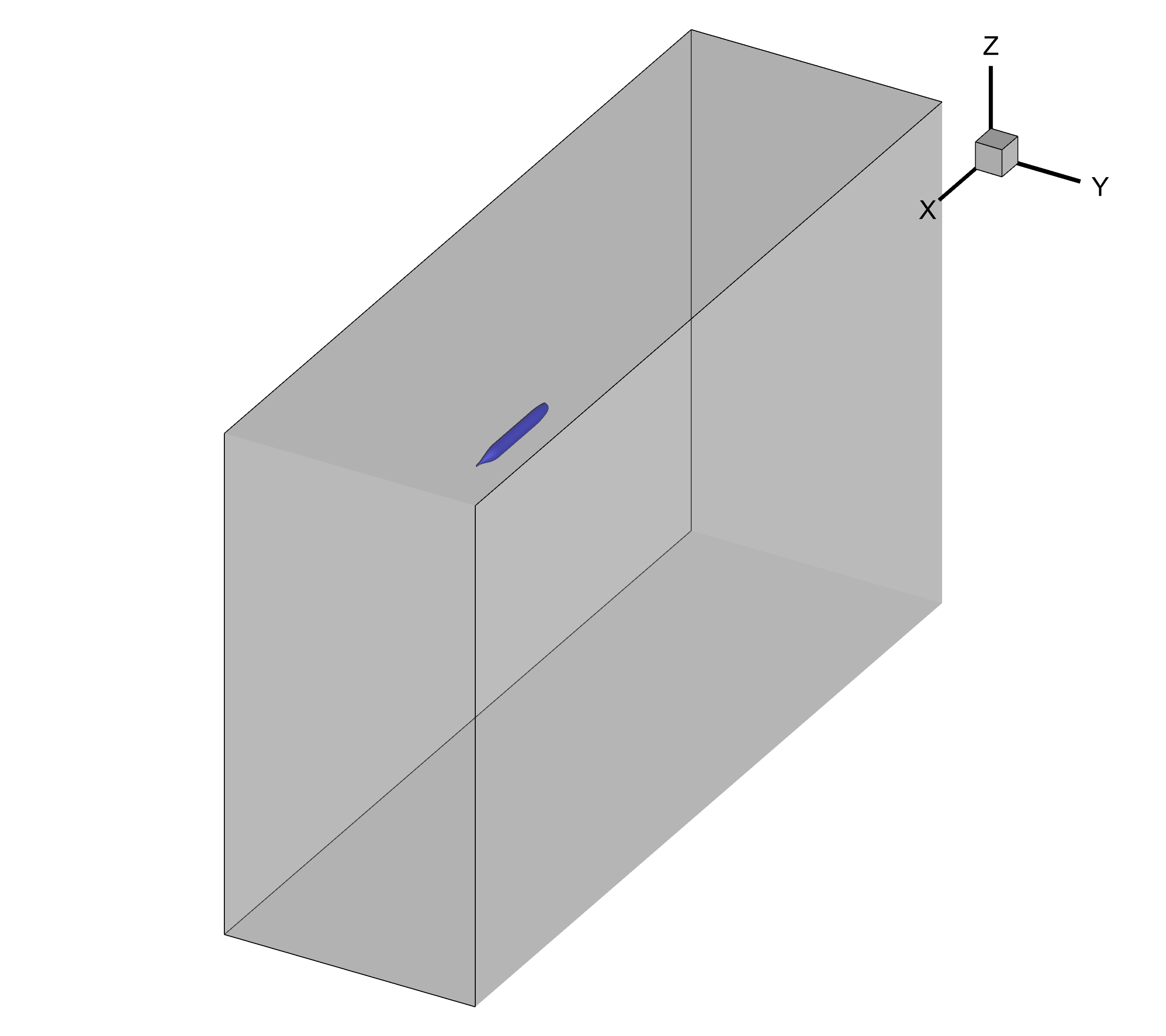} 
    \end{subfigure}
\caption{\label{Fig:suboff2} Schematic diagram of the computational domain for the suboff model.}
\end{figure*}
To validate the acceleration performance of the DDFI strategy proposed in this paper for CFD calculations,  we consider the vertical-plane oblique (ascent/descent) motions of the bare suboff hull under different inlet velocities and angles of attack, and the simulation of this vertical maneuvering motion is more challenging than that of straight-line sailing, since it involves more complex nonlinear phenomena. 

Fig.~\ref{Fig:suboff1} presents the bare hull model of the suboff without additional sails and rudders adopted in this study and illustrate the mesh distribution on the wall of the bare suboff hull. This model has a total length of 4.356 m, a parallel middle hull section length of 2.229 m, and a maximum diameter of 0.508 m \citep{groves1989geometric}. Fig.~\ref{Fig:suboff2} further illustrates the configuration of its computational domain.

The hull motion is confined to the vertical plane, which means that for the inlet velocity $\mathbf{V}^{*}$, its $y$-component is always zero, i.e. $V^{*}\equiv 0$, while its $x$ and $z$ components $U^{*}$ and $W^{*}$ vary with the angle of attack $\alpha$. As illustrated in Fig.~\ref{Fig:suboff1} and Fig.~\ref{Fig:suboff2}, $\alpha$ is defined as the angle between the inlet velocity vector and the positive direction of $x$-axis, with  $\alpha$ assigned a positive value when the $W^{*}$ coincides with the positive $z$-axis direction.

\begin{figure*}[htbp]
\centering
    \begin{subfigure}[b]{0.95\textwidth}
    \centering
    \includegraphics[height=9cm]{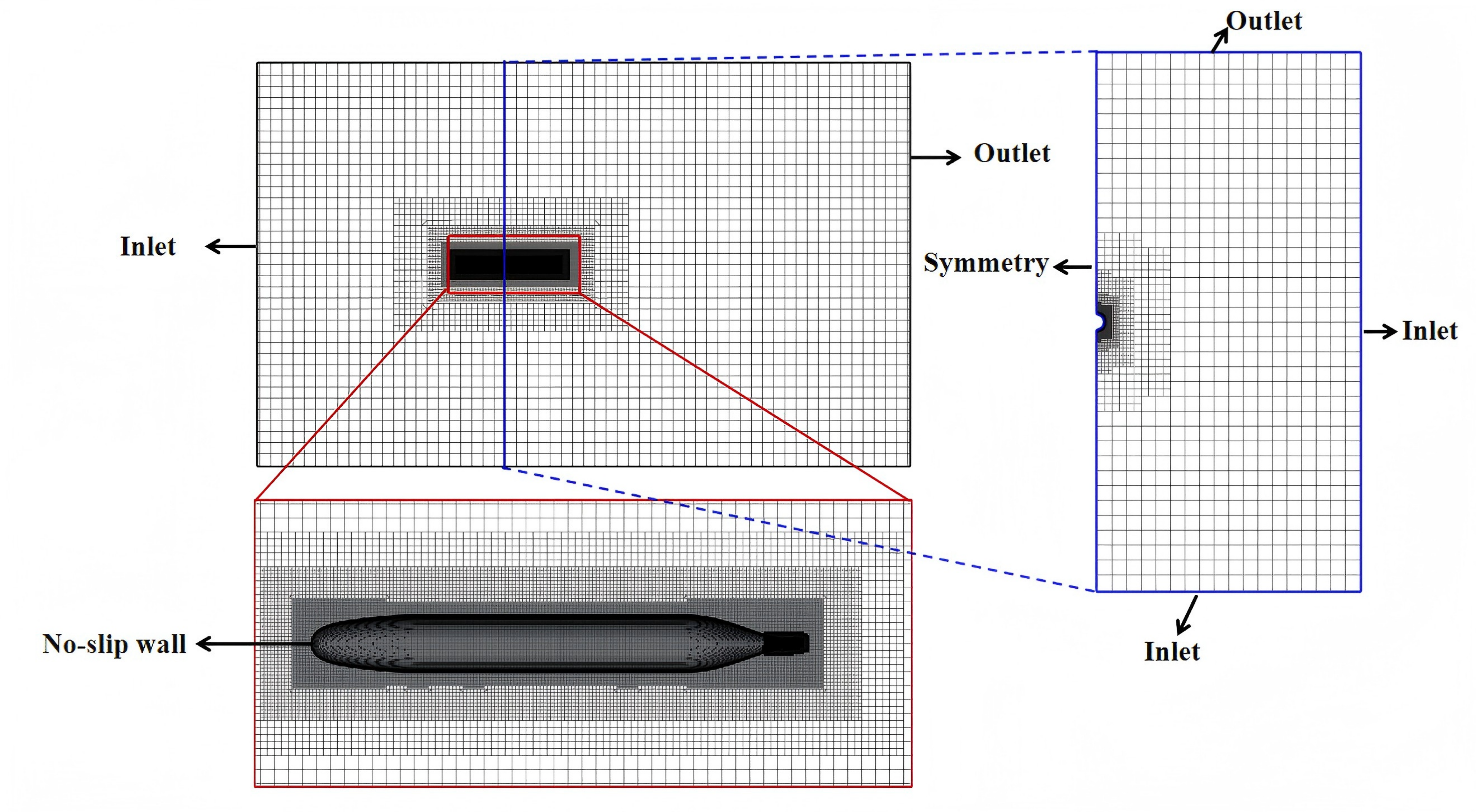}
    \end{subfigure}
\caption{\label{Fig:suboff3} Schematic diagram of the mesh distribution in the computational domain and the setting of boundary conditions for the suboff model.}
\end{figure*}

The boundary conditions of the computational domain are illustrated in Fig.~\ref{Fig:suboff3}. For this hexahedral computational domain, three of its faces are designated as velocity inlets with a static velocity of $\mathbf{V}=\mathbf{V}^{*}$ and two of its faces are configured as pressure outlets with a static pressure $P=0$.  Additionally, one symmetric plane is established, on which the normal component of velocity and the normal gradient of pressure are both equal to zero. Furthermore, the no-slip boundary condition is imposed on the suboff hull surface.

\begin{table*}[htbp]
    \centering
    \small
    \caption{Distribution of training and testing sets in velocity-angle parameter space. Symbols: $\Huge\bullet$ (training set), $T_{\text{i}}$ (testing set), where $i=1,...,5$. Empty cells denote no cases.}
    \label{tab:train_test_distribution}
    \begin{tabular*}{\textwidth}{@{\extracolsep{\fill}} c c c c c c@{}}
        \toprule
        \bfseries Angle ($^\circ$) & \multicolumn{5}{c}{\bfseries Velocity (m/s)} \\
        \cmidrule(lr){2-6} 
        & \bfseries 2.000 & \bfseries 2.850 & \bfseries 3.765 & \bfseries 4.250 & \bfseries 5.000 \\
        \midrule
        \bfseries 3.0 & \Huge$\bullet$ &\Huge$\bullet$ &  & &\Huge$\bullet$ \\
        \bfseries 4.0 & & & & \large$T_{\text{4}}$ & \\
        \bfseries 5.0 &\Huge$\bullet$ & \Huge$\bullet$ &  &\Huge$\bullet$ & \Huge$\bullet$ \\
        \bfseries 6.0 &  & & \large$T_{3}$ &  & \large$T_{5}$\\
        \bfseries 7.0 & \Huge$\bullet$ & &  & \Huge$\bullet$ & \\
        \bfseries 8.0 &  \large$T_{1}$ &\large$T_{2}$ &  & & \\
        \bfseries 9.0 & &  \Huge$\bullet$  &  &\Huge$\bullet$  & \Huge$\bullet$  \\
        \bottomrule
    \end{tabular*}
\end{table*}

In configuring the dataset operating conditions, the magnitude of the inlet velocity  $|\mathbf{V}^{*}|$ and angle of attack $\alpha$ are sampled from the sets $\big\{2.0,2.85,4.25,5.0\big\}$m/s and $\big\{3.0,5.0,7.0,9.0\big\}$$^{\circ}$, respectively. In the selection of test sets, it is considered that variations in the angle of attack tend to have a more pronounced impact on the flow field than variations in the inlet velocity. To enhance the challenge of the test sets, we select one angle not included in the training set for each inlet velocity. Furthermore,  an additional test case is included where both the inlet velocity and angle of attack are not present in the training set, specifically with $|\mathbf{V^{*}}|=3.765$m/s and $\alpha=6^{\circ}$. The aforementioned training set enables a comprehensive evaluation of the DDFI strategy. Following the approach proposed by \citet{zhou2025neural}, the test set also can be classified into three categories: interpolation ($T_{3}$), weak extrapolation ($T_{2},\,T_{4}$), and extrapolation ($T_{1},\,T_{5}$), based on its distributional relationship with the training set. Overall, as presented in Table.~\ref{tab:train_test_distribution}, the dataset comprises 12 flow field operating conditions for the training set and 5 for the test set.
\subsection{Dataset Acquisition and Training Setup}

In this study, the Reynolds-Averaged Navier-Stokes Equations for incompressible fluids \citep{reynolds1895iv} are employed to acquire the turbulent flow field of the suboff under ascending and descending motions, which are expressed as follows:

\begin{equation}\label{conti}
\nabla\cdot(\rho \overline{\mathbf{V}})=0,
\end{equation}

\begin{equation}\label{N_S}
\nabla\cdot\{\rho \overline{\mathbf{V}}\overline{\mathbf{V}} \}=-\nabla\overline{p}+\nabla\cdot(\overline{\bm{\tau}}-\rho\overline{\mathbf{V}^{'}\mathbf{V}^{'}})+\overline{\mathbf{f}_{b}}
\end{equation}

where the overline and the prime symbol in the superscript denote the time-averaged and the time-fluctuating quantity components, respectively, $\{\cdot\}$ denotes the dyadic product of vectors, the time term in Eq.~(\ref{N_S}) is neglected to correspond to the steady-state condition. Accordingly, $\overline{\mathbf{V}}$ and $\mathbf{V}^{'}$ represent the time-averaged quantity and the time-fluctuating parts of velocity vector, $\overline{\bm{\tau}}=\mu(\nabla\overline{\mathbf{V}}+(\nabla\overline{\mathbf{V}})^{T})$ denotes the viscous stress term,  $\mu$ denotes the dynamic viscosity, $\overline{p}$ denotes the time-averaged component of pressure, and $\overline{\mathbf{f}_{b}}$ denotes the time-averaged body force source term. It is noted that $\rho\overline{\mathbf{V}^{'}\mathbf{V}^{'}}$ denotes the Reynolds stress, which characterizes  the effects of turbulence. However, it also introduces additional unknowns, leading to a non-closed system of equations. To achieve closure, the two-equation $k-\epsilon$ turbulence model proposed by \citet{jones1972prediction} is employed, which is based on the Boussinesq assumption \citep{boussinesq1877essai,schlichting1961boundary}.

\begin{figure*}[htbp]
\centering
    \begin{subfigure}[b]{0.9\textwidth}
    \centering
    \includegraphics[width=0.52\textwidth]{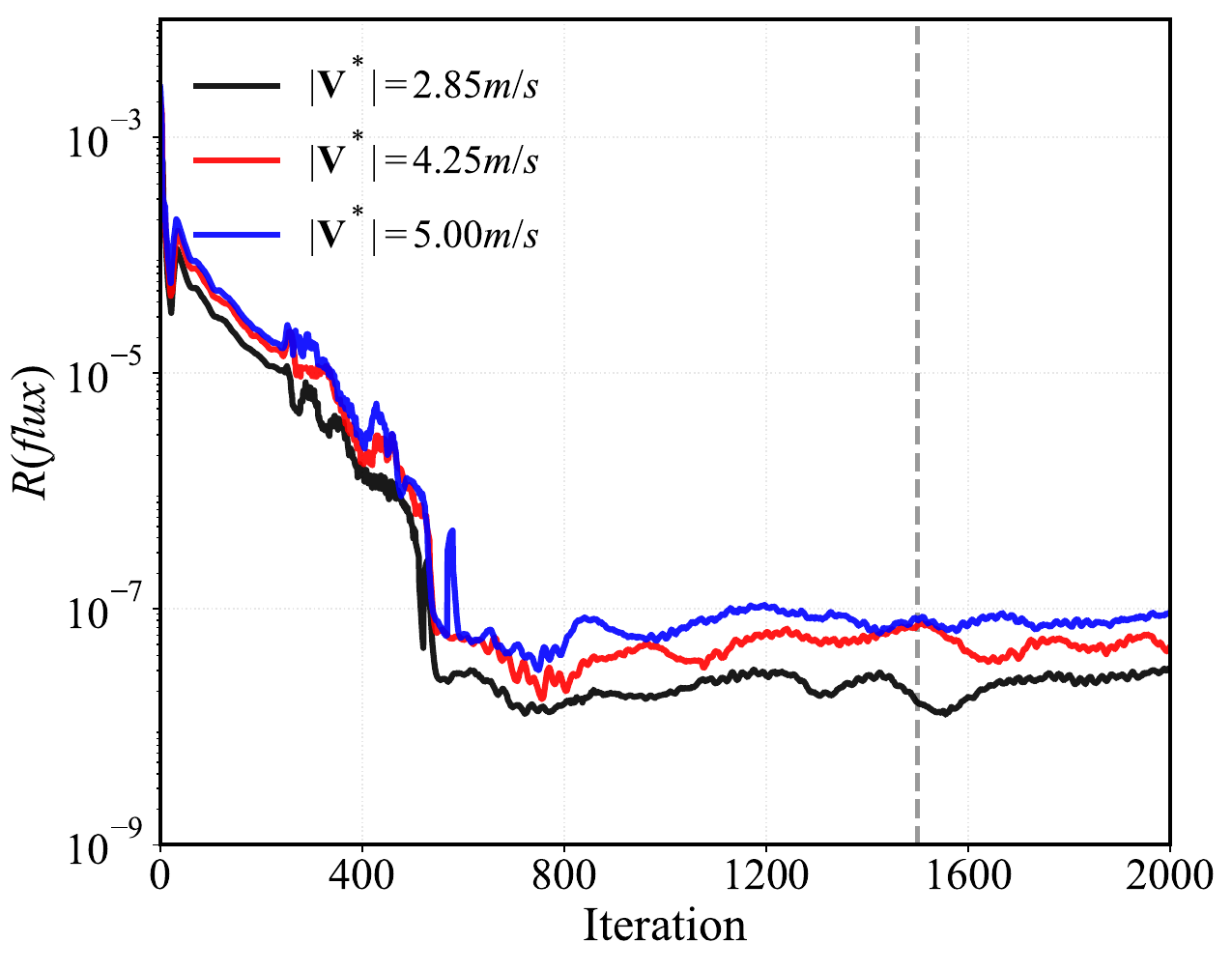}
    \end{subfigure}
\caption{Comparison of flux residual $R(flux)$ convergence curves at three inlet velocities ($|\mathbf{V}^{*}|=2.85,\,4.25,\,5.00$m/s) when the angle of attack $\alpha$ is $9^{\circ}$.}
\label{Fig:angle9_ini} 
\end{figure*}

\begin{figure*}[!t]
\centering
\vspace{0.3cm}
\raisebox{5\height}{\makebox[0.08\textwidth][c]{{\large $U$}}}
\begin{subfigure}[b]{0.30\textwidth}
    \centering
    \includegraphics[width=0.95\textwidth]{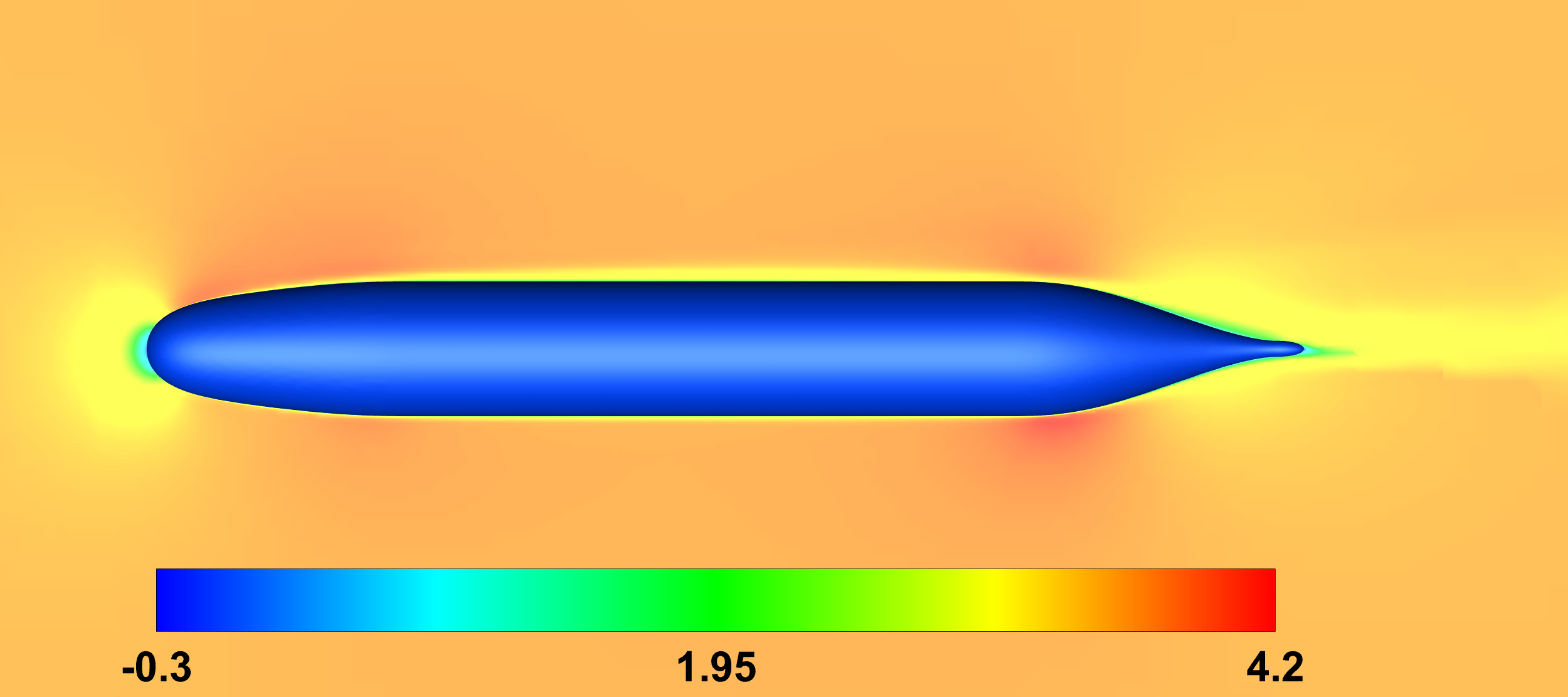}
\end{subfigure}
\hfill
\begin{subfigure}[b]{0.30\textwidth}
    \centering
    \includegraphics[width=0.95\textwidth]{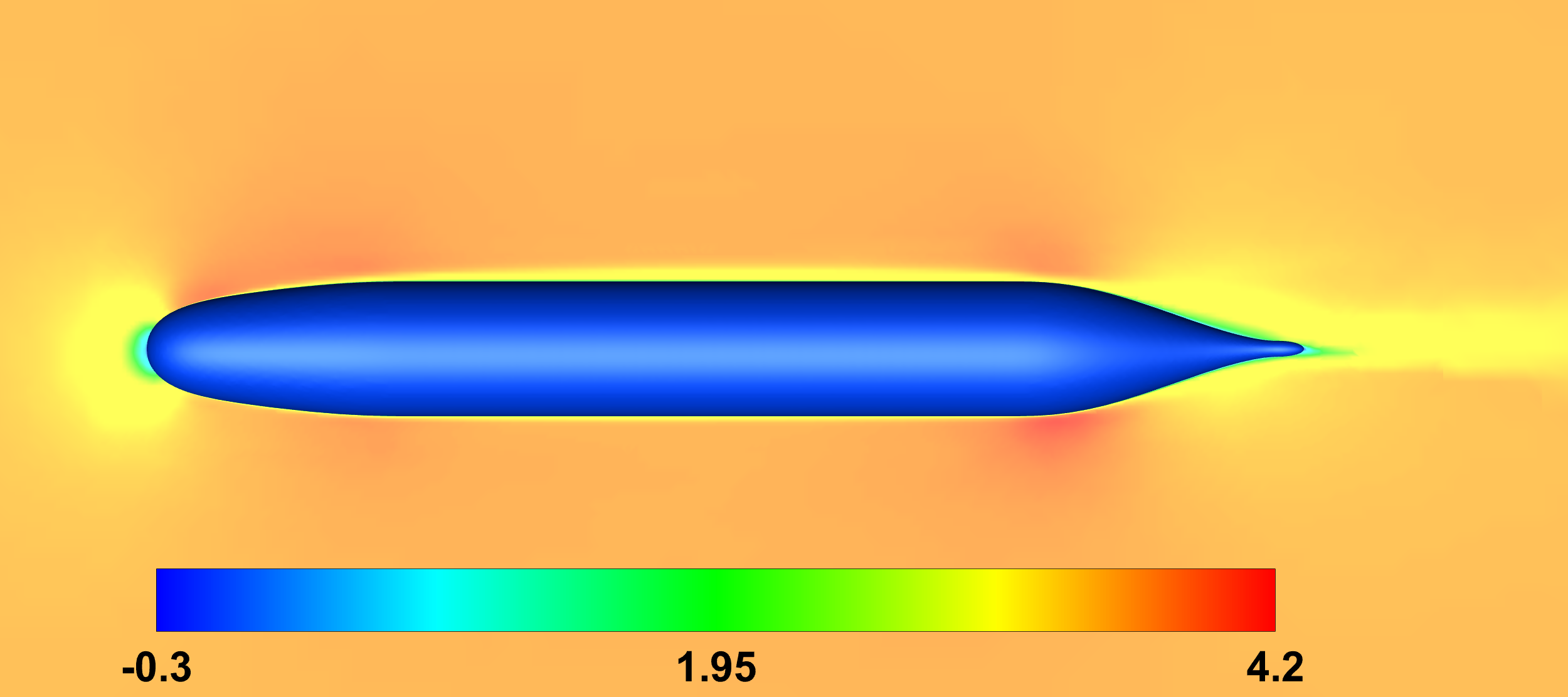}
\end{subfigure}
\hfill
\begin{subfigure}[b]{0.30\textwidth}
    \centering
    \includegraphics[width=0.95\textwidth]{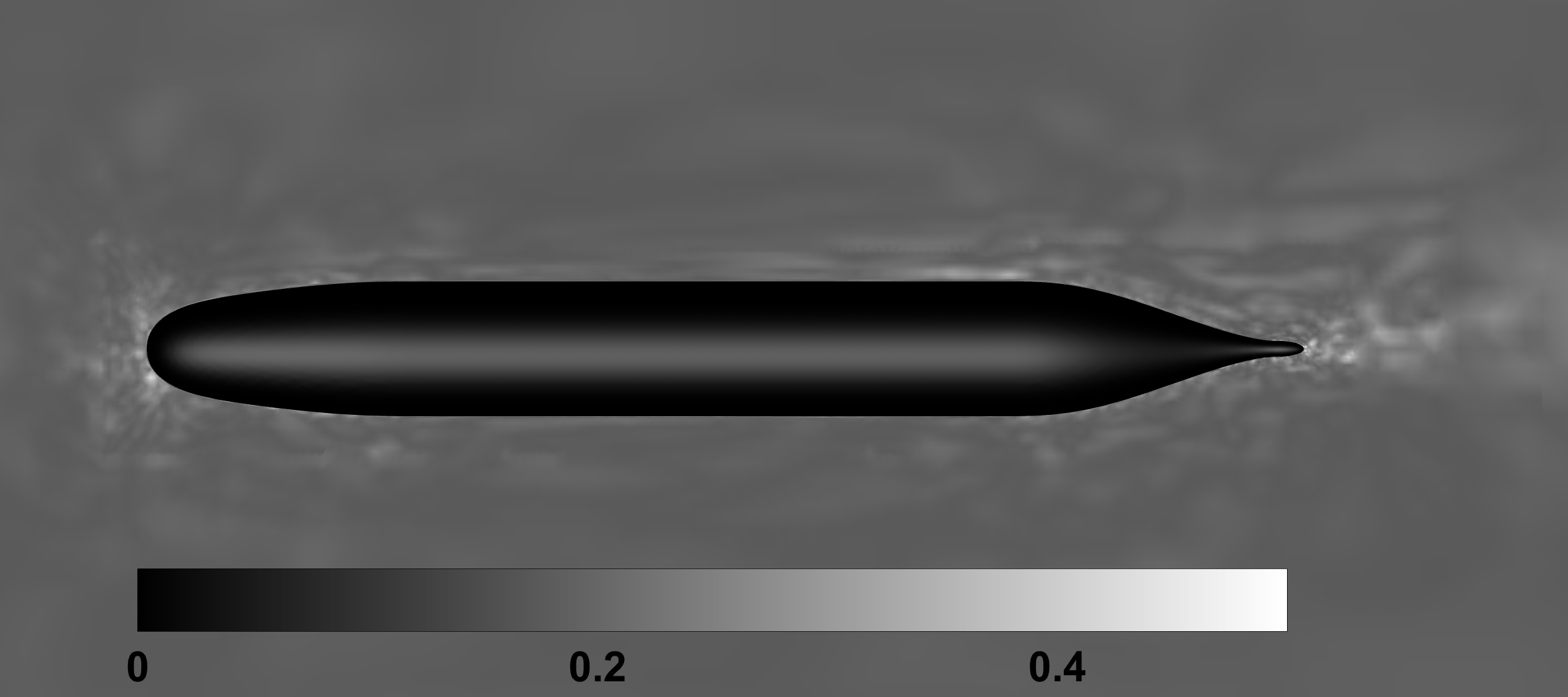}
\end{subfigure}

\vspace{0.3cm}

\raisebox{5\height}{\makebox[0.08\textwidth][c]{{\large $W$}}}
\begin{subfigure}[b]{0.30\textwidth}
    \centering
    \includegraphics[width=0.95\textwidth]{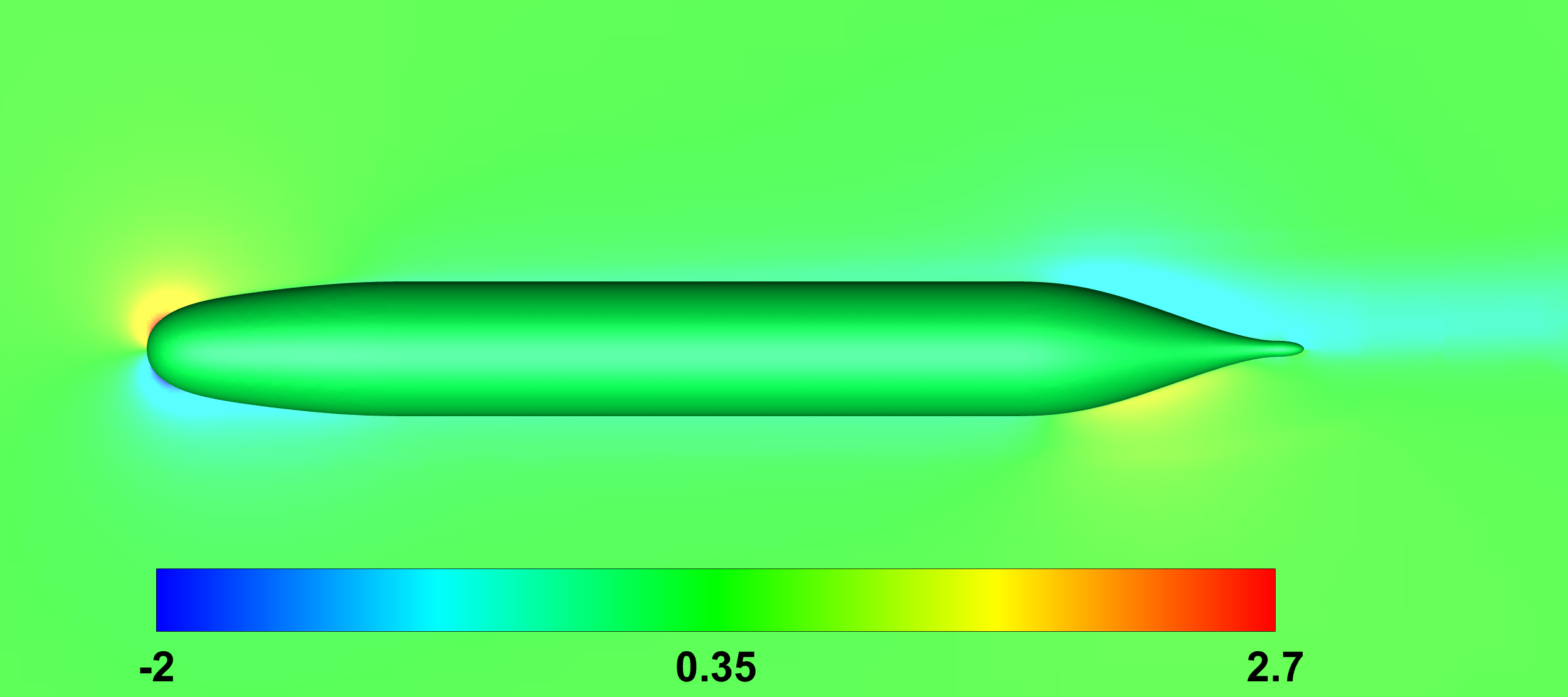}
\end{subfigure}
\hfill
\begin{subfigure}[b]{0.30\textwidth}
    \centering
    \includegraphics[width=0.95\textwidth]{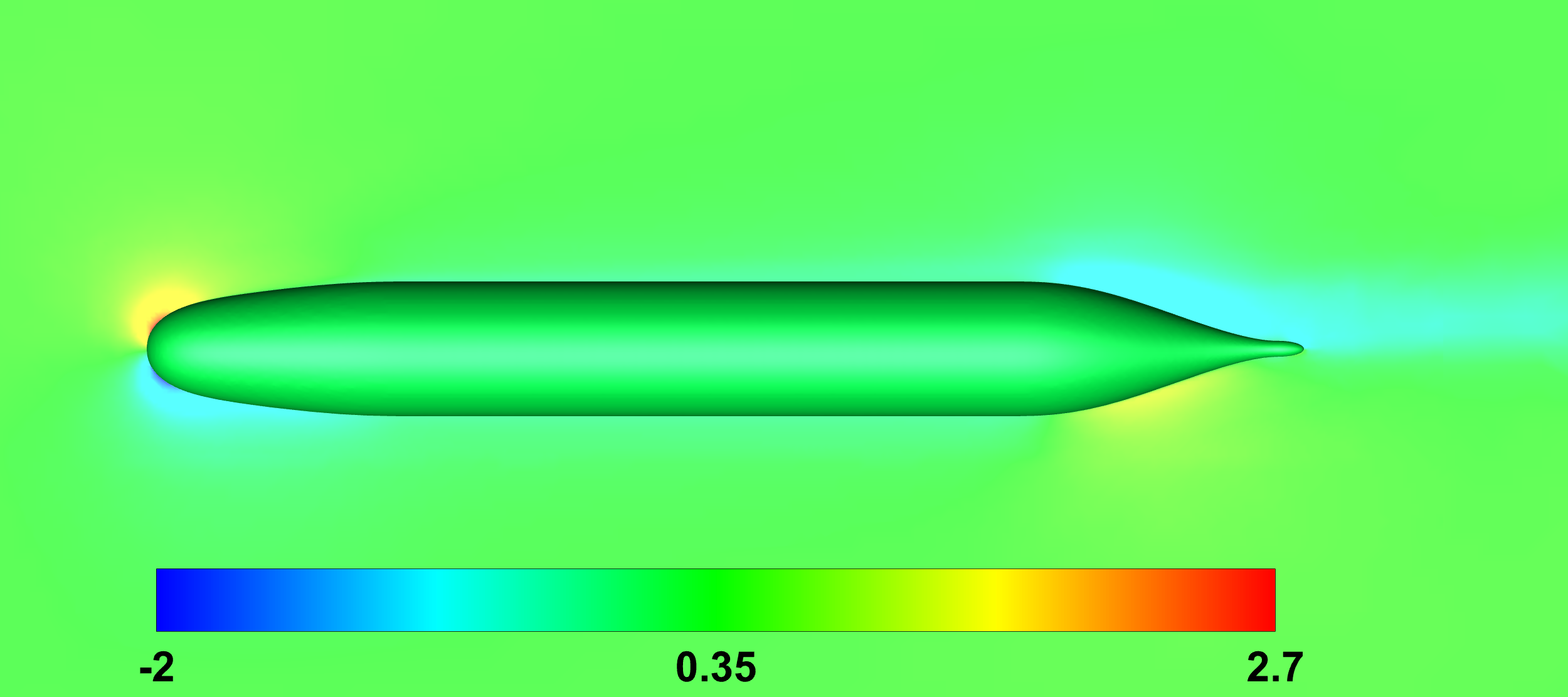}
\end{subfigure}
\hfill
\begin{subfigure}[b]{0.30\textwidth}
    \centering
    \includegraphics[width=0.95\textwidth]{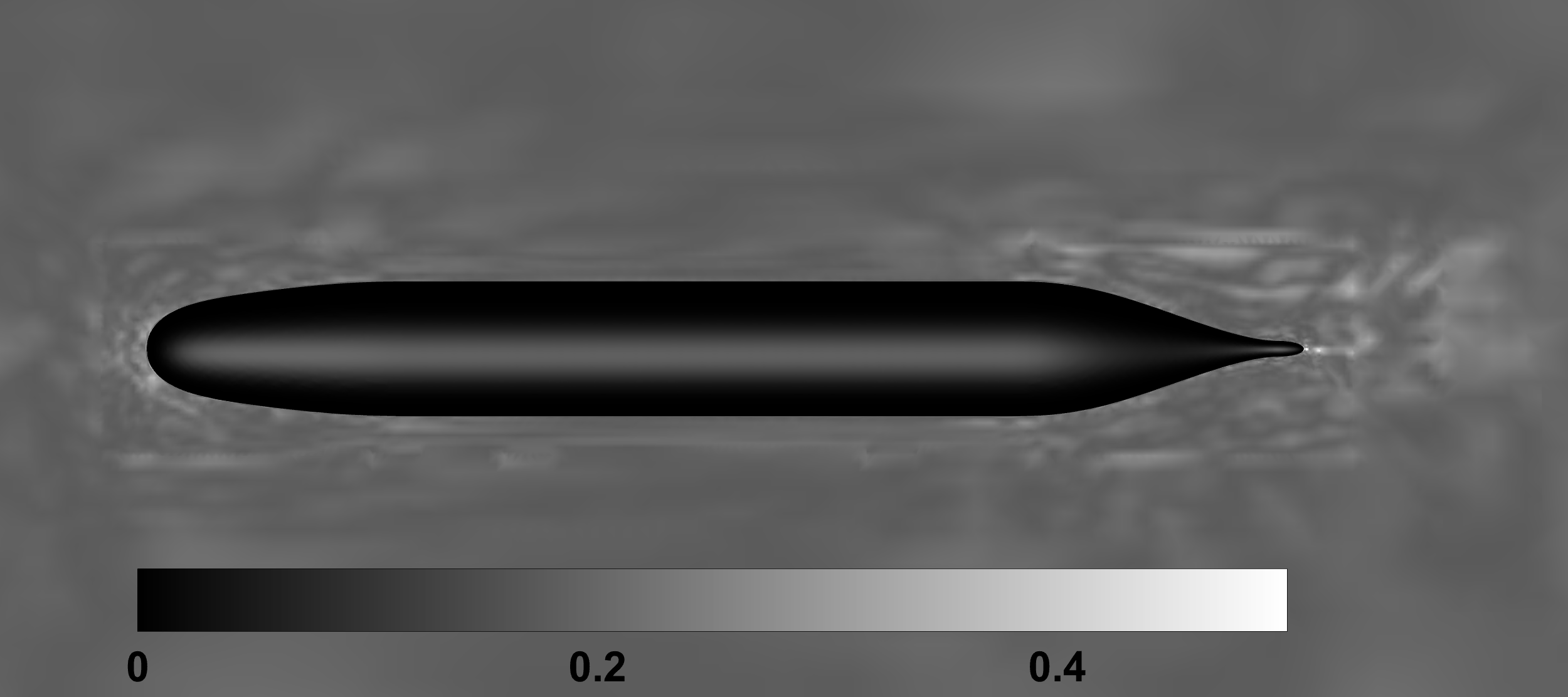}
\end{subfigure}

\vspace{0.3cm}

\raisebox{5\height}{\makebox[0.08\textwidth][c]{{\large $P$}}}
\begin{subfigure}[b]{0.30\textwidth}
    \centering
    \includegraphics[width=0.95\textwidth]{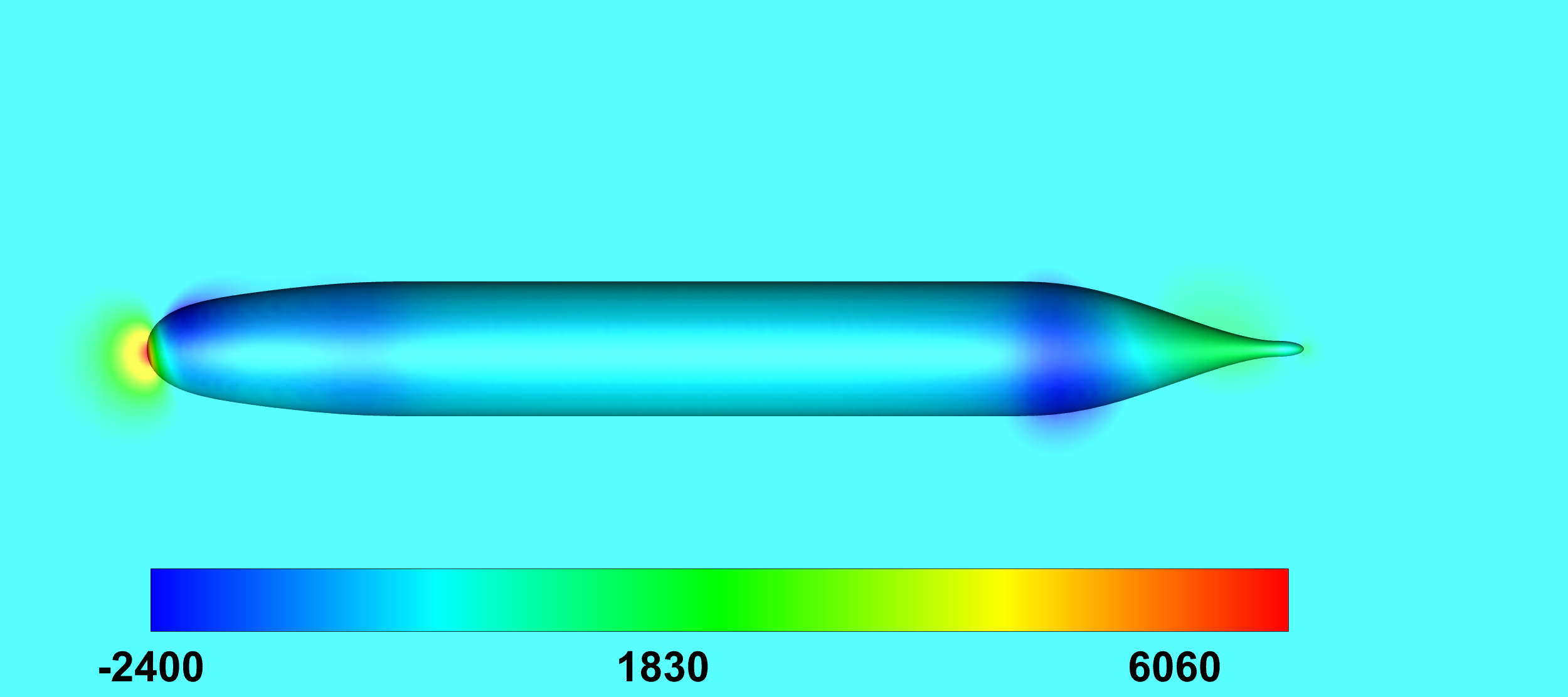}
\end{subfigure}
\hfill
\begin{subfigure}[b]{0.30\textwidth}
    \centering
    \includegraphics[width=0.95\textwidth]{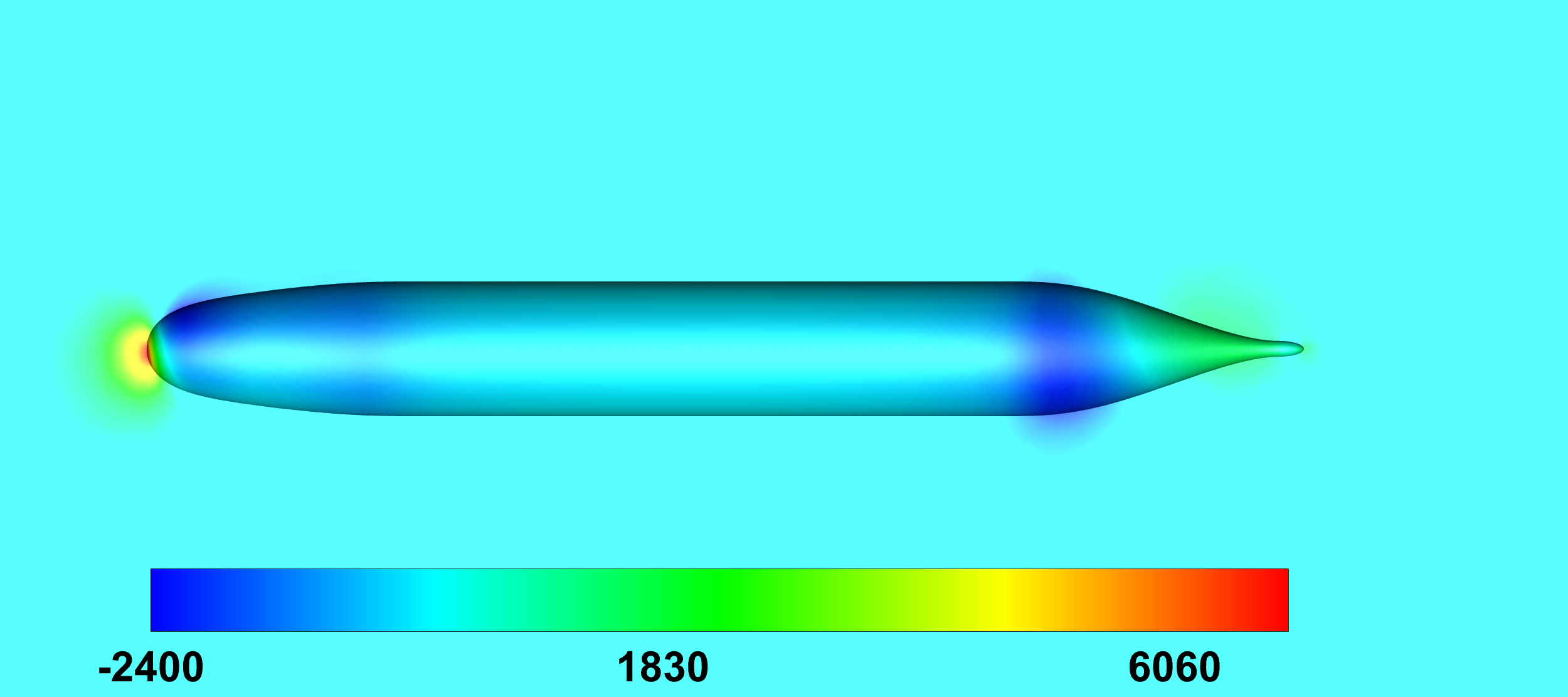}
\end{subfigure}
\hfill
\begin{subfigure}[b]{0.30\textwidth}
    \centering
    \includegraphics[width=0.95\textwidth]{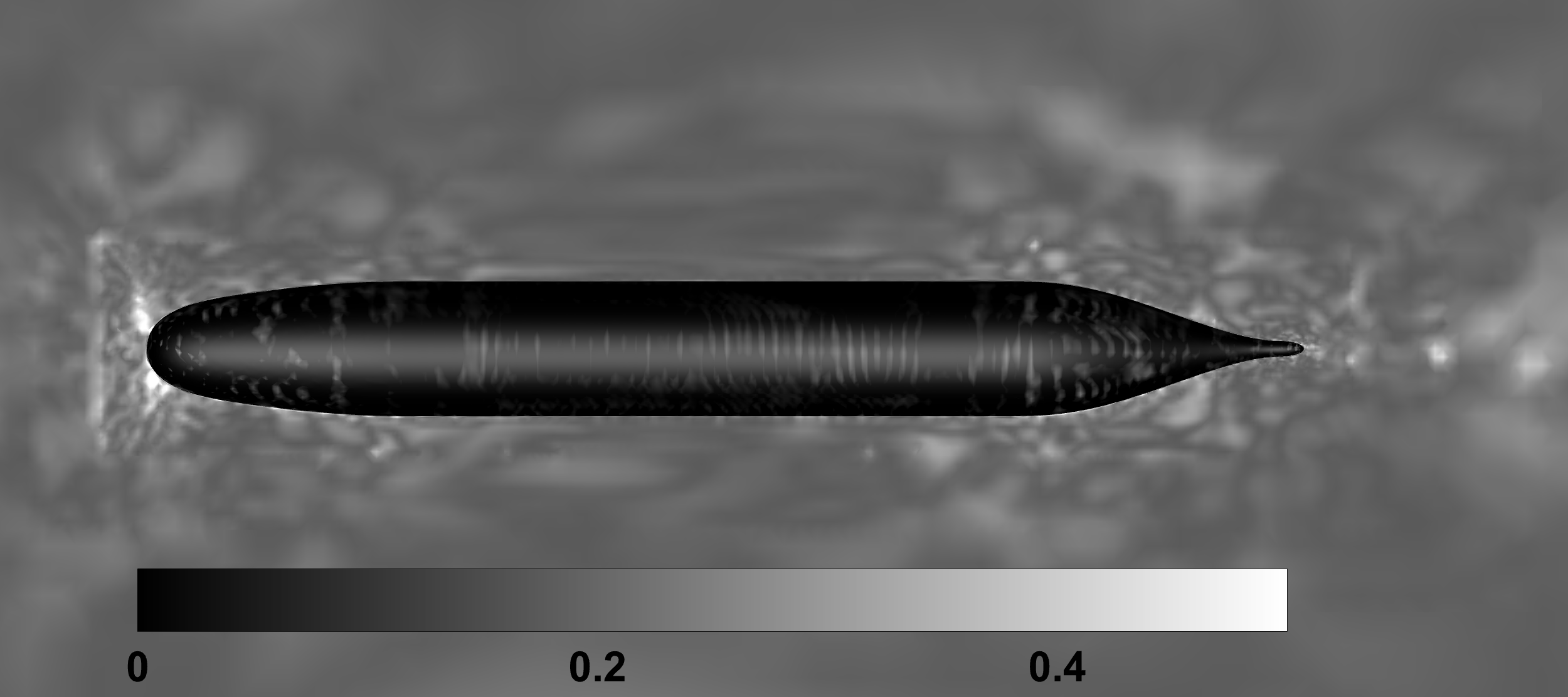}
\end{subfigure}

\vspace{0.3cm}

\raisebox{5\height}{\makebox[0.08\textwidth][c]{{\large $K$}}}
\begin{subfigure}[b]{0.30\textwidth}
    \centering
    \includegraphics[width=0.95\textwidth]{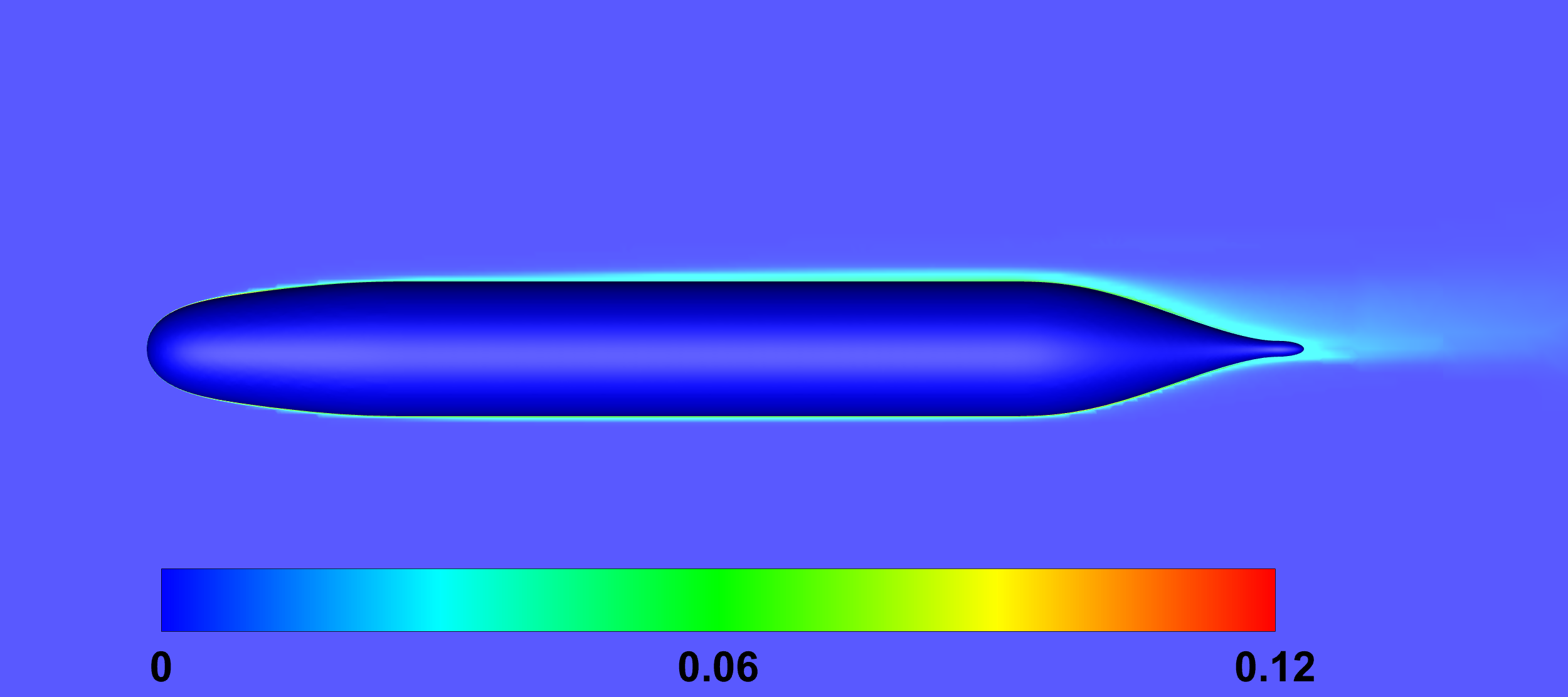}
         \caption*{\large CFD}
\end{subfigure}
\hfill
\begin{subfigure}[b]{0.30\textwidth}
    \centering
    \includegraphics[width=0.95\textwidth]{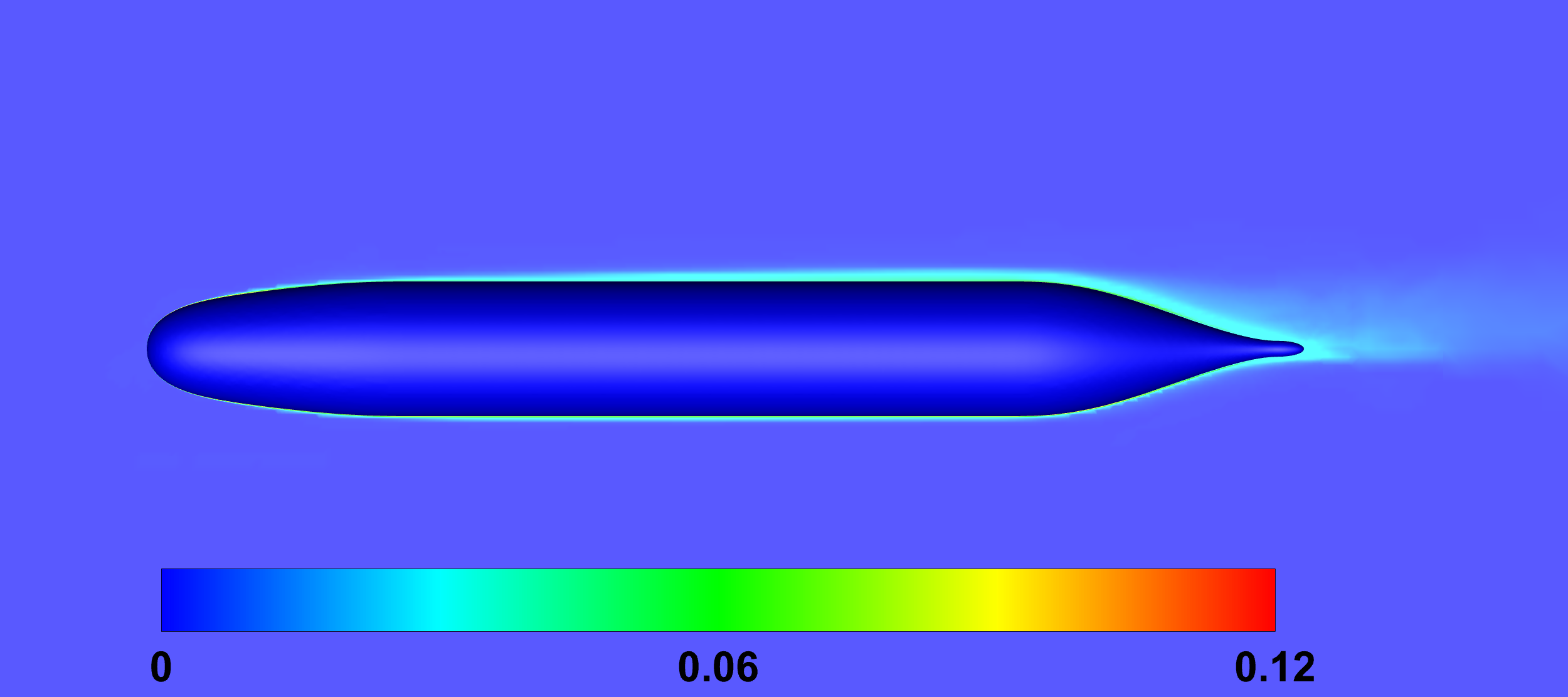}
         \caption*{\large NN-Predicted}
\end{subfigure}
\hfill
\begin{subfigure}[b]{0.30\textwidth}
    \centering
    \includegraphics[width=0.95\textwidth]{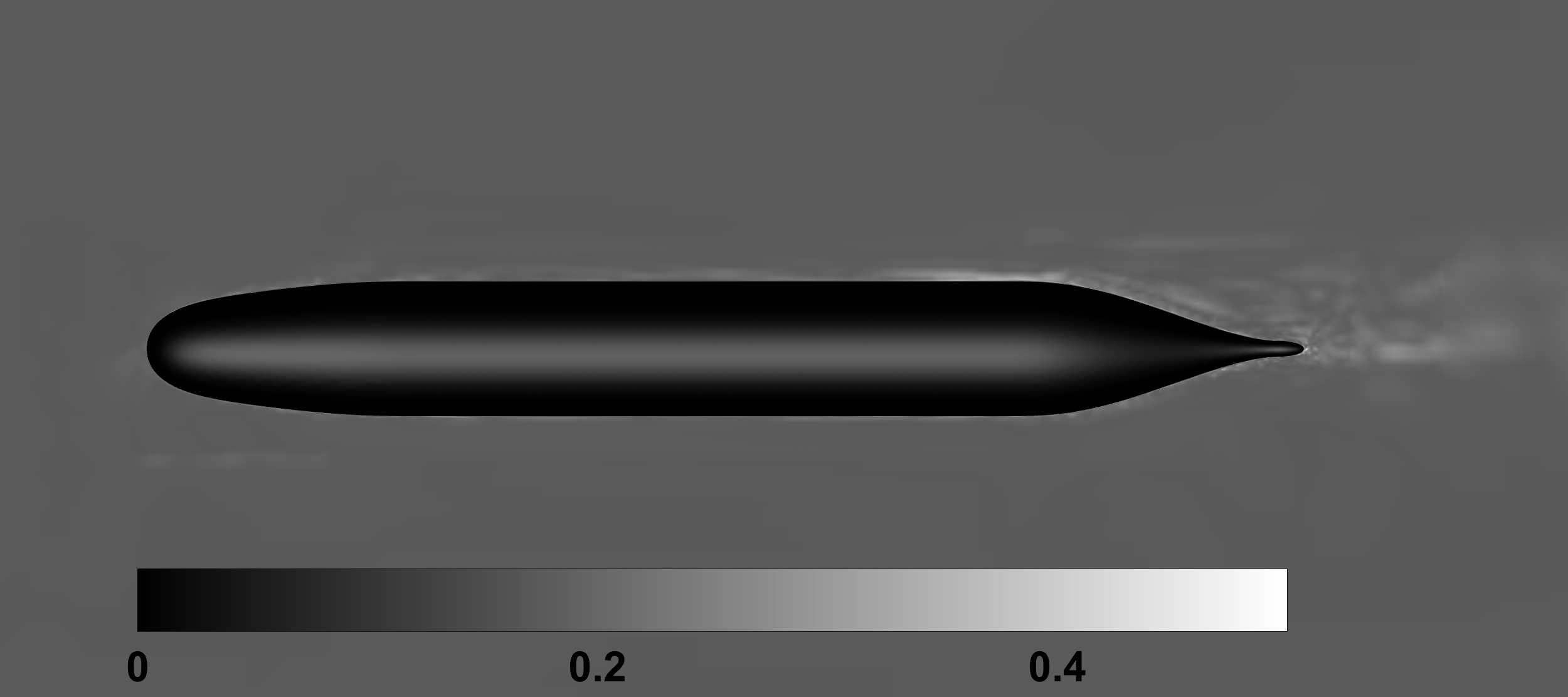}
        \caption*{\large Normalized Error}
\end{subfigure}

\caption{\label{Fig:v3.765} Comparison diagram of the near-wall flow field distribution of suboff model under test case $T_{3}$($|\mathbf{V}^{*}|=3.765$m/s,$\,\alpha=6^{\circ}$).}
\end{figure*}

All calculations in this paper are performed  using a single core with the  steady-state solver of the self-developed incompressible viscous flow software MarineFlow, with spatial discretization implemented based on the finite volume method (FVM) and the velocity-pressure equations solved in a decoupled manner via the semi-implicit method for pressure linked equations (SAMPLE) algorithm. As shown in Fig.~\ref{Fig:suboff2}, the total number of meshes in the computational domain is $4.45\times10^{5}$. The fluid density $\rho$ and dynamic viscosity $\mu$ in all simulations are set to $999.1$kg/m$^{3}$ and $0.001145N\cdot$s/m$^{2}$, which are consistent with the experimental conditions. It is worth noting that while operating condition configurations are referenced to experimental data, the present study takes conventional CFD results as the benchmark and focuses primarily on comparing the efficiency enhancement of the DDFI strategy relative to traditional CFD approaches.

\begin{figure*}[!h]
\centering
\vspace{0.3cm}
\raisebox{5\height}{\makebox[0.08\textwidth][c]{{\large $U$}}}
\begin{subfigure}[b]{0.30\textwidth}
    \centering
    \includegraphics[width=0.95\textwidth]{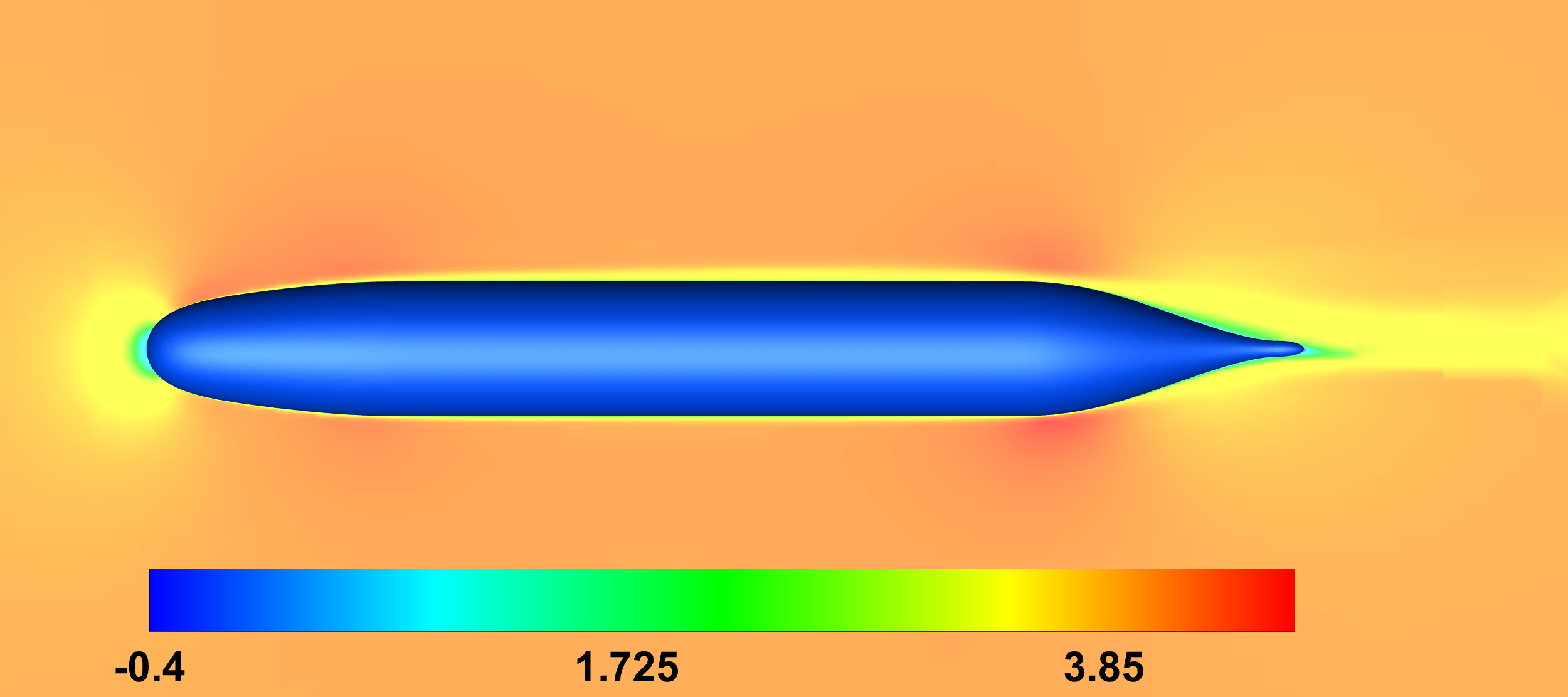}
\end{subfigure}
\hfill
\begin{subfigure}[b]{0.30\textwidth}
    \centering
    \includegraphics[width=0.95\textwidth]{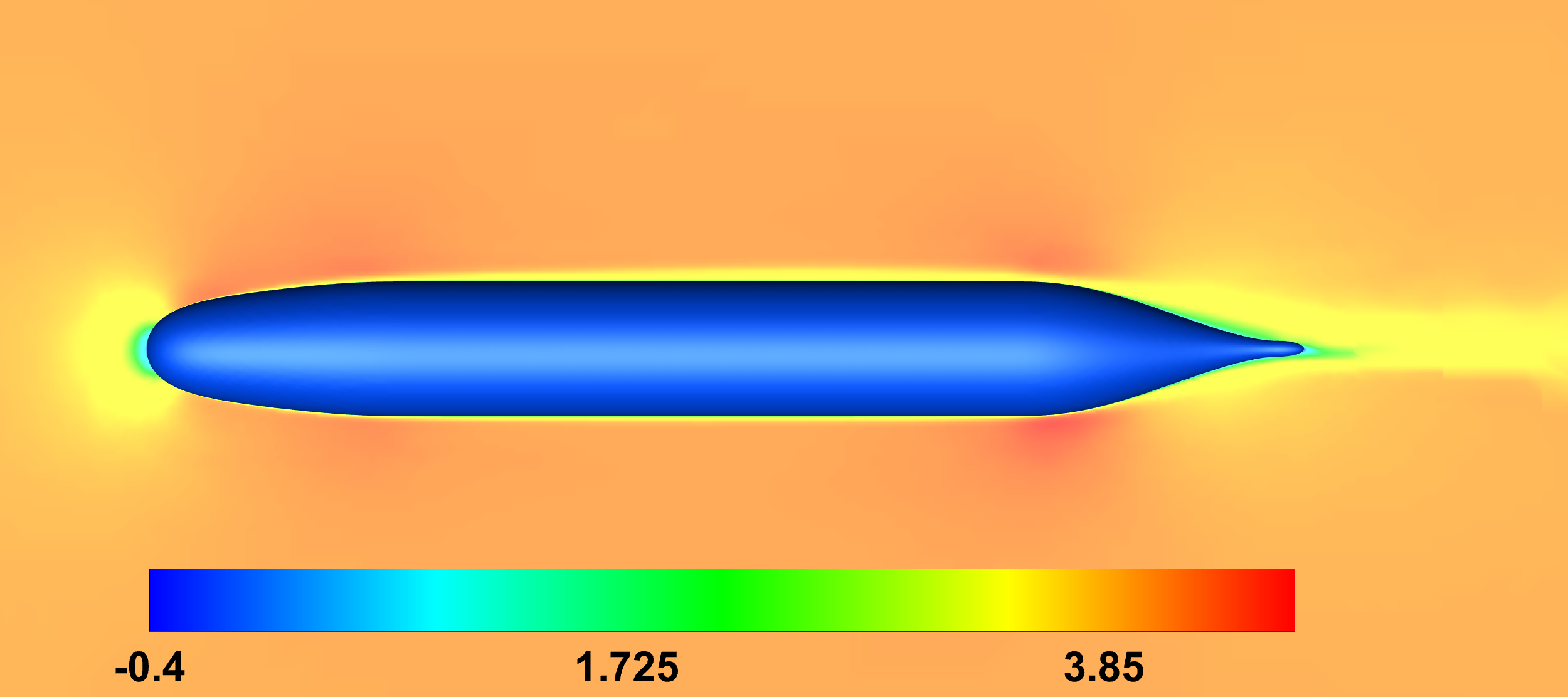}
\end{subfigure}
\hfill
\begin{subfigure}[b]{0.30\textwidth}
    \centering
    \includegraphics[width=0.95\textwidth]{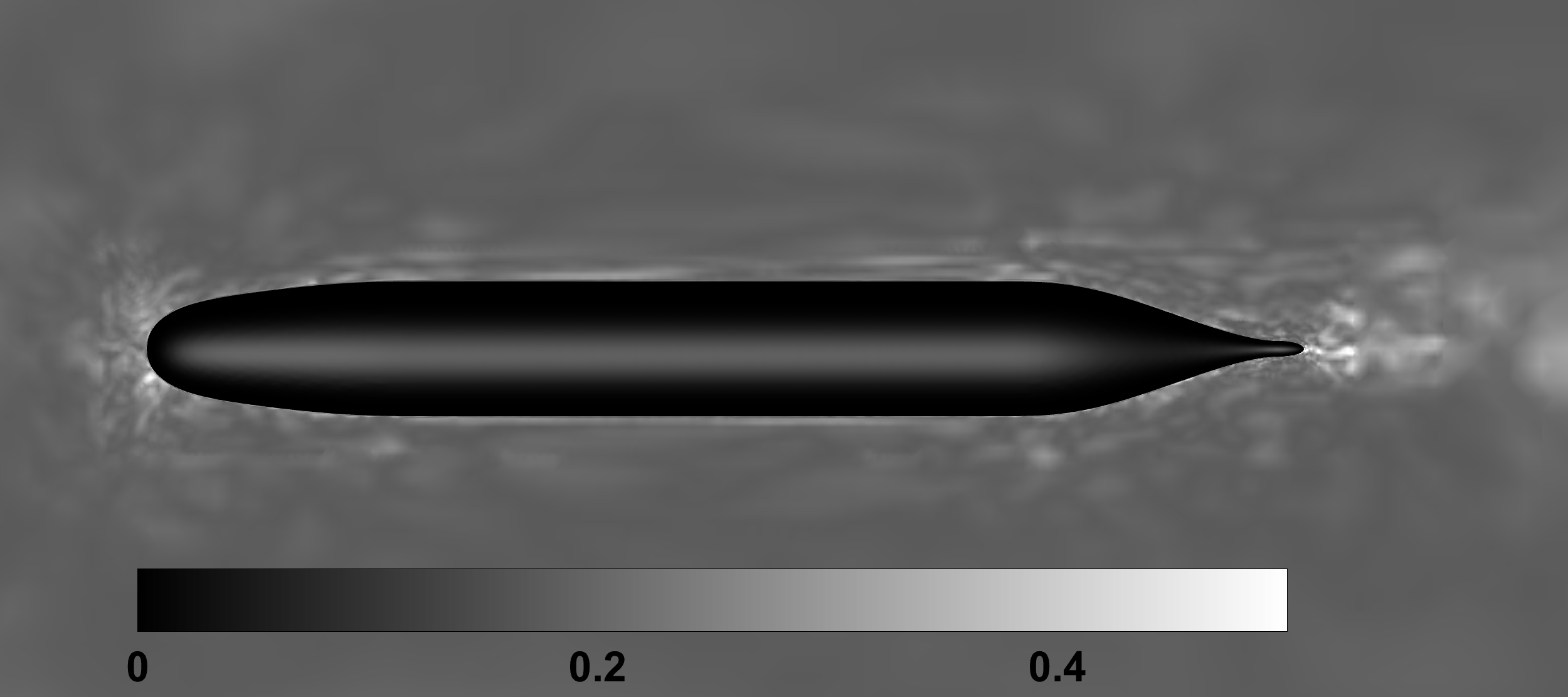}
\end{subfigure}

\vspace{0.3cm}

\raisebox{5\height}{\makebox[0.08\textwidth][c]{{\large $W$}}}
\begin{subfigure}[b]{0.30\textwidth}
    \centering
    \includegraphics[width=0.95\textwidth]{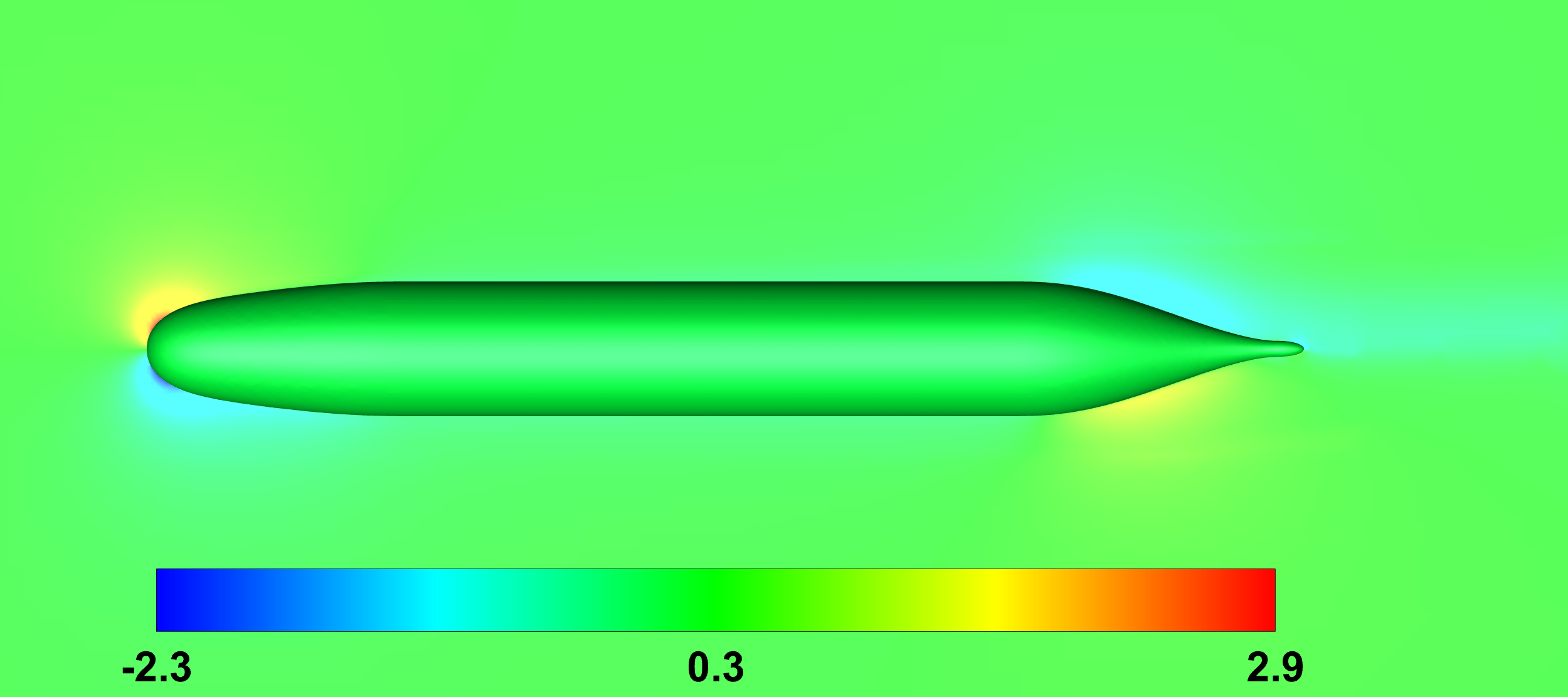}
\end{subfigure}
\hfill
\begin{subfigure}[b]{0.30\textwidth}
    \centering
    \includegraphics[width=0.95\textwidth]{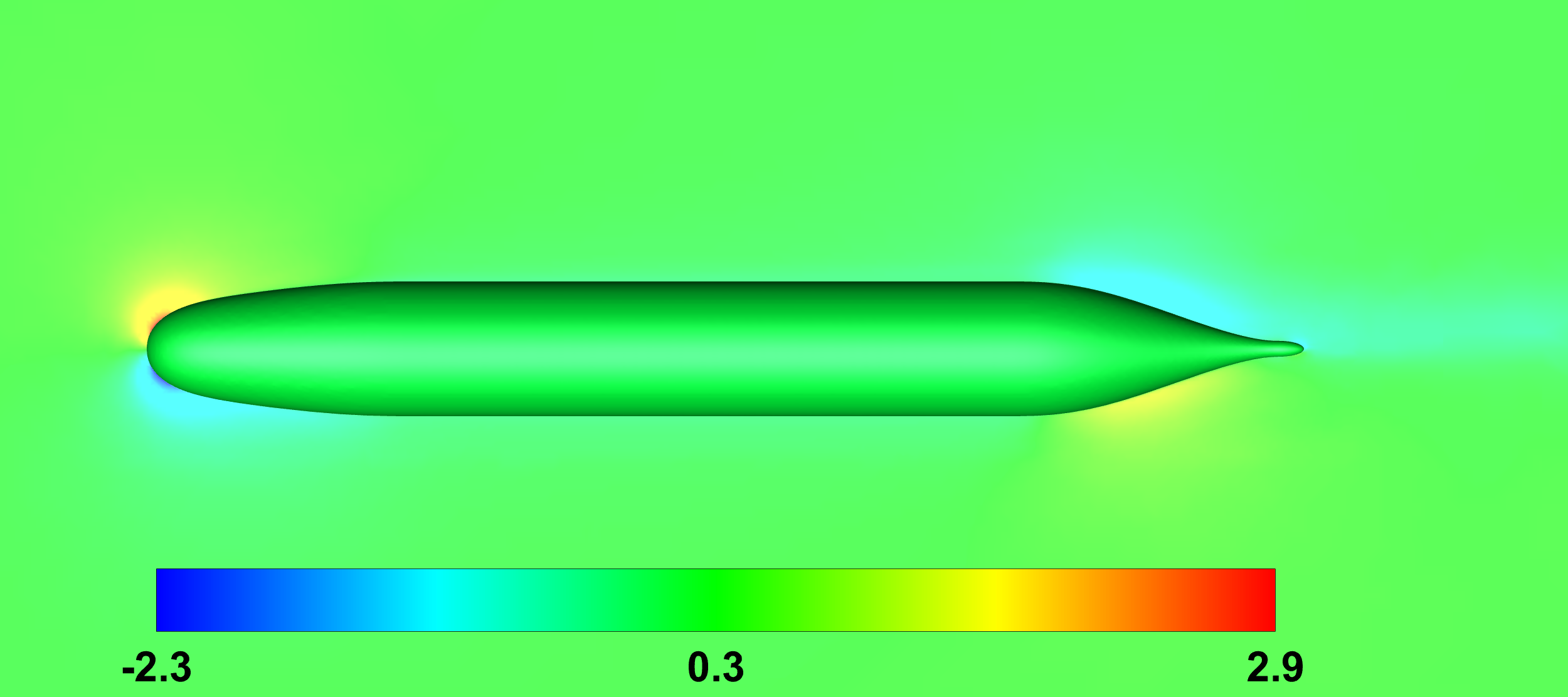}
\end{subfigure}
\hfill
\begin{subfigure}[b]{0.30\textwidth}
    \centering
    \includegraphics[width=0.95\textwidth]{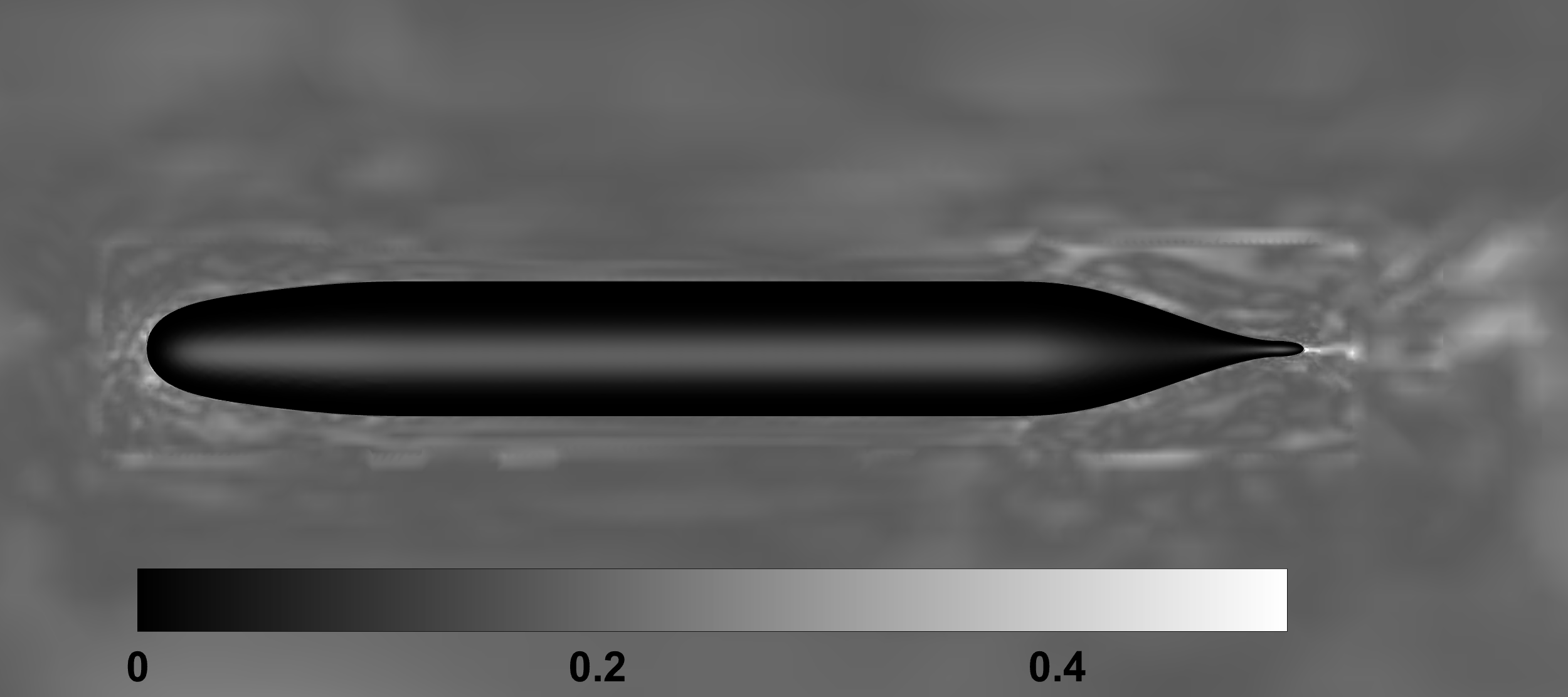}
\end{subfigure}

\vspace{0.3cm}

\raisebox{5\height}{\makebox[0.08\textwidth][c]{{\large $P$}}}
\begin{subfigure}[b]{0.30\textwidth}
    \centering
    \includegraphics[width=0.95\textwidth]{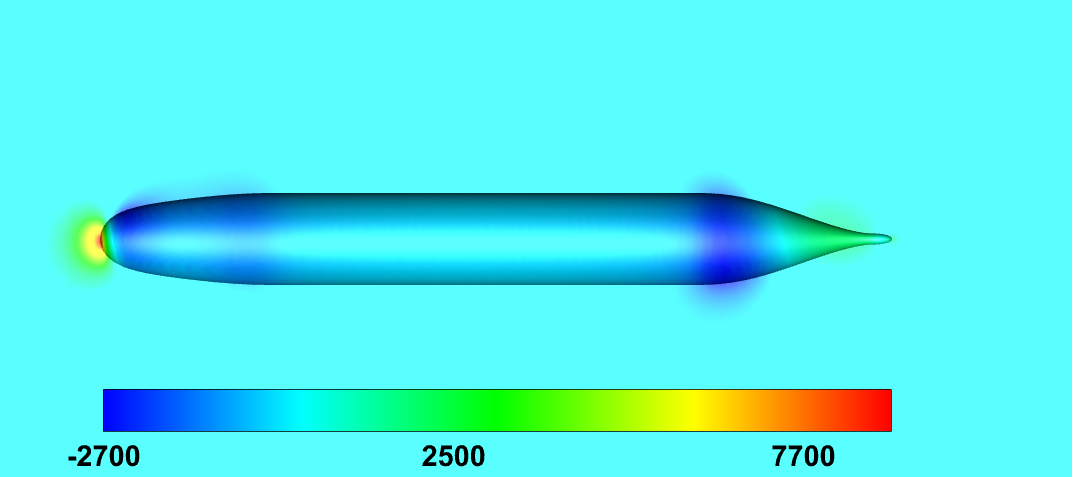}
\end{subfigure}
\hfill
\begin{subfigure}[b]{0.30\textwidth}
    \centering
    \includegraphics[width=0.95\textwidth]{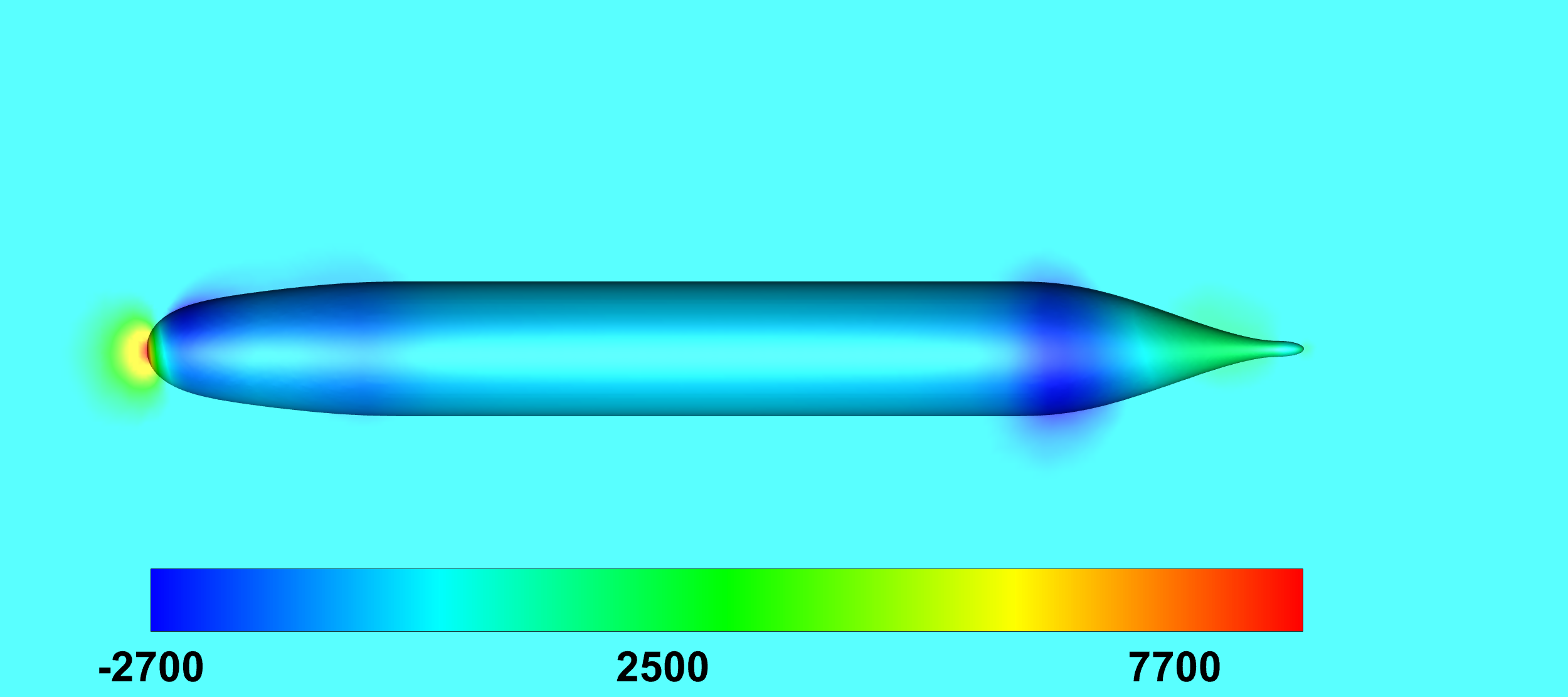}
\end{subfigure}
\hfill
\begin{subfigure}[b]{0.30\textwidth}
    \centering
    \includegraphics[width=0.95\textwidth]{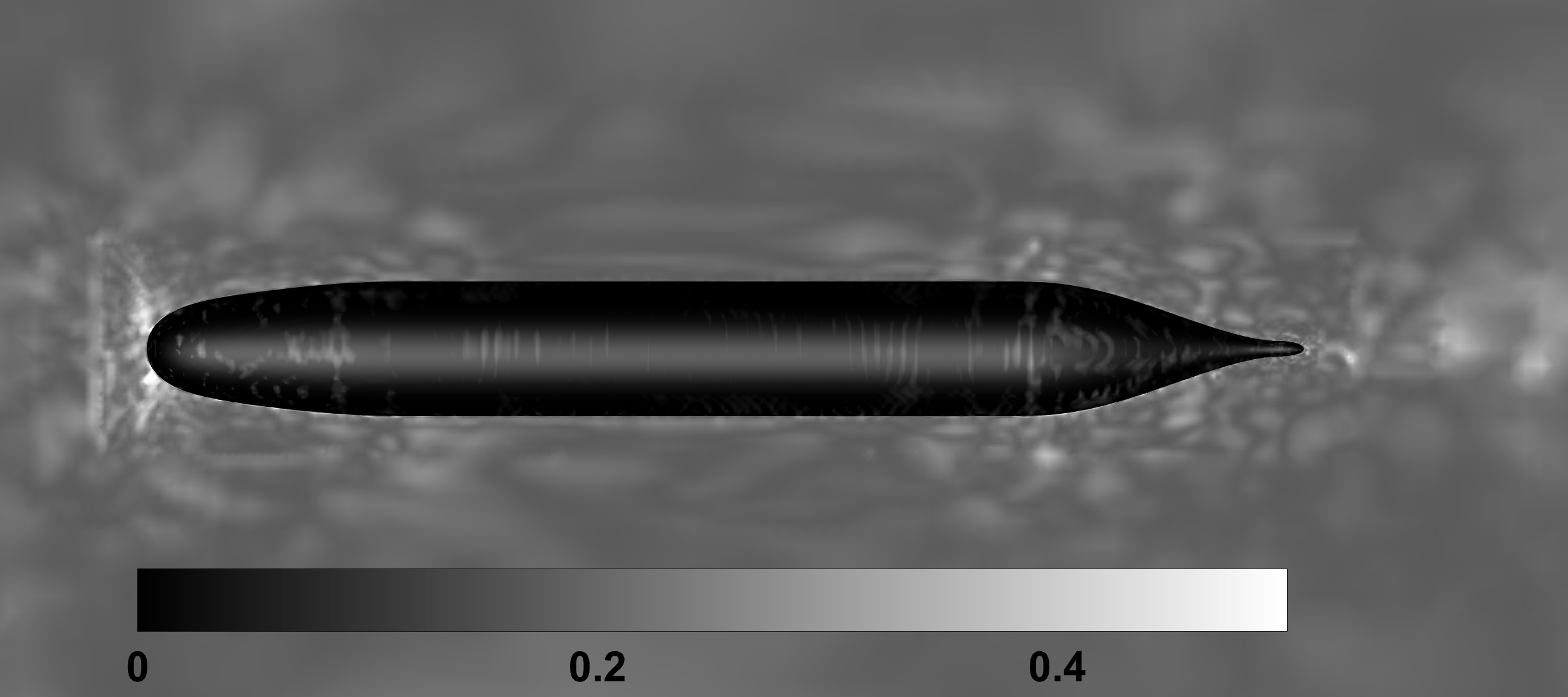}
\end{subfigure}

\vspace{0.3cm}
\raisebox{5\height}{\makebox[0.08\textwidth][c]{{\large $K$}}}
\begin{subfigure}[b]{0.30\textwidth}
    \centering
    \includegraphics[width=0.95\textwidth]{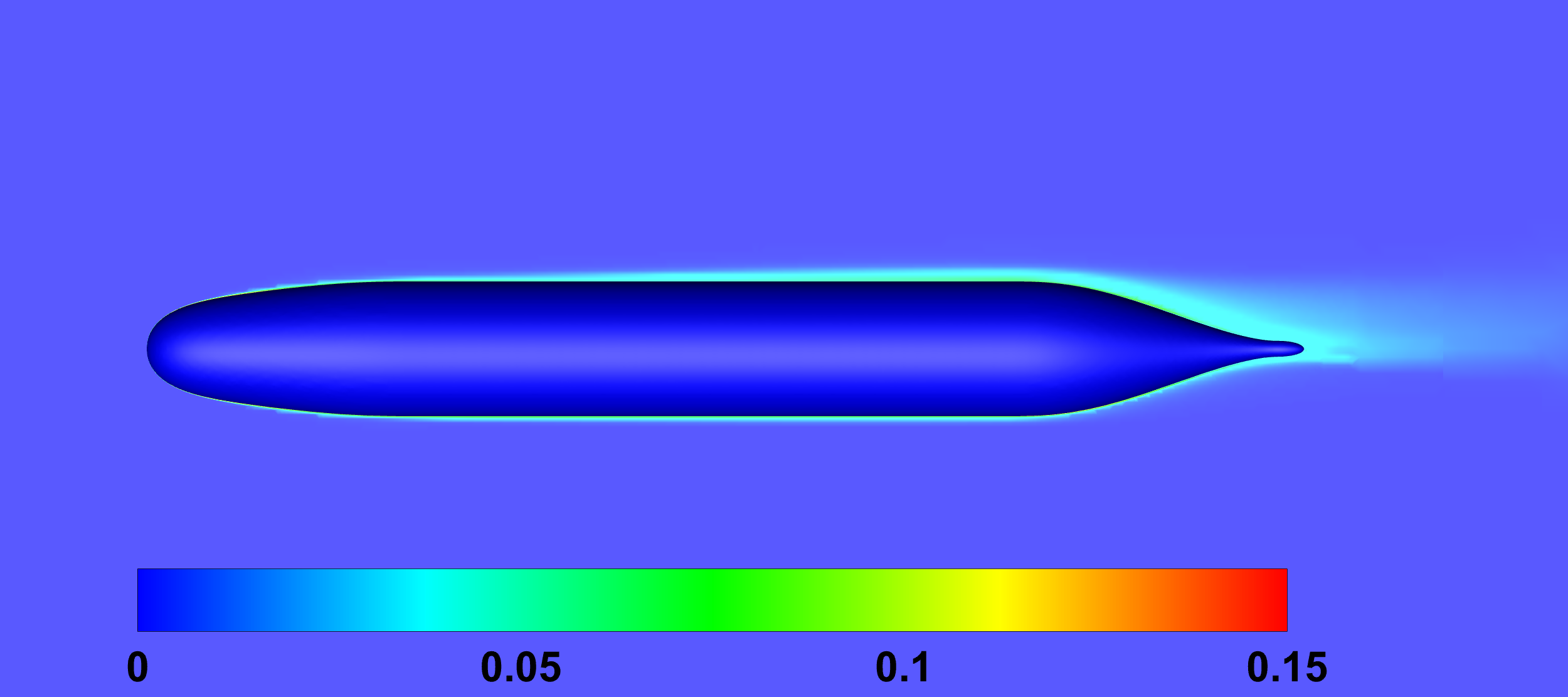}
         \caption*{\large CFD}
\end{subfigure}
\hfill
\begin{subfigure}[b]{0.30\textwidth}
    \centering
    \includegraphics[width=0.95\textwidth]{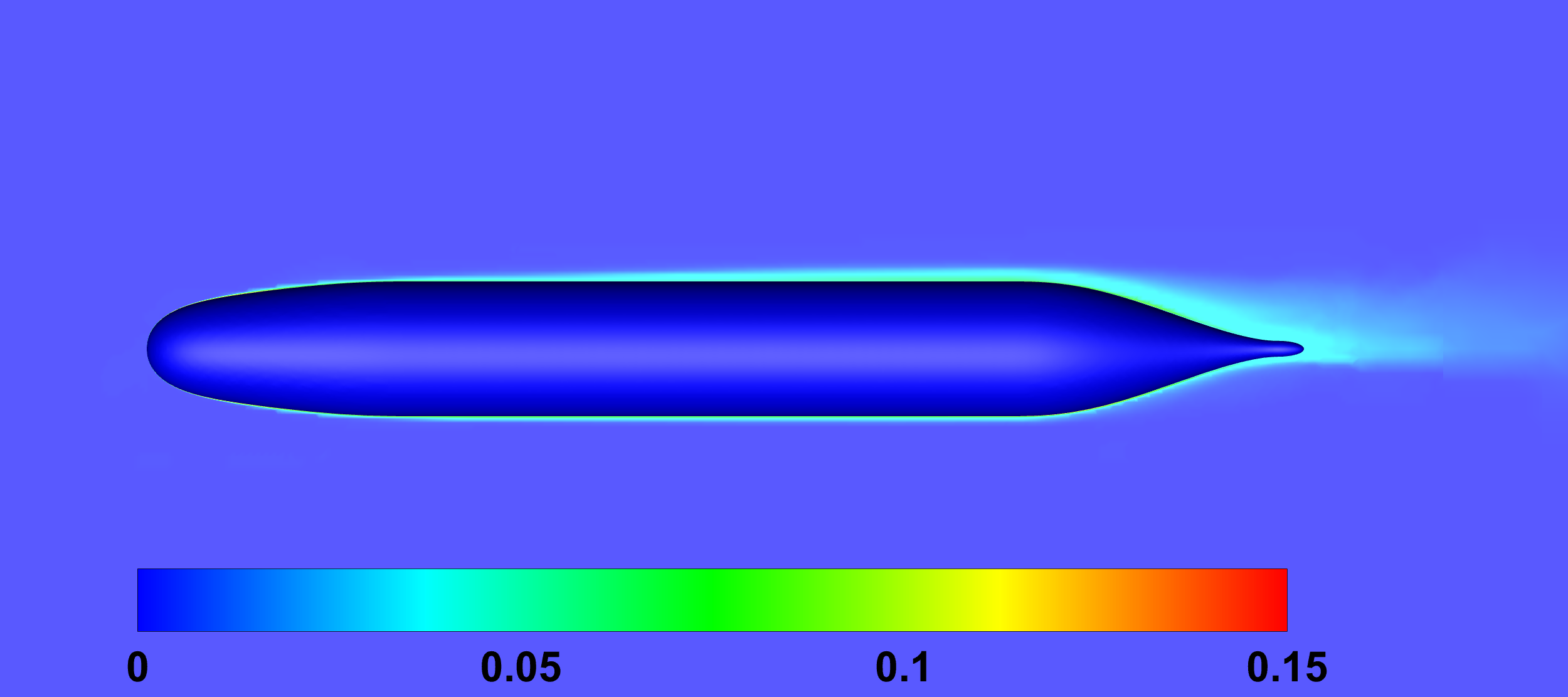}
         \caption*{\large NN-Predicted}
\end{subfigure}
\hfill
\begin{subfigure}[b]{0.30\textwidth}
    \centering
    \includegraphics[width=0.95\textwidth]{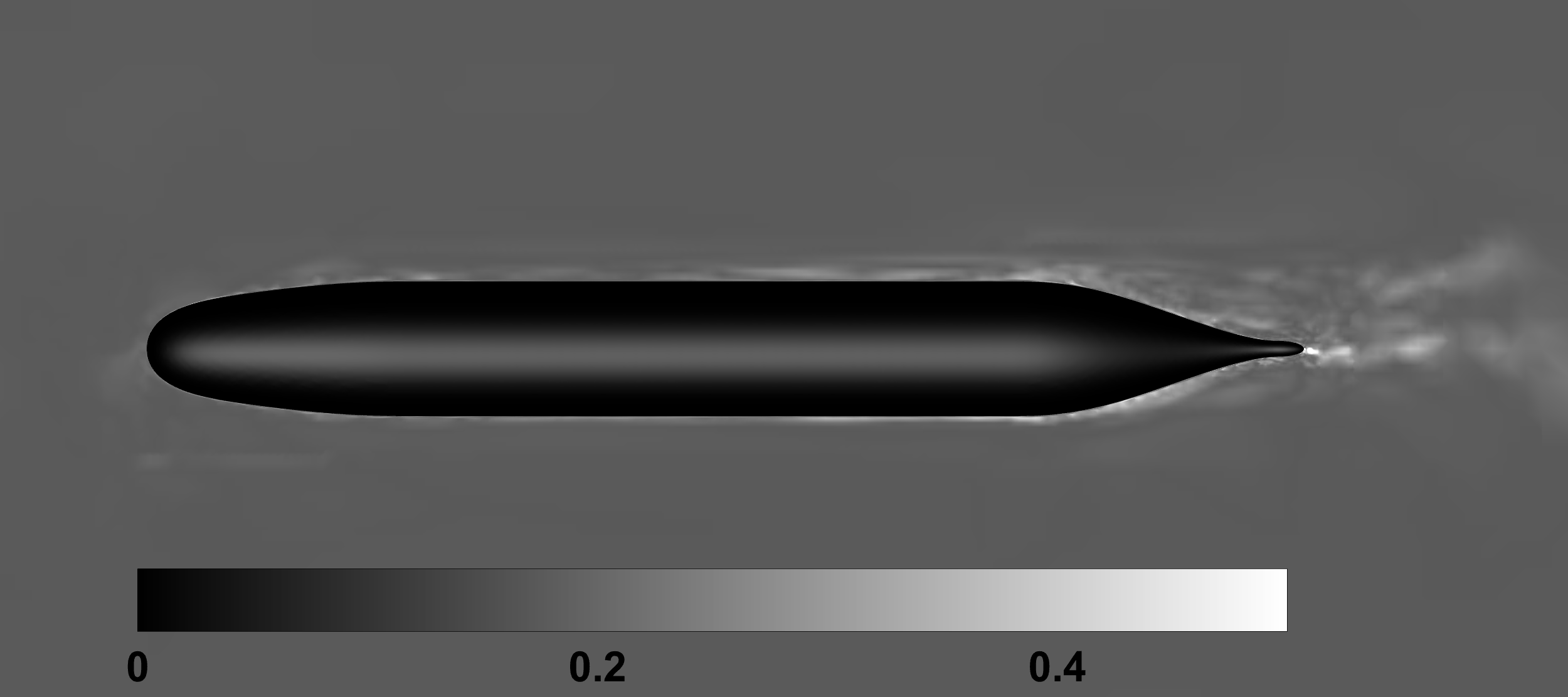}
        \caption*{\large Normalized Error}
\end{subfigure}

\caption{\label{Fig:v4.25} Comparison diagram of the near-wall flow field distribution of suboff model under test case $T_{4}$($|\mathbf{V}^{*}|=4.25$m/s,$\,\alpha=4^{\circ}$).}
\end{figure*}

\begin{figure*}[!hh]
\centering
\vspace{0.3cm}
\raisebox{5\height}{\makebox[0.08\textwidth][c]{{\large $U$}}}
\begin{subfigure}[b]{0.30\textwidth}
    \centering
    \includegraphics[width=0.95\textwidth]{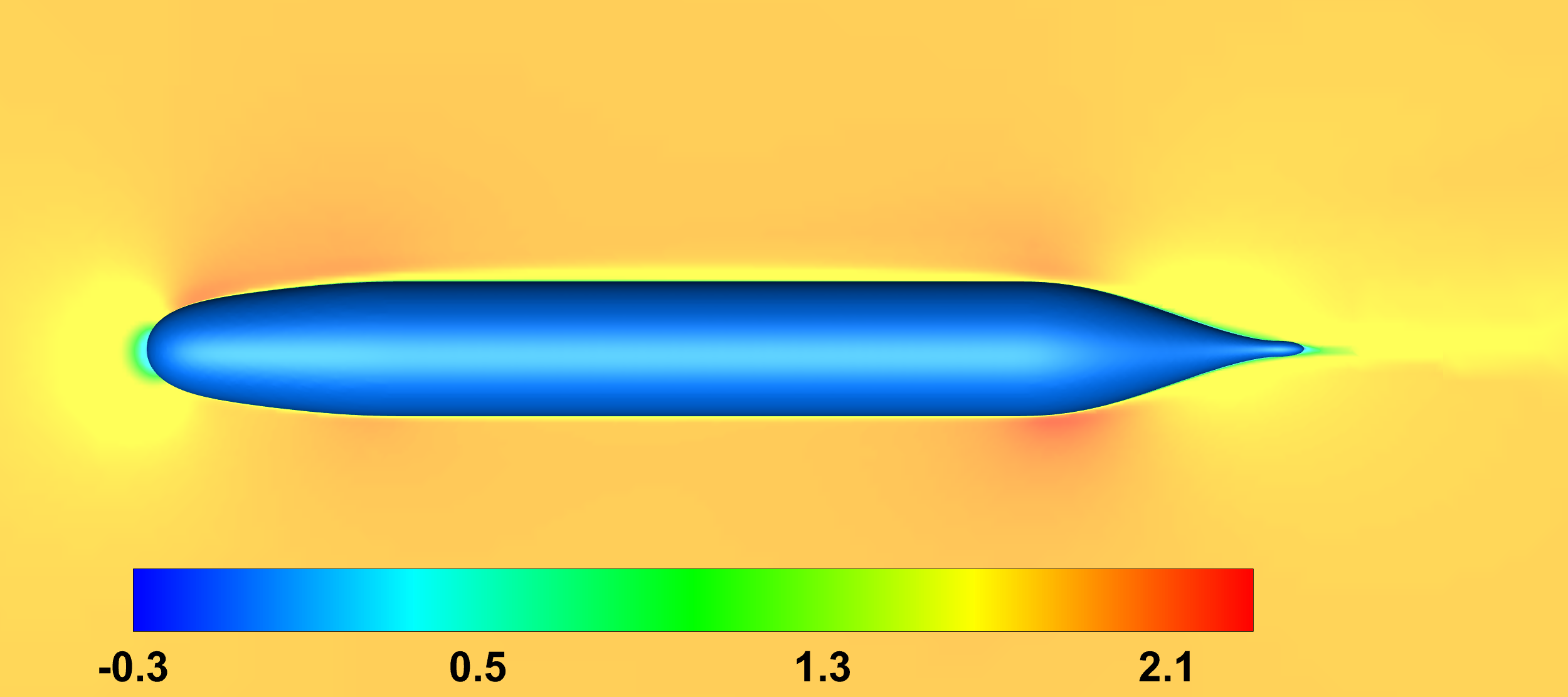}
\end{subfigure}
\hfill
\begin{subfigure}[b]{0.30\textwidth}
    \centering
    \includegraphics[width=0.95\textwidth]{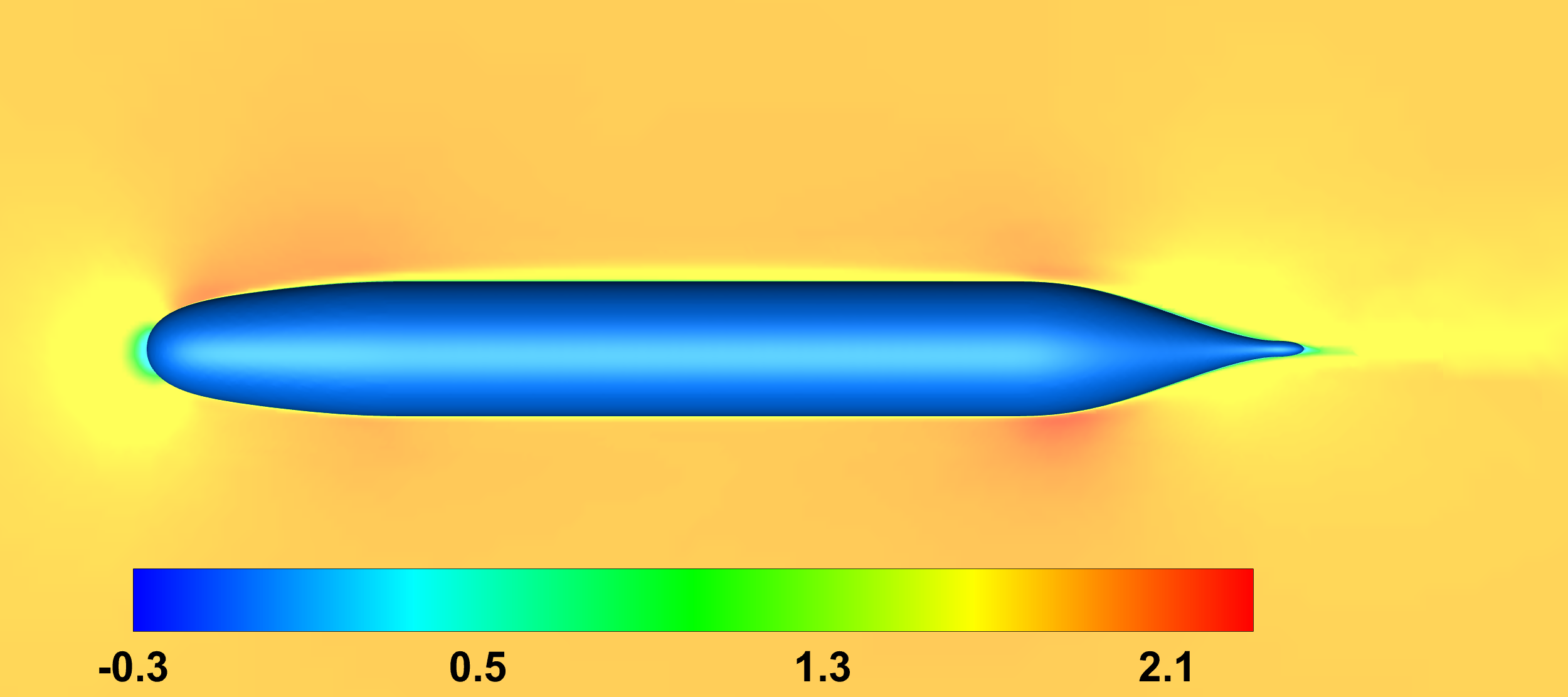}
\end{subfigure}
\hfill
\begin{subfigure}[b]{0.30\textwidth}
    \centering
    \includegraphics[width=0.95\textwidth]{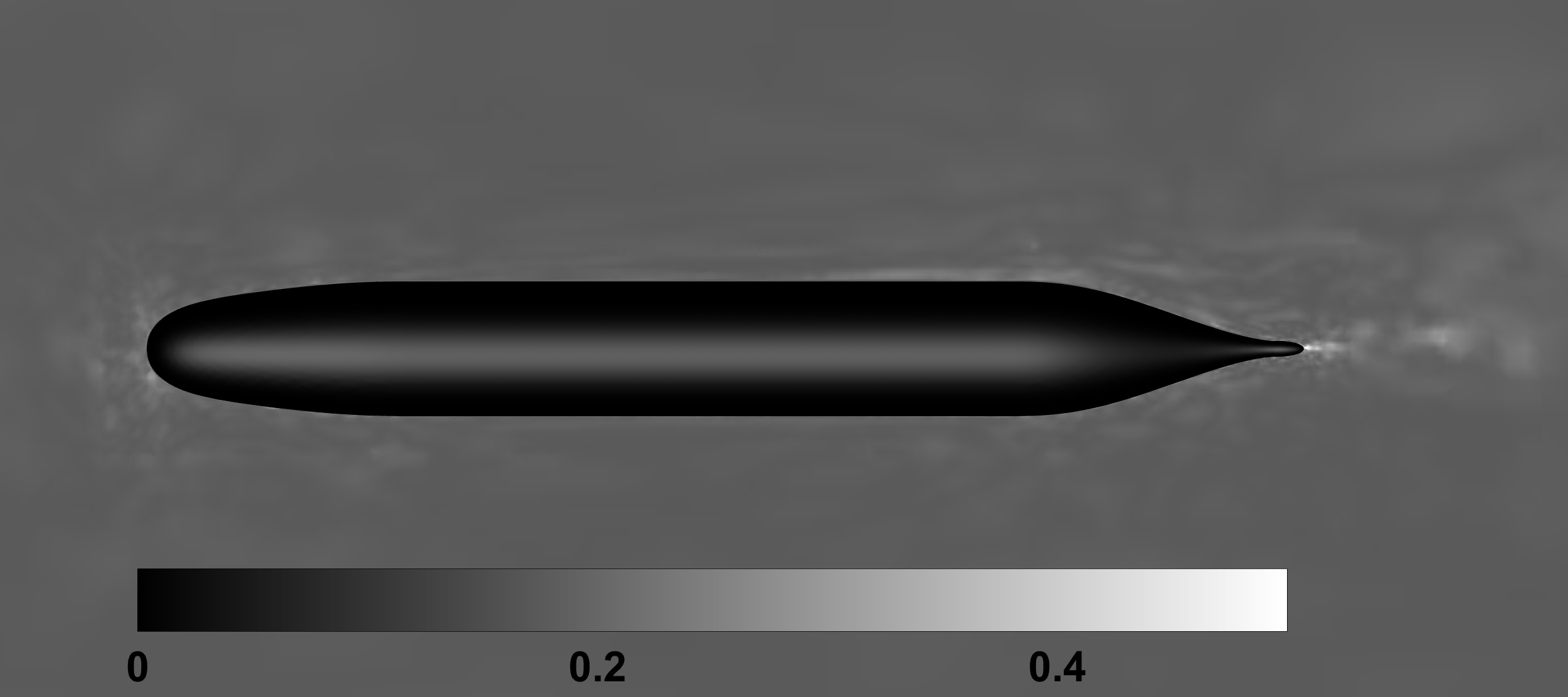}
\end{subfigure}

\vspace{0.3cm}

\raisebox{5\height}{\makebox[0.08\textwidth][c]{{\large $W$}}}
\begin{subfigure}[b]{0.30\textwidth}
    \centering
    \includegraphics[width=0.95\textwidth]{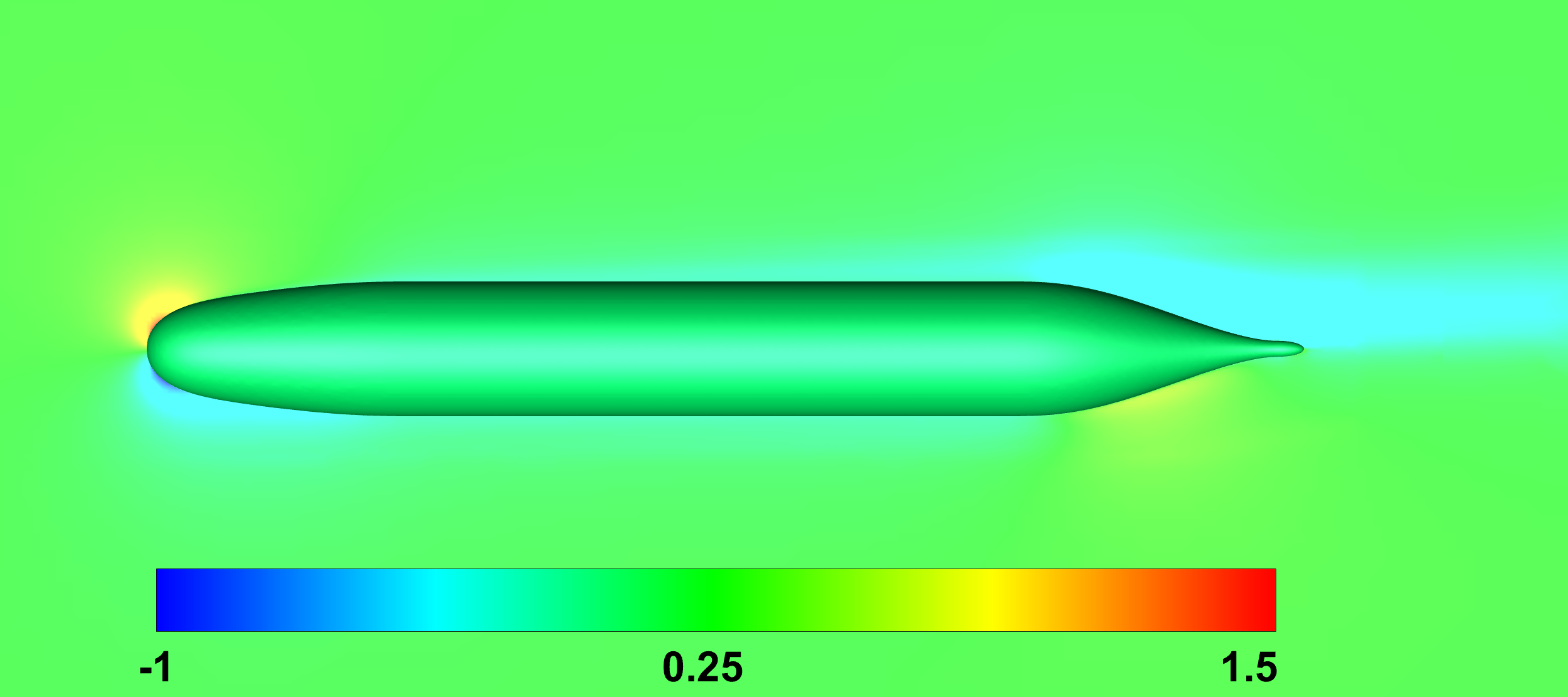}
\end{subfigure}
\hfill
\begin{subfigure}[b]{0.30\textwidth}
    \centering
    \includegraphics[width=0.95\textwidth]{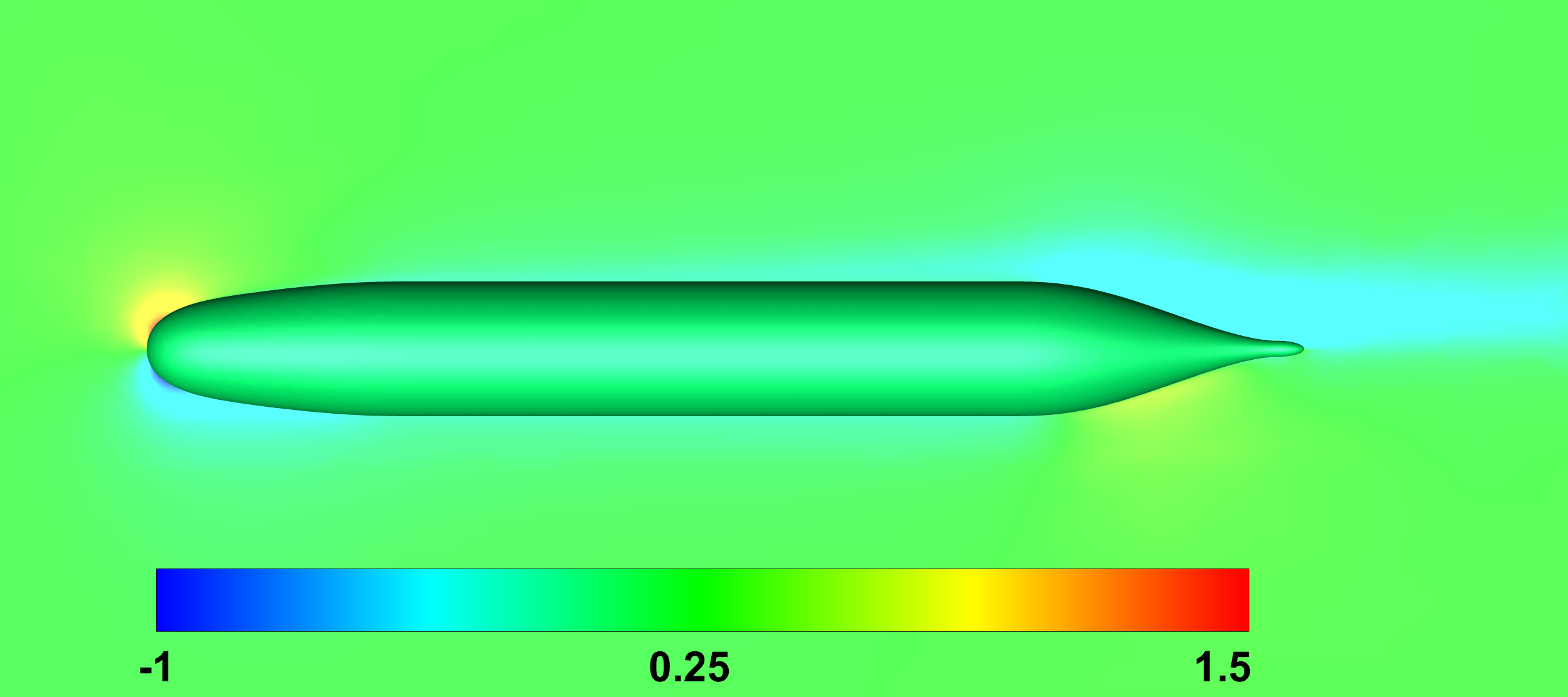}
\end{subfigure}
\hfill
\begin{subfigure}[b]{0.30\textwidth}
    \centering
    \includegraphics[width=0.95\textwidth]{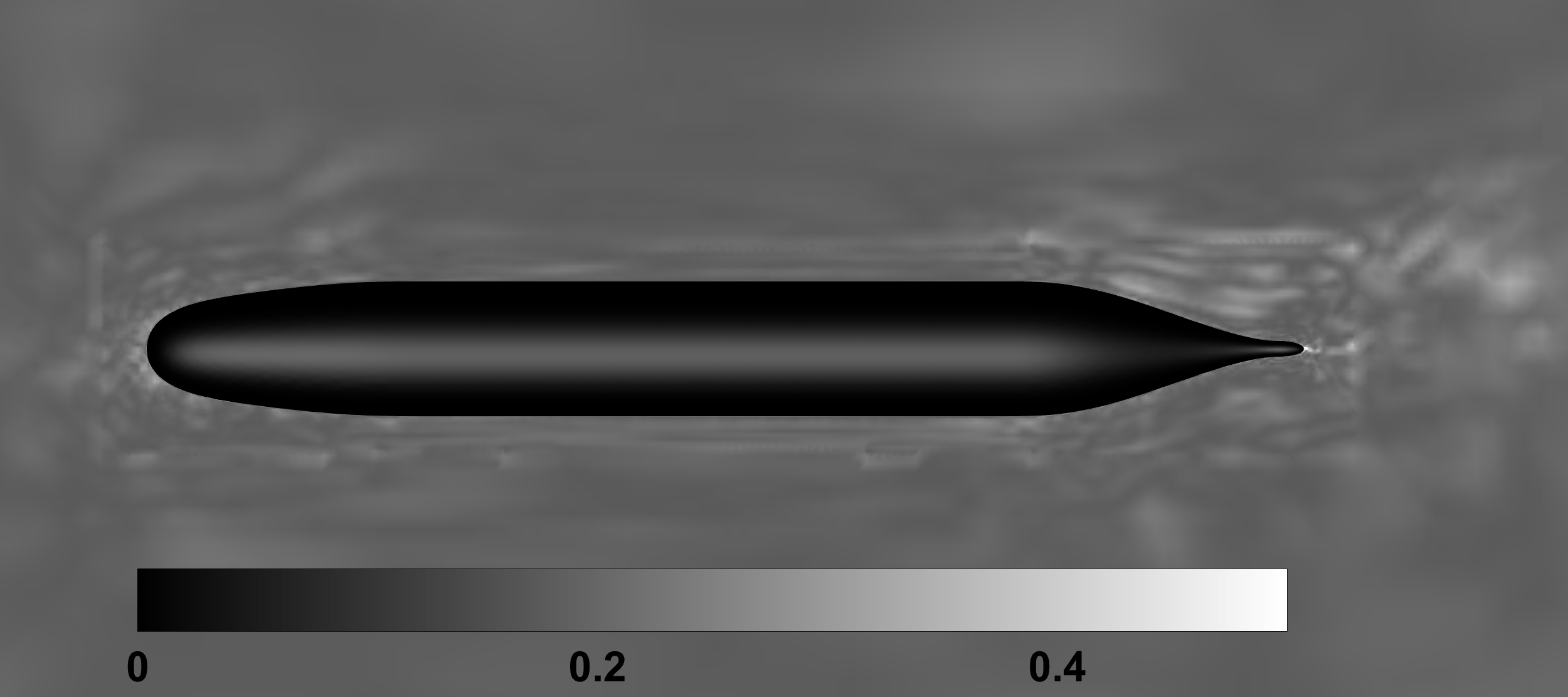}
\end{subfigure}

\vspace{0.3cm}

\raisebox{5\height}{\makebox[0.08\textwidth][c]{{\large $P$}}}
\begin{subfigure}[b]{0.30\textwidth}
    \centering
    \includegraphics[width=0.95\textwidth]{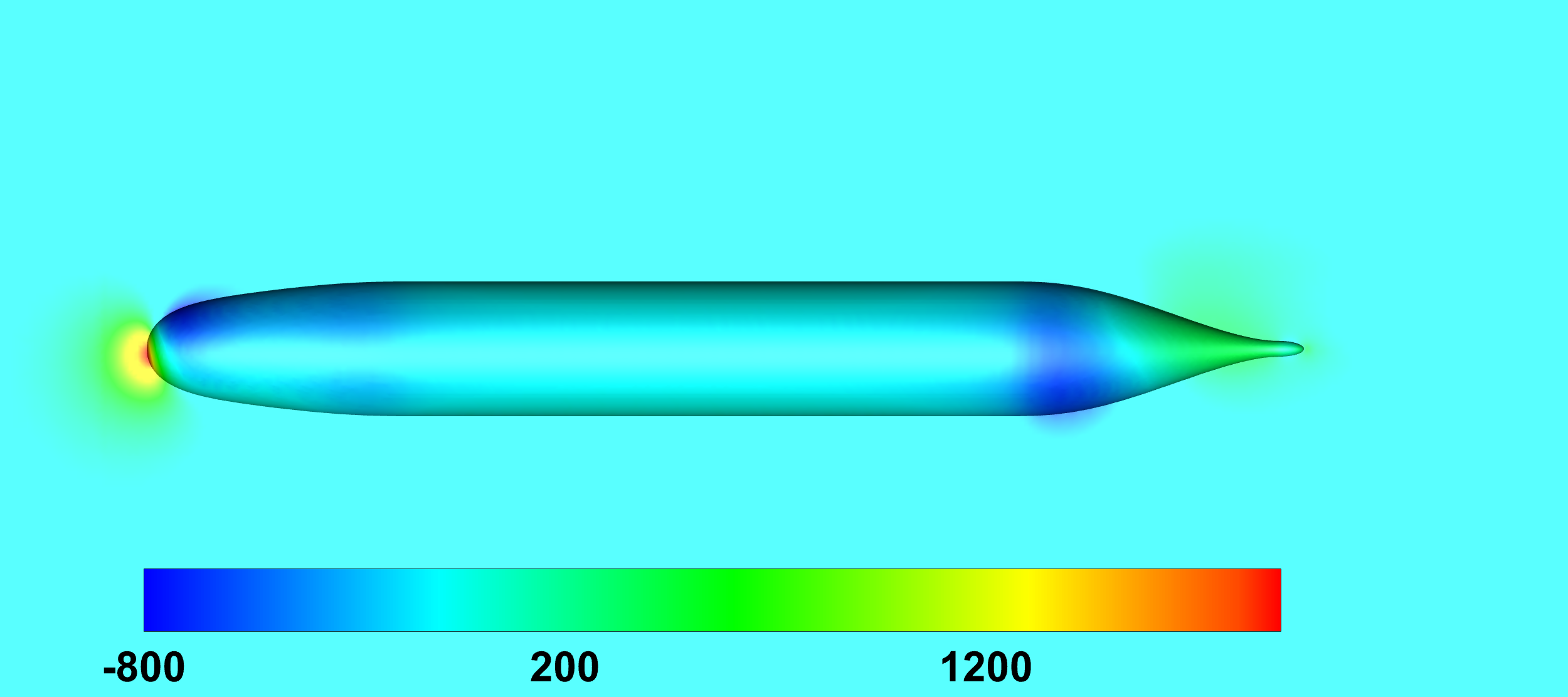}
\end{subfigure}
\hfill
\begin{subfigure}[b]{0.30\textwidth}
    \centering
    \includegraphics[width=0.95\textwidth]{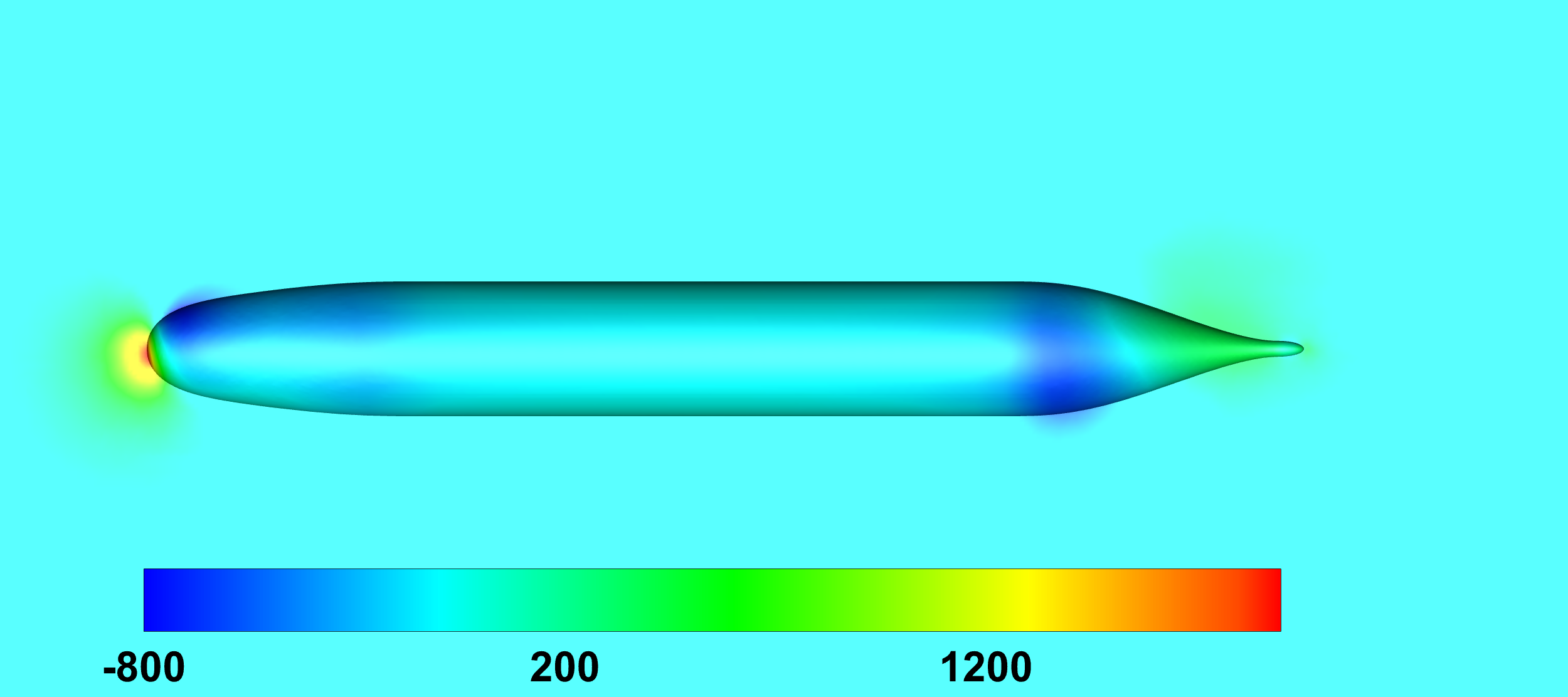}
\end{subfigure}
\hfill
\begin{subfigure}[b]{0.30\textwidth}
    \centering
    \includegraphics[width=0.95\textwidth]{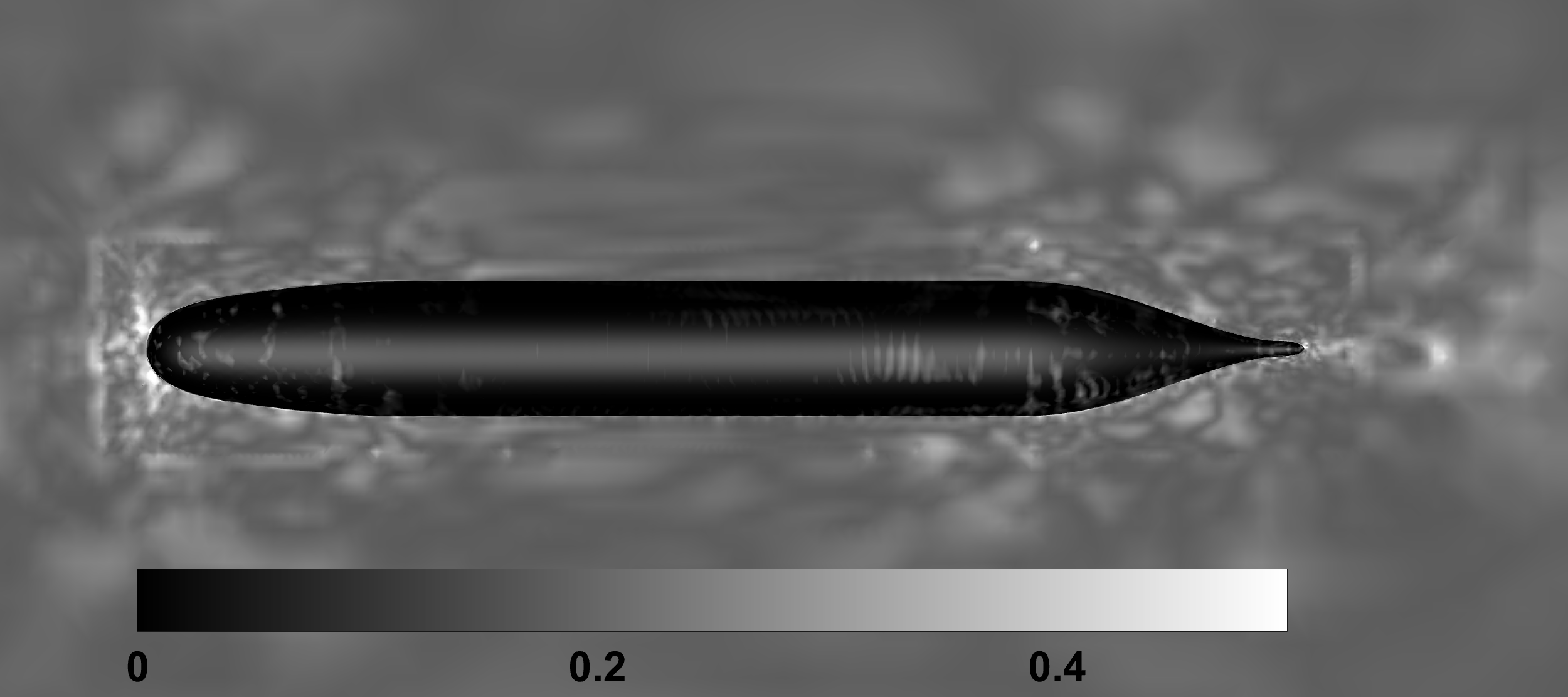}
\end{subfigure}

\vspace{0.3cm}

\raisebox{5\height}{\makebox[0.08\textwidth][c]{{\large $K$}}}
\begin{subfigure}[b]{0.30\textwidth}
    \centering
    \includegraphics[width=0.95\textwidth]{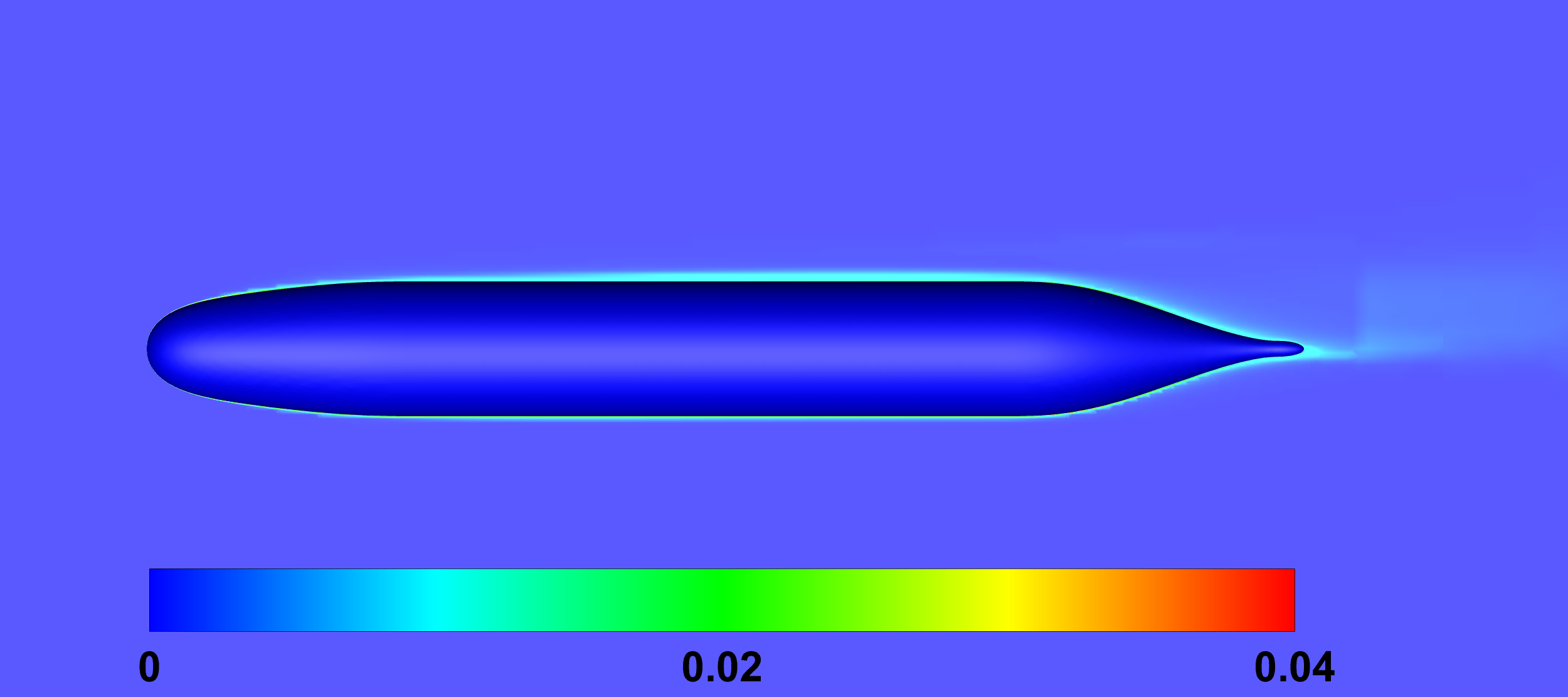}
         \caption*{\large CFD}
\end{subfigure}
\hfill
\begin{subfigure}[b]{0.30\textwidth}
    \centering
    \includegraphics[width=0.95\textwidth]{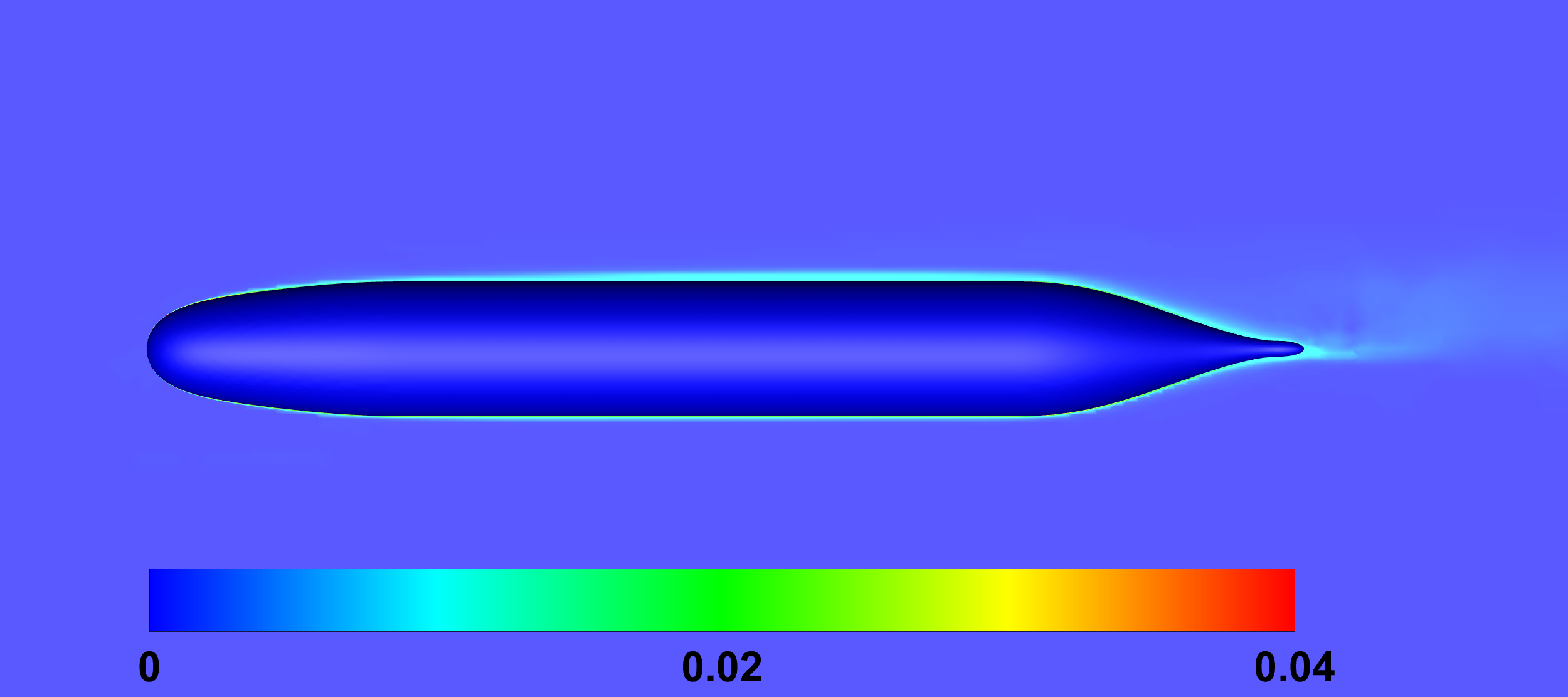}
         \caption*{\large NN-Predicted}
\end{subfigure}
\hfill
\begin{subfigure}[b]{0.30\textwidth}
    \centering
    \includegraphics[width=0.95\textwidth]{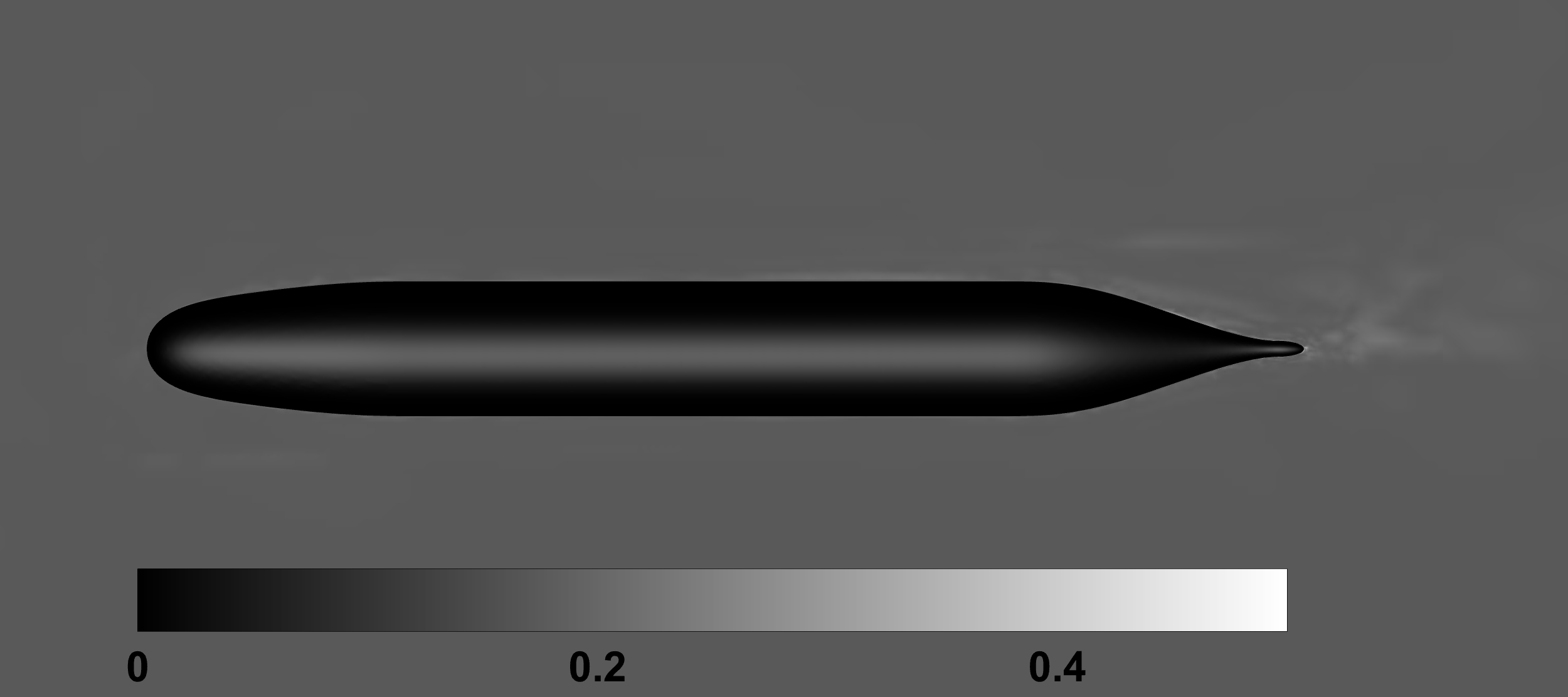}
        \caption*{\large Normalized Error}
\end{subfigure}

\caption{\label{Fig:v2} Comparison diagram of the near-wall flow field distribution of suboff model under test case $T_{2}$($|\mathbf{V}^{*}|=2$m/s,$\,\alpha=8^{\circ}$).}
\end{figure*}

Furthermore, all result data are obtained after completing 1500 iteration steps.  Fig.~\ref{Fig:angle9_ini} illustrates the convergence process of flow field flux residuals $R(flux)$ under several inlet velocities at an angle of attack of $9^{\circ}$. It can be observed that the flow field essentially reaches a steady state after 600 iteration steps. Thus,  extracting the results after 1500 iteration steps is fully reliable.

\begin{figure*}[htbp]
\centering
    \begin{subfigure}[b]{0.9\textwidth}
    \centering
    \includegraphics[width=0.5\textwidth]{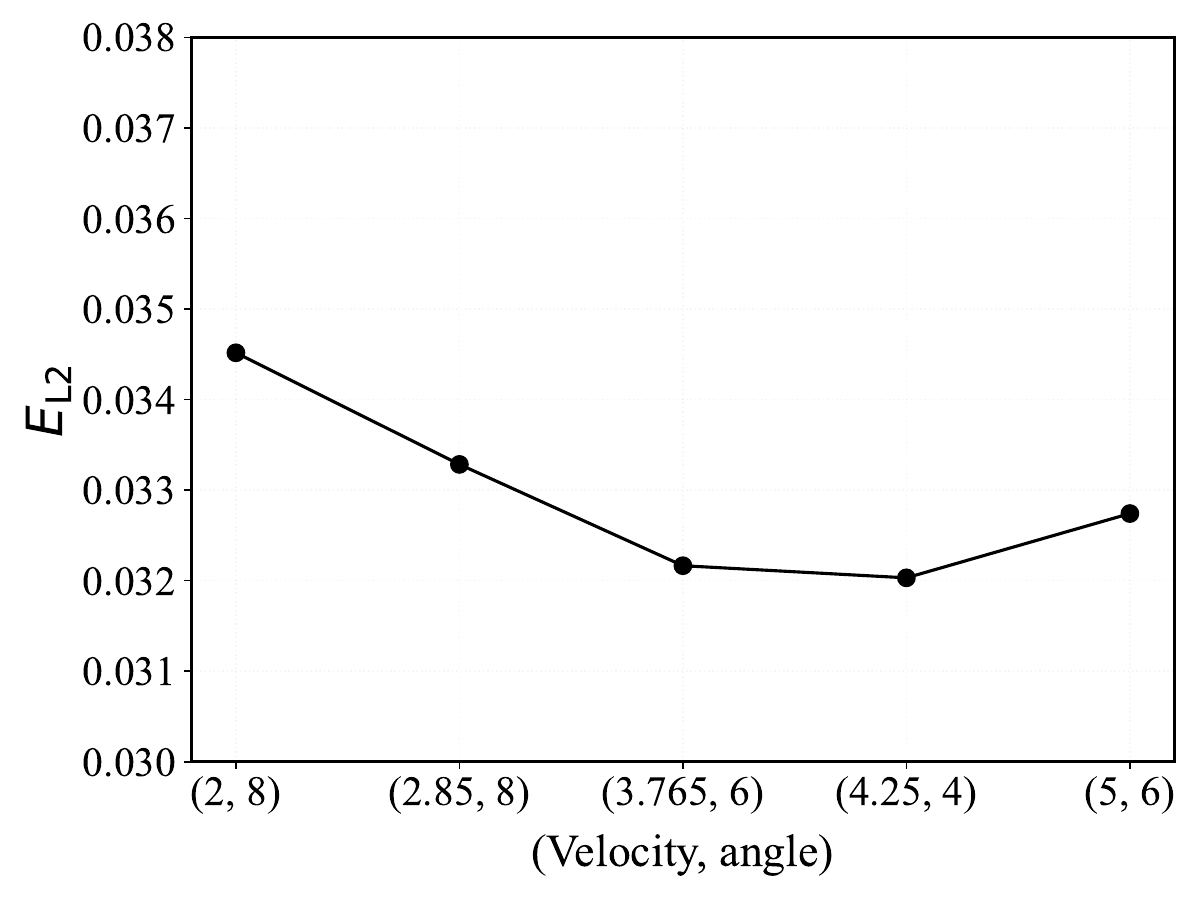}
    \end{subfigure}
\caption{The relative L2 error $E_{L2}$ (defined in Eq.~(\ref{E_L2_1})) between NN-predicted results and CFD results for the test cases $T_{i}$, where $i=1,...,5$.}
\label{Fig:E_L2_1} 
\end{figure*}

In the training phase, unlike most existing studies that primarily focus on the wake flow field or near-wall pre-sampled flow field, the focus of this work lies on the global flow field, with the aim of acquiring data covering the entire computational domain. The training framework is constructed based on PyTorch, where the Adam optimizer is adopted to minimize the loss function. The initial learning rate is set to $0.001$, and a step-decay learning rate strategy is employed, specifically, the learning rate is scaled by a factor of $0.75$ every $800$ training epochs. Given the massive total number of samples, a relatively large batch size of 16384 is utilized. The training process runs for a total of 6000 epochs and takes approximately 16.8 hours on a single Ascend 910 chip.

\subsection{Data-Driven Flow Field Accelerated Computing}
In this section, comparison plots between CFD convergence results and NN-predicted results for three representative test cases, i.e. interpolation $T_{3}$, weak extrapolation $T_{4}$, and extrapolation cases  $T_{1}$, are first presented to demonstrate the prediction accuracy of the DNN for flow fields. Specifically, Fig.~\ref{Fig:v3.765} depicts  the flow field comparison in the near-wall region on the symmetry plane of the suboff computational domain. The comparison corresponds to the interpolation test case operating under $|\mathbf{V^{\ast}}|=3.765$m/s and $\alpha=6^{\circ}$. In this figure, the first column shows the converged CFD  results, the second column presents the NN-predicted results, and the third column is the normalized error between these two columns, with the normalization method defined as:
\begin{equation}\label{E_n}
E_{n}^{(i)}=\frac{\left|\phi^{(i)}_{pred}-\phi^{(i)}_{true}\right|}{max(\left|\phi^{(i)}_{true}\right|)},\quad\quad\quad\quad i \in \Omega.
\end{equation}

As can be seen from Fig.~\ref{Fig:v3.765}, the neural network proposed in this study is capable of generating prediction results that are visually highly consistent with the converged CFD results for the velocity field, pressure field, and turbulent kinetic energy field. Notably, the velocity and angle of this test case are not included in the training set. However, the neural network still accurately captures the inherent law governing flow variations with initial operating conditions. Furthermore, the results of the  weak extrapolation $T_{4}$ ($\left|\textbf{V}^*\right|=4.25$m/s,$\alpha=4^\circ$) and the extrapolation  $T_{2}$ ($\left|\textbf{V}^*\right|=2$m/s,$\alpha=8^\circ$) presented in Fig.~\ref{Fig:v4.25} and Fig.~\ref{Fig:v2} further validate the aforementioned conclusion. For instance, compared with the $\alpha=4^\circ$ case, the velocity field $U$ in the wake region exhibits a more pronounced asymmetry under the $\alpha=8^\circ$, and the prediction results of the neural network can also accurately reflect this variation characteristic.

\begin{figure*}[!t]
\centering
\setlength{\tabcolsep}{1pt}
\renewcommand{\figurename}{Fig.}
\makebox[0.4\textwidth]{\large \rmfamily\fontfamily{ptm}\selectfont CFD}  
\vspace{0.2cm}  
\hspace{0.05\textwidth}
\makebox[0.4\textwidth]{\large \rmfamily\fontfamily{ptm}\selectfont NN-Predicted}
\vspace{0.3cm}  
\begin{subfigure}[b]{0.4\textwidth}  
    \centering
    \includegraphics[width=0.85\textwidth]{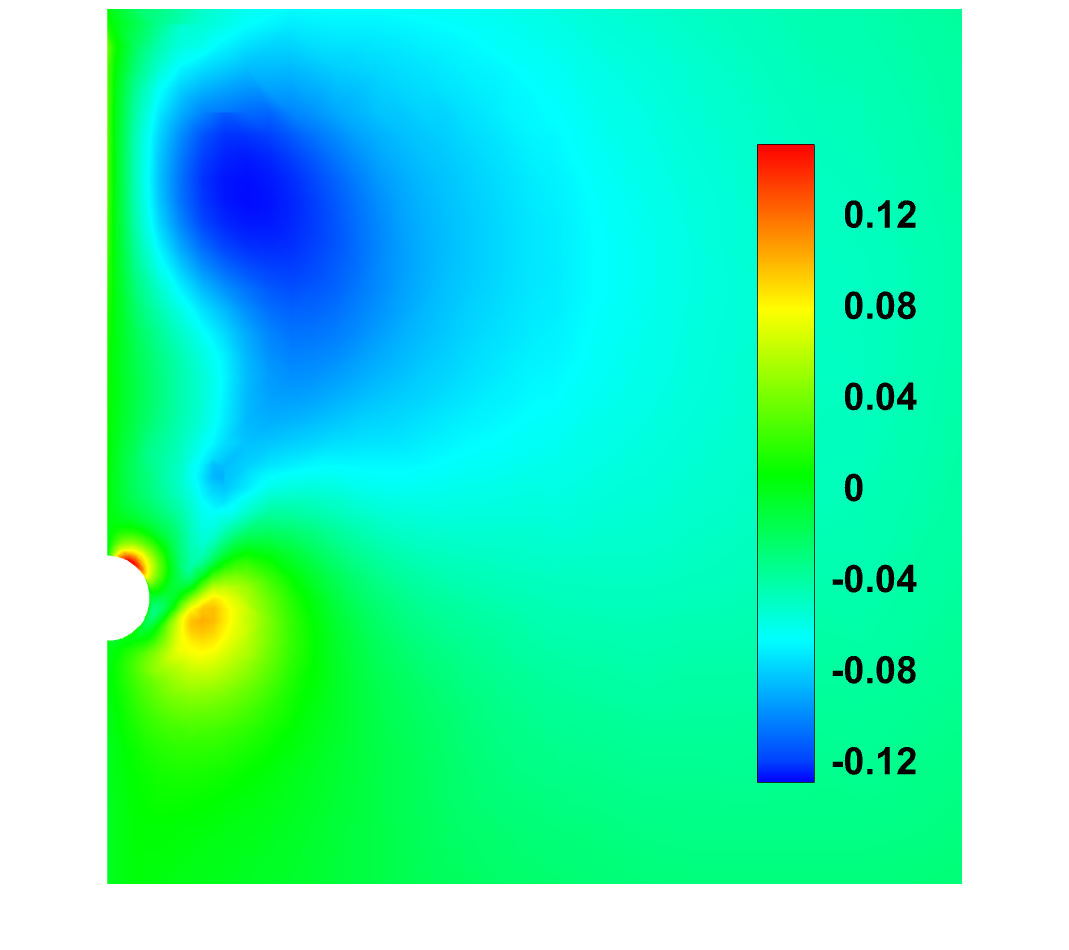}
    \label{subfig:case1_1}  
\end{subfigure}
\hspace{0.05\textwidth}
\begin{subfigure}[b]{0.4\textwidth}
    \centering
    \includegraphics[width=0.85\textwidth]{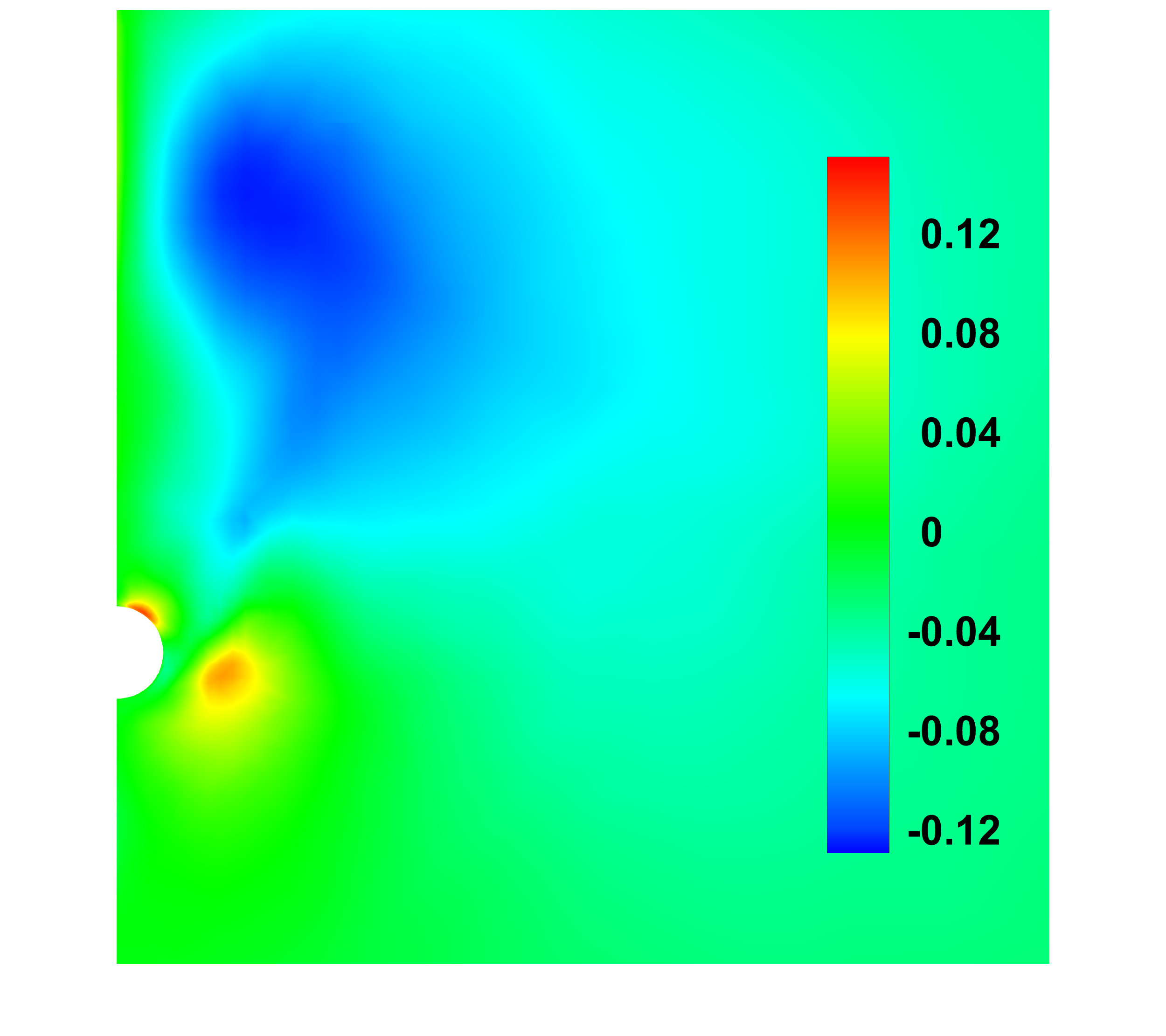}
    \label{subfig:case1_2}
\end{subfigure}
\par\centering\textbf{\large ($|\mathbf{V}^*|=2$m/s,$\alpha=8^\circ$)}\par
\vspace{0.3cm}  

\begin{subfigure}[b]{0.4\textwidth}
    \centering
    \includegraphics[width=0.85\textwidth]{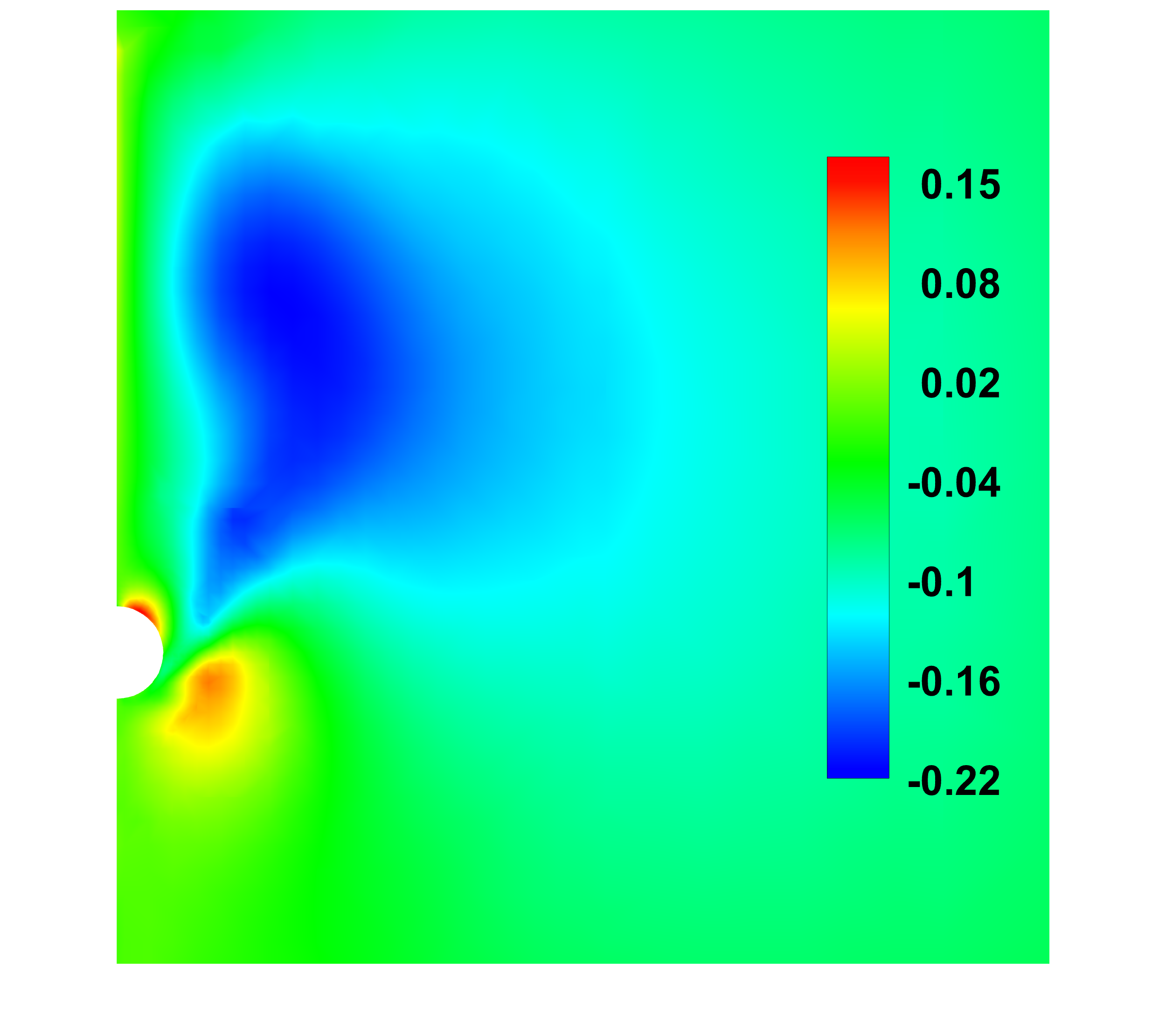}
    \label{subfig:case2_1}
\end{subfigure}
\hspace{0.05\textwidth}
\begin{subfigure}[b]{0.4\textwidth}
    \centering
    \includegraphics[width=0.85\textwidth]{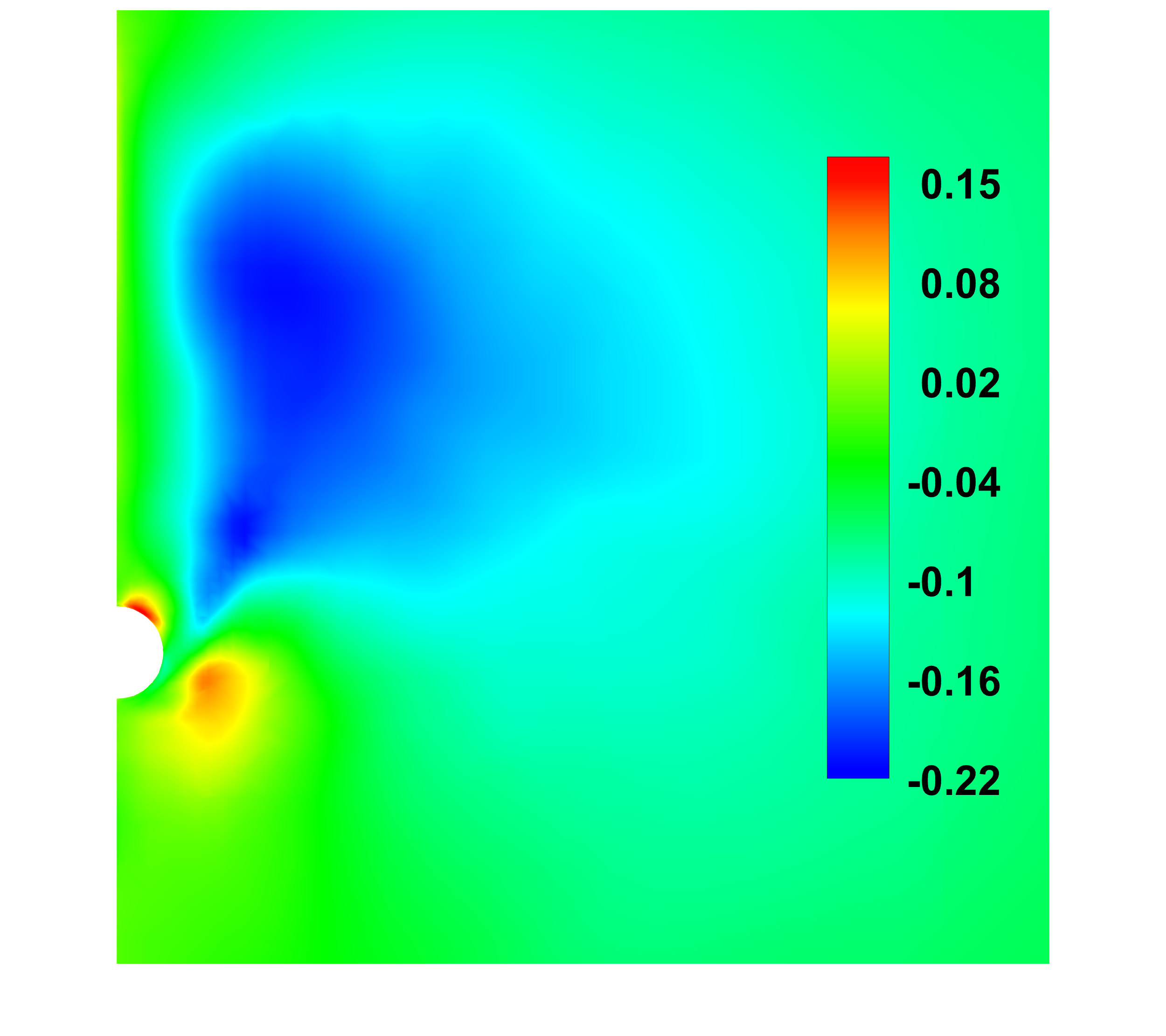}
    \label{subfig:case2_2}
\end{subfigure}
\par\centering\textbf{\large ($|\mathbf{V}^*|=3.765$m/s,$\alpha=6^\circ$)}\par
\vspace{0.2cm}

\begin{subfigure}[b]{0.4\textwidth}
    \centering
    \includegraphics[width=0.85\textwidth]{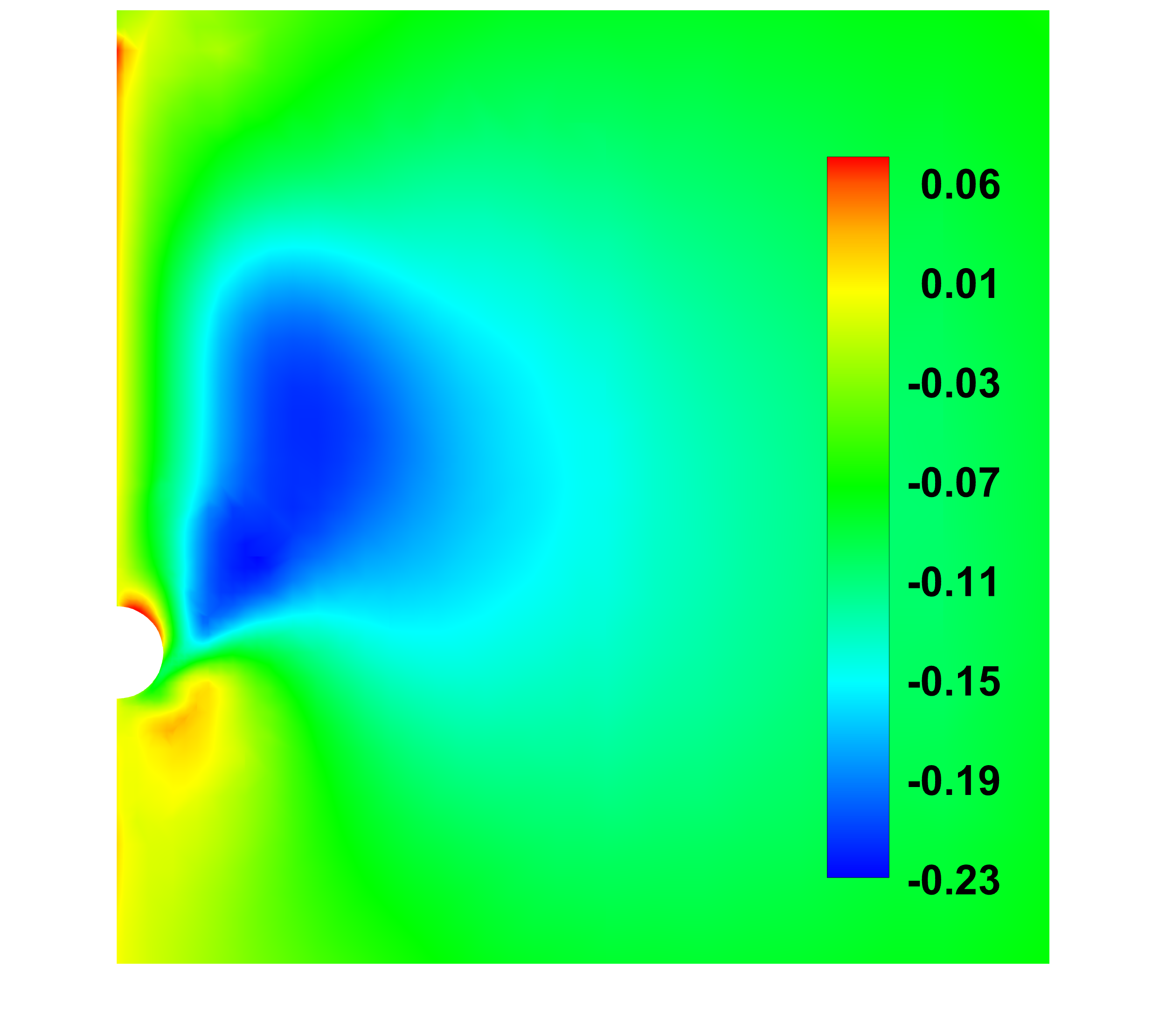}
    \label{subfig:case3_1}
\end{subfigure}
\hspace{0.05\textwidth}
\begin{subfigure}[b]{0.4\textwidth}
    \centering
    \includegraphics[width=0.85\textwidth]{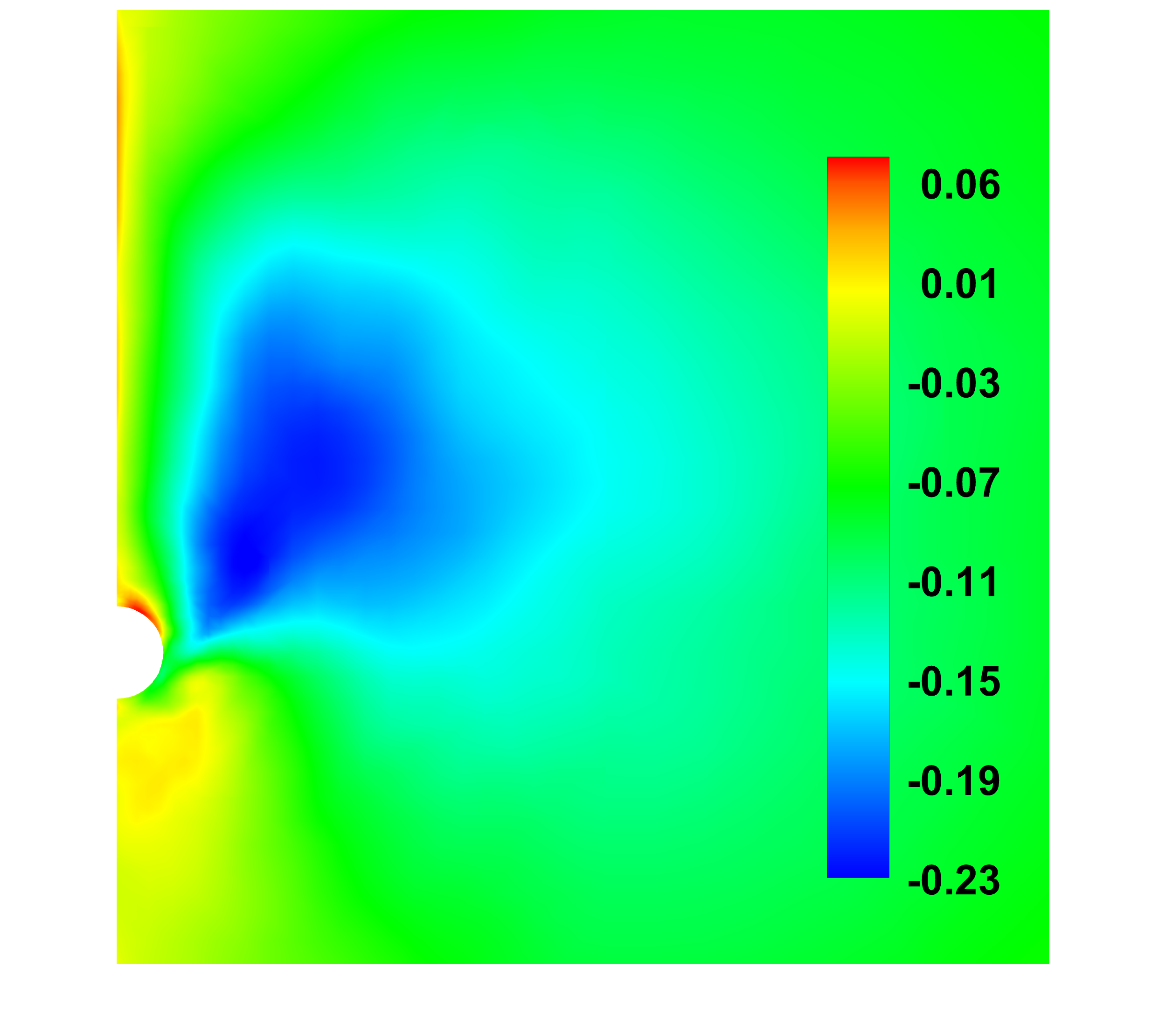}
    \label{subfig:case3_2}
\end{subfigure}
\par\centering\textbf{\large ($|\mathbf{V}^*|=4.25$m/s,$\alpha=4^\circ$)}\par
\caption{\label{Fig:V_compare} 
Comparison of distribution of $y$-direction velocity $V$ at the $x/L=0.978$ cross-section for the suboff model under three test cases. Left: CFD results, right: NN-predicted results.}
\end{figure*}

Furthermore, Fig.~\ref{Fig:E_L2_1} illustrates the relative L2 errors  between the predicted values and the CFD results for five test cases, with the formula given by:
\begin{equation}\label{E_L2_1}
E_{L2}=\frac{\sqrt{\sum_{i\in\Omega}(\phi^{(i)}_{pred}-\phi^{(i)}_{true})^{2}}}{\sqrt{\sum_{i\in\Omega}(\phi^{(i)}_{true})^{2}}},\quad\quad i \in \Omega.
\end{equation}

\begin{figure*}[htbp]
\centering
\setlength{\tabcolsep}{0pt}
\renewcommand{\figurename}{Fig.}
\begin{subfigure}[b]{0.475\textwidth}  
    \centering
    \includegraphics[width=0.975\textwidth]{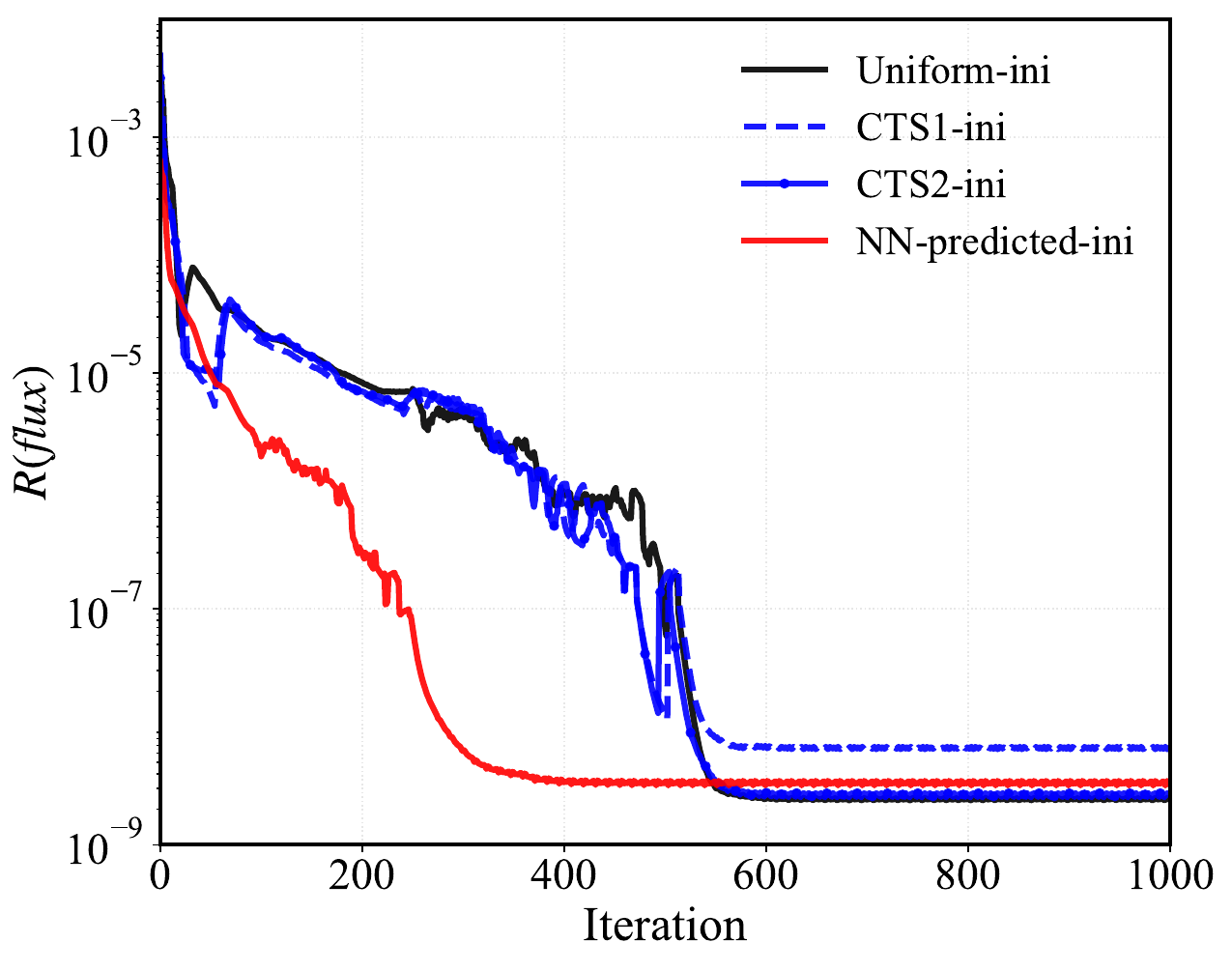}  
\end{subfigure}
\begin{subfigure}[b]{0.475\textwidth}
    \centering
    \includegraphics[width=0.975\textwidth]{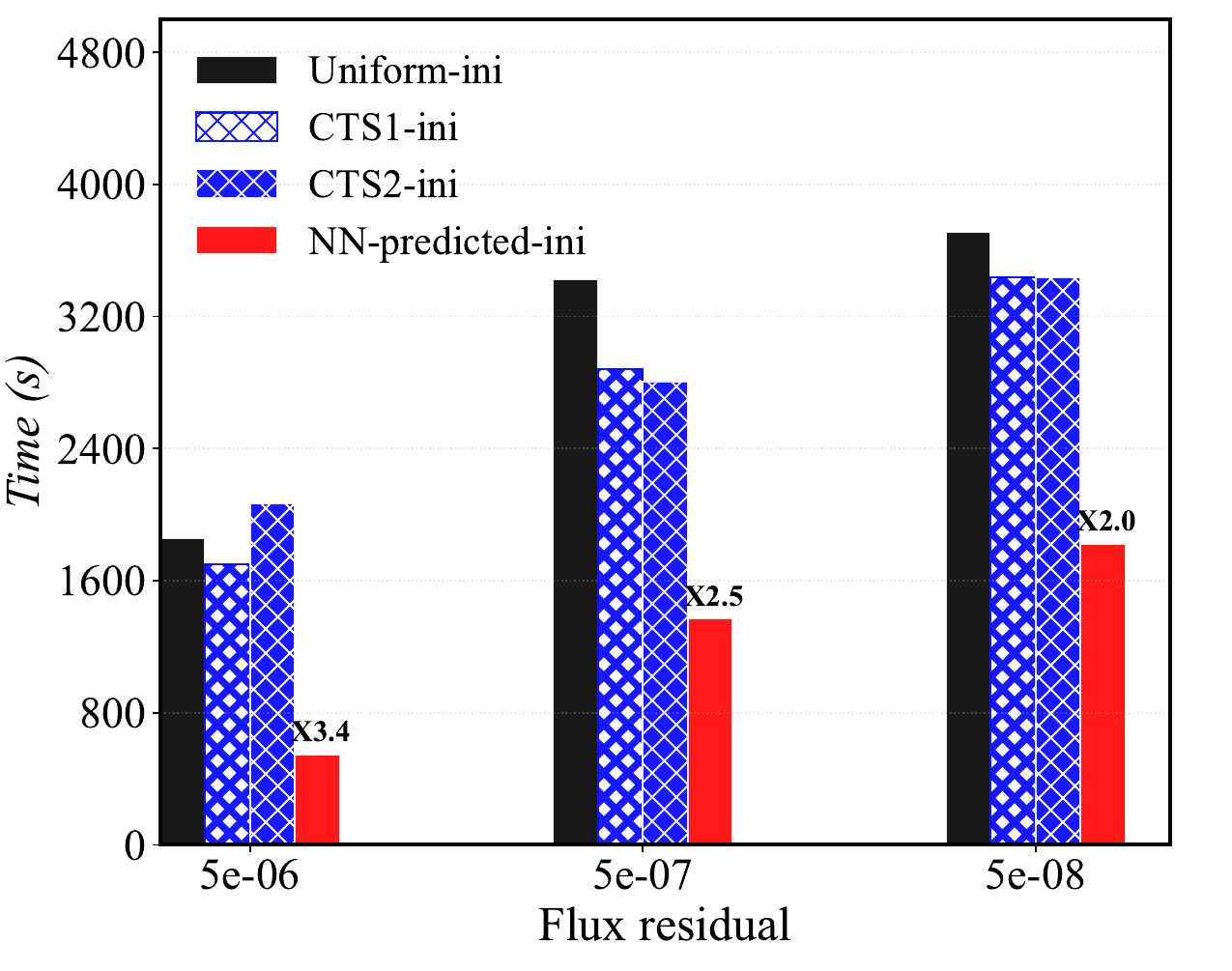}
\end{subfigure}
\\ 
\makebox[\textwidth][c]{{\large (a)}} 
\vspace{0.3cm}  
\begin{subfigure}[b]{0.475\textwidth}
    \centering
    \includegraphics[width=0.975\textwidth]{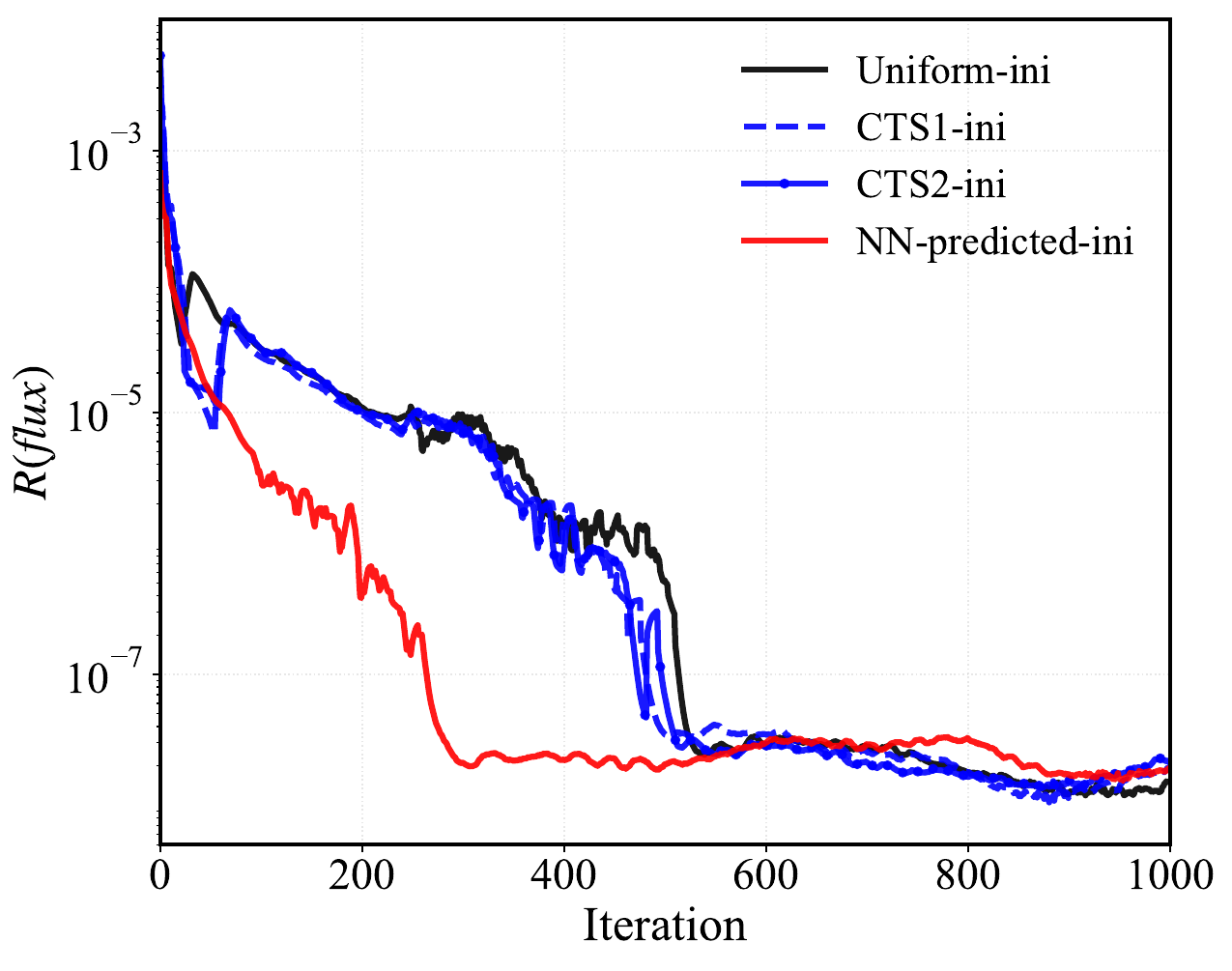}
\end{subfigure}
\begin{subfigure}[b]{0.475\textwidth}
    \centering
    \includegraphics[width=0.975\textwidth]{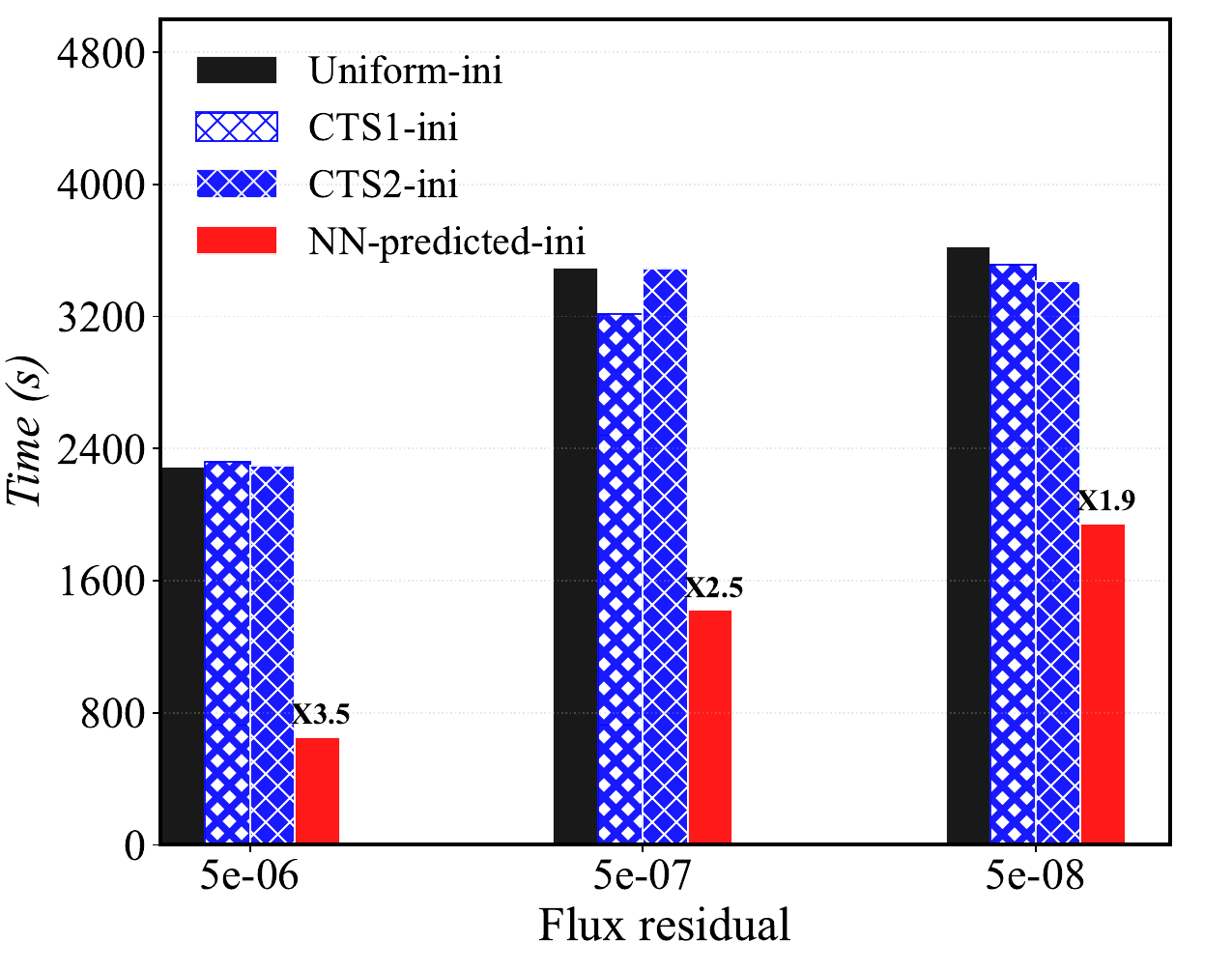}
\end{subfigure}
\\
\makebox[\textwidth][c]{{\large (b)}}
\vspace{0.3cm}

\begin{subfigure}[b]{0.475\textwidth}
    \centering
    \includegraphics[width=0.975\textwidth]{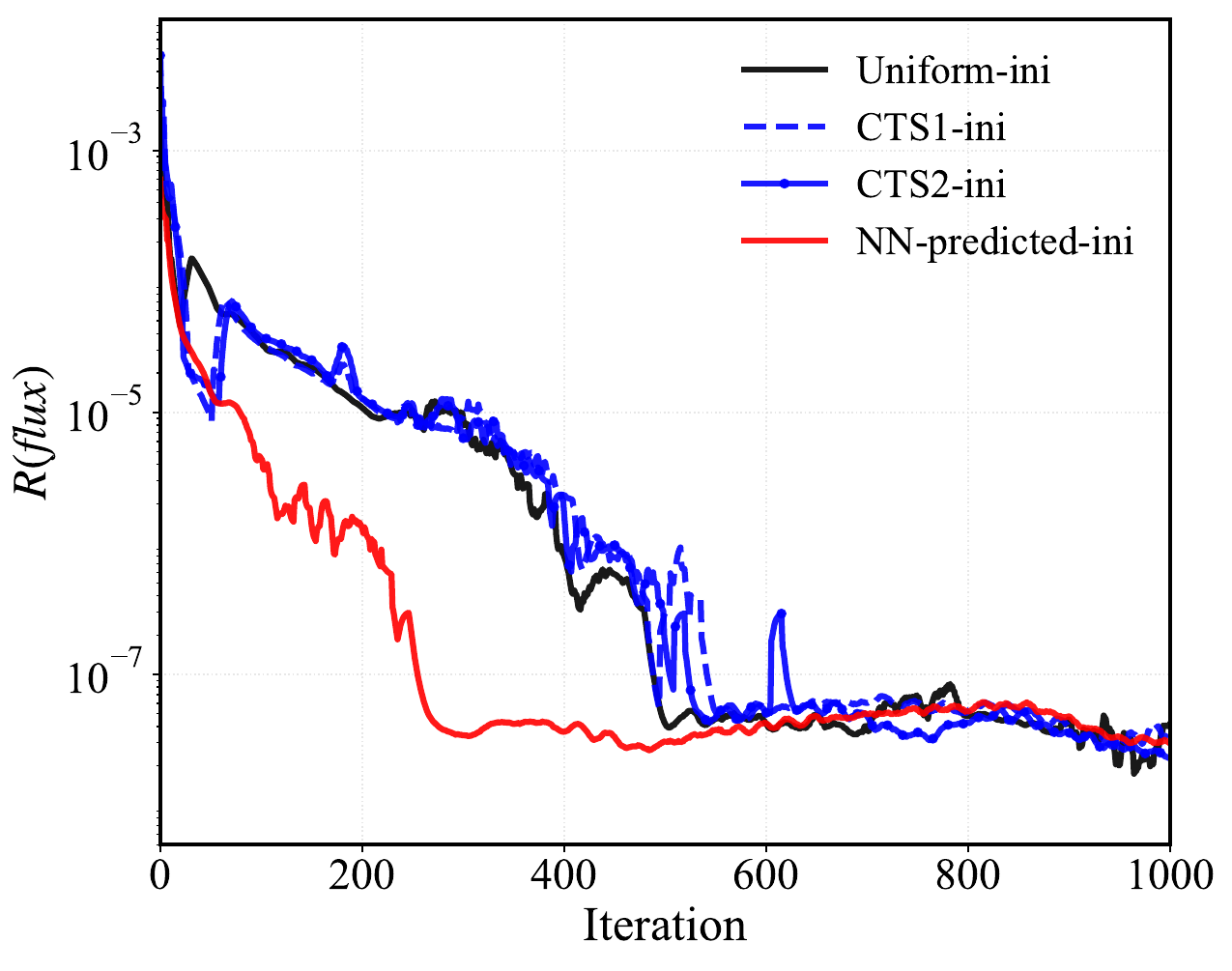}
\end{subfigure}
\begin{subfigure}[b]{0.475\textwidth}
    \centering
    \includegraphics[width=0.975\textwidth]{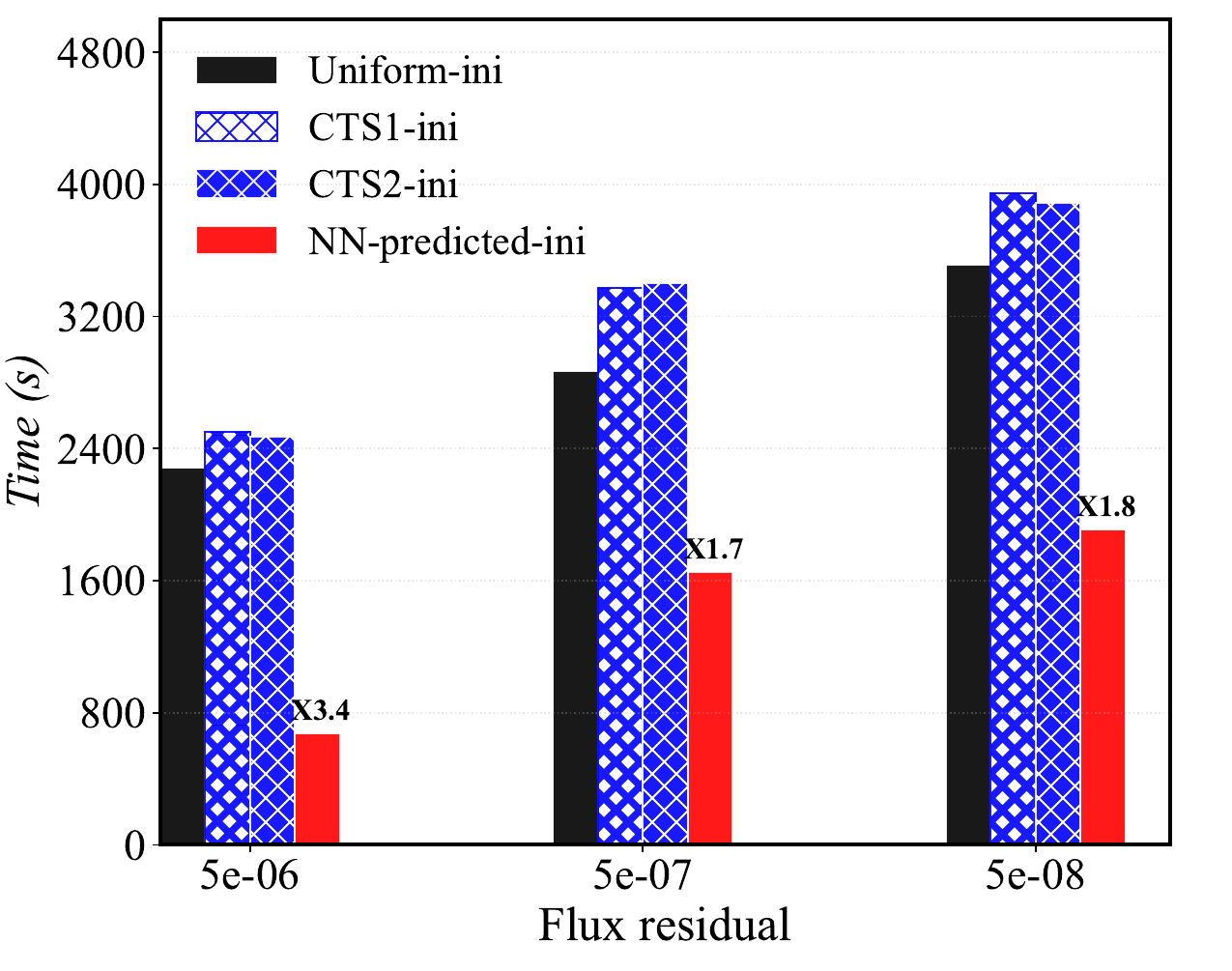}
\end{subfigure}
\\
\makebox[\textwidth][c]{{\large (c)}}
\vspace{0.2cm}
\caption{\label{Fig:acc_abc} 
Comparison of flux residual convergence curves (left) and computation time required to reach the convergence threshold of $5\times10^{-6}$, $5\times10^{-7}$ and $5\times10^{-8}$ (right) respectively under four initialization conditions: uniform, CTS1,CTS2 and NN-predicted (based on a 10-step moving average). (a) Test case $T_{1}$($2.0$m/s,$\,8^{\circ}$),  CTS1= ($2.0$m/s,$\,7^{\circ}$), CTS2=($2.85$m/s,$\,9^{\circ}$). (b) Test case $T_{2}$($2.85$m/s,$\,8^{\circ}$), CTS1=($2.0$m/s,$\,7^{\circ}$),  CTS2=($2.85$m/s,$\,9^{\circ}$). (c) Test case $T_{3}$($3.765$m/s,$\,6^{\circ}$), CTS1=($2.85$m/s,$\,5^{\circ}$), CTS2=($4.25$m/s,$\,7^{\circ}$).  }
\end{figure*}

\begin{figure*}[!t]
\centering

\setlength{\tabcolsep}{0pt}
\renewcommand{\figurename}{Fig.}

\begin{subfigure}[b]{0.475\textwidth}
    \centering
    \includegraphics[width=0.975\textwidth]{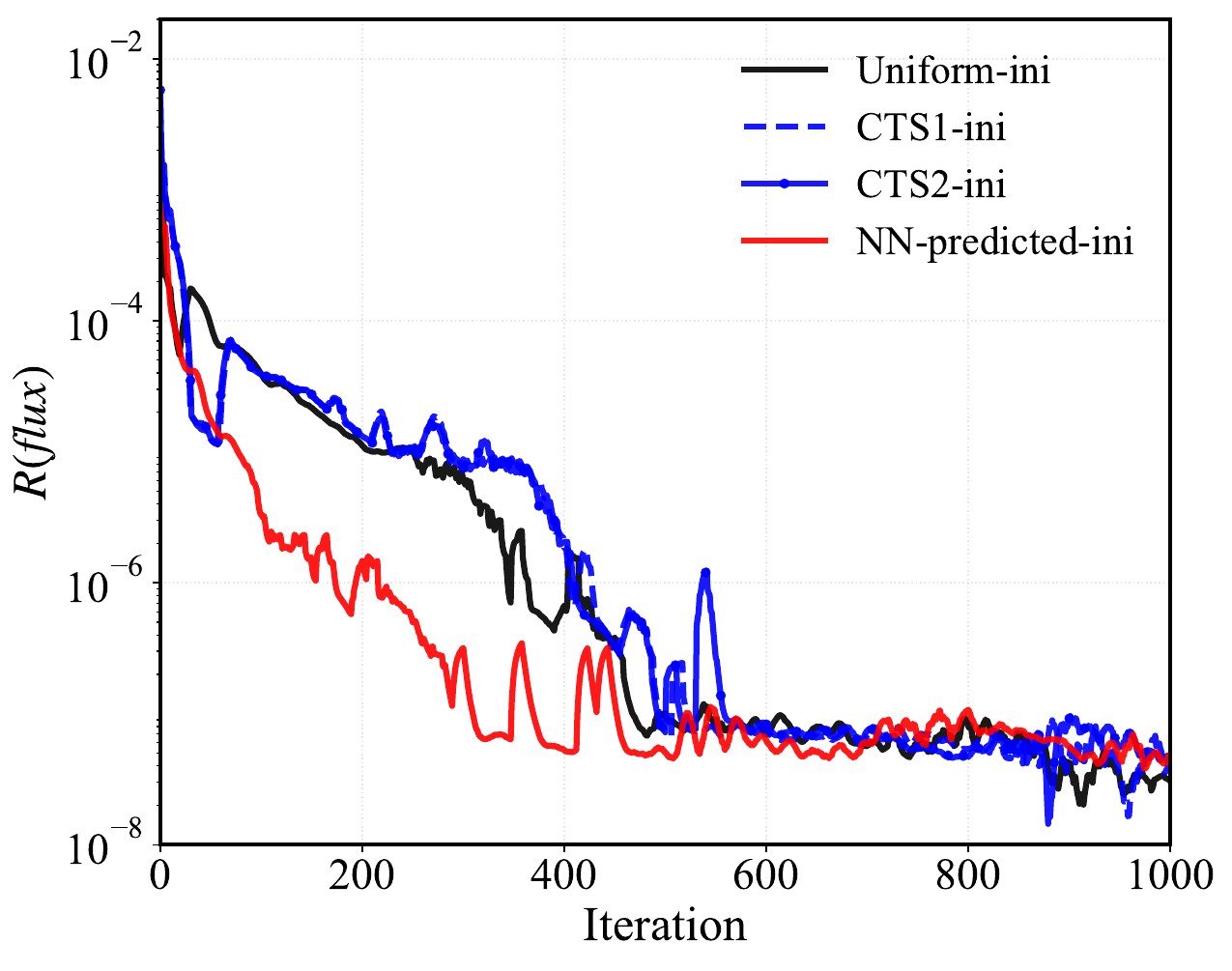}
\end{subfigure}
\begin{subfigure}[b]{0.475\textwidth}
    \centering
    \includegraphics[width=0.975\textwidth]{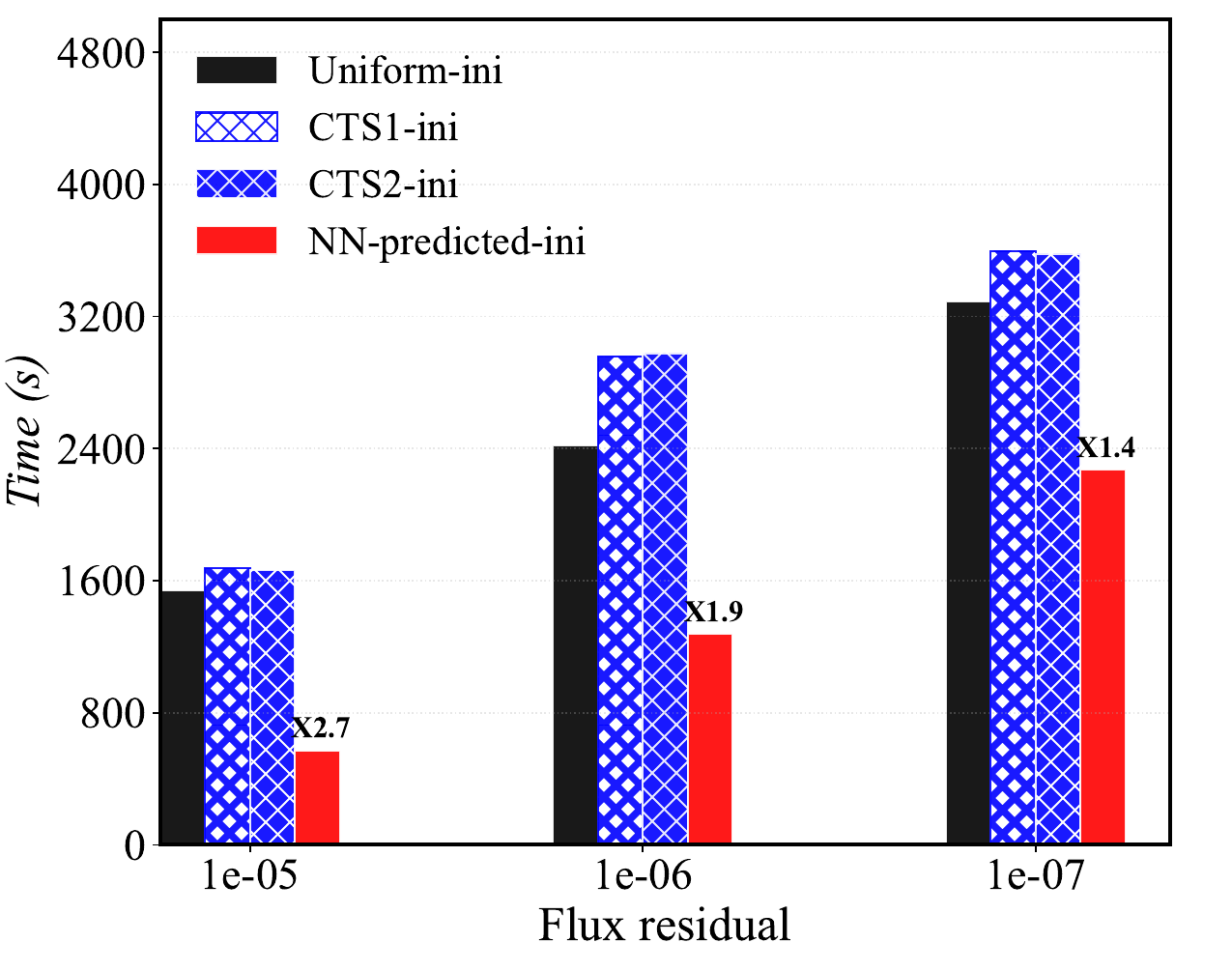}
\end{subfigure}
\\
\makebox[\textwidth][c]{{\large (d)}}
\vspace{0.5cm}

\begin{subfigure}[b]{0.475\textwidth}
    \centering
    \includegraphics[width=0.975\textwidth]{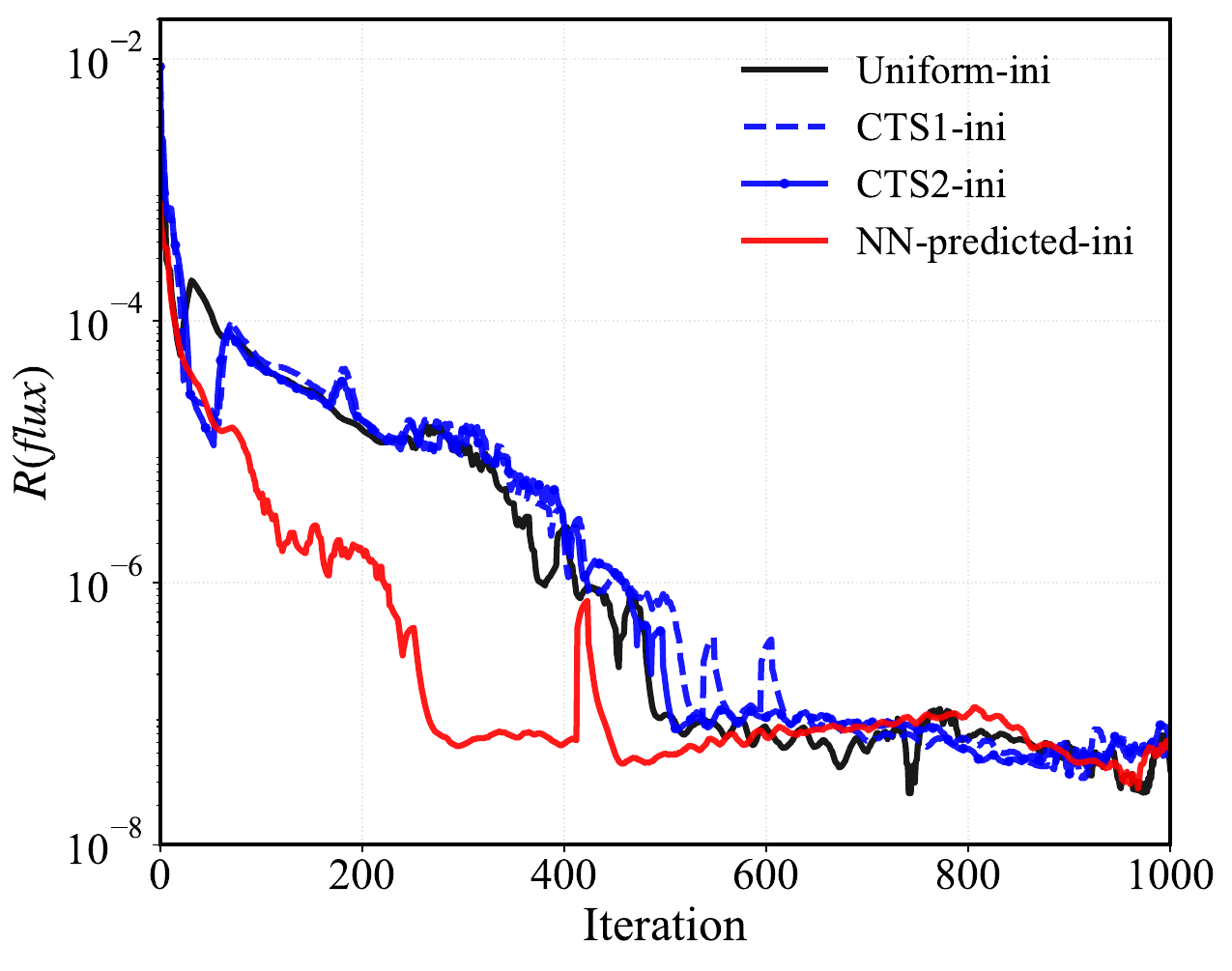}
\end{subfigure}
\begin{subfigure}[b]{0.475\textwidth}
    \centering
    \includegraphics[width=0.975\textwidth]{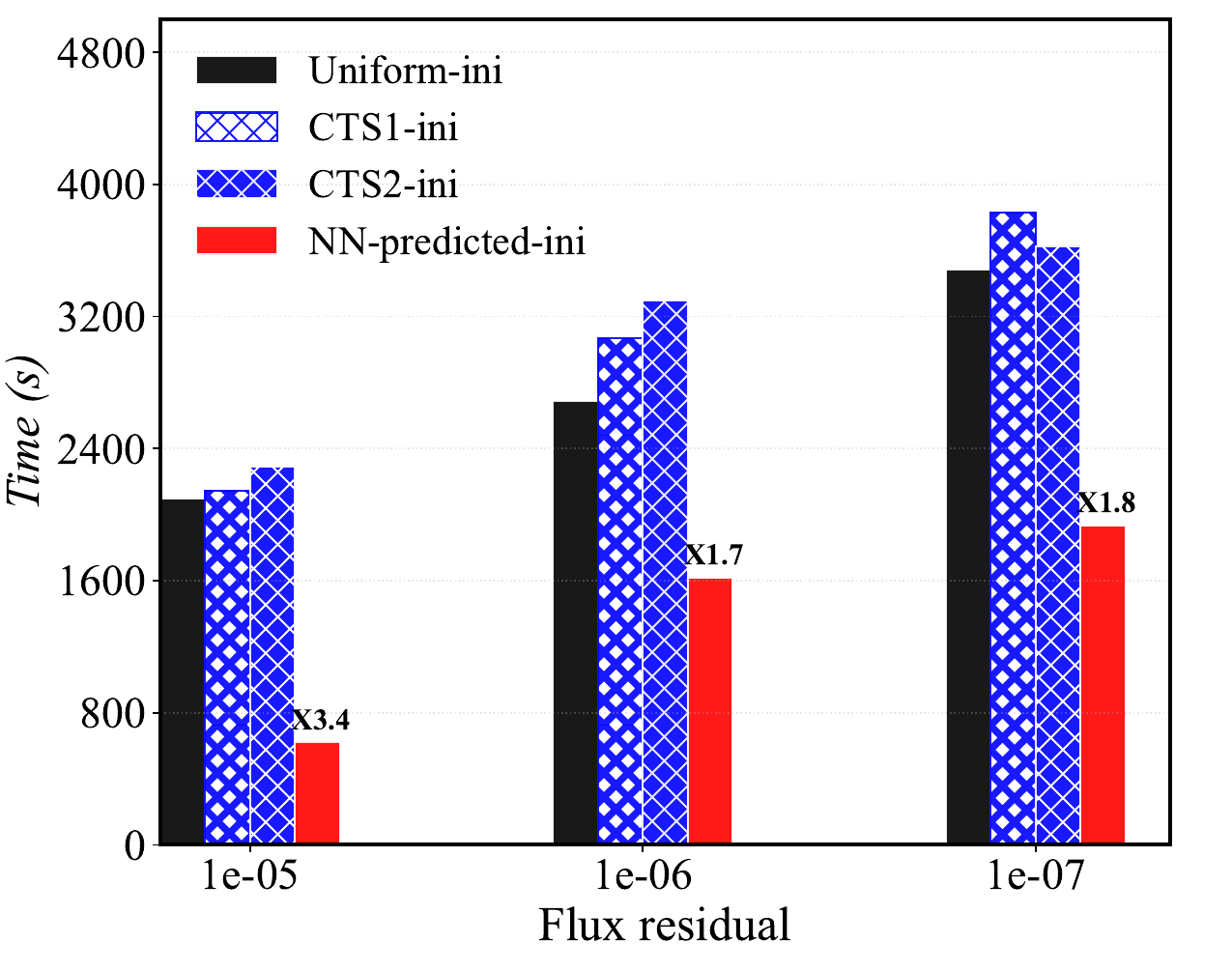}
\end{subfigure}
\\
\makebox[\textwidth][c]{{\large (e)}}

\caption{\label{Fig:acc_de} 
Comparison of flux residual convergence curves (left) and computation time required to reach the convergence threshold of $5\times10^{-6}$, $5\times10^{-7}$ and $5\times10^{-8}$ (right) respectively under four initialization conditions: uniform, CTS1,CTS2 and NN-predicted (based on a 10-step moving average). (d) Test case $T_{4}$($4.25$m/s,$\,4^{\circ}$),  CTS1= ($5.0$m/s,$\,3^{\circ}$), CTS2=($4.25$m/s,$\,5^{\circ}$). (e) Test case $T_{5}$($5.0$m/s,$\,6^{\circ}$), CTS1=($4.25$m/s,$\,7^{\circ}$),  CTS2=($5.0$m/s,$\,5^{\circ}$).  }
\end{figure*}

Fig.~\ref{Fig:V_compare} presents the contour plots of the velocity field $V$ near the propeller disk plane of the suboff model. It can be observed that the neural network can accurately output the distribution of high and low velocity regions consistent with the CFD results, and this distribution consistency provides a core guarantee for the implementation of the DDFI acceleration strategy.

Notably, although data-driven neural network models are capable of generating results visually highly similar to real flow fields, they still face significant limitations when directly applied to practical engineering scenarios. As shown in the third column of Figs.~\ref{Fig:v3.765} to \ref{Fig:v2}, in key characteristic regions of the suboff model (such as the bow and stern), almost all flow field variables (especially the pressure field) exhibit more pronounced prediction deviations. These errors directly reflect the physical inconsistency of the neural network's prediction results in local regions. This non-smoothness and non-conservation, on the one hand, cause the local flow field to deviate from real physical mechanisms, and on the other hand, significantly affect the reliability of hydrodynamic performance indicators. Therefore, a traditional CFD solver should be incorporated to constrain and correct the NN-predicted results through the basic physical conservation laws embodied in Eq.~(\ref{conti}) and Eq.~(\ref{N_S}), thereby simultaneously balancing the efficiency of the neural network and the reliability of results enabled by physical equation constraints.

\begin{figure*}[t]  
\centering

\begin{subfigure}[b]{0.4\textwidth}
    \centering
    \includegraphics[width=0.95\textwidth]{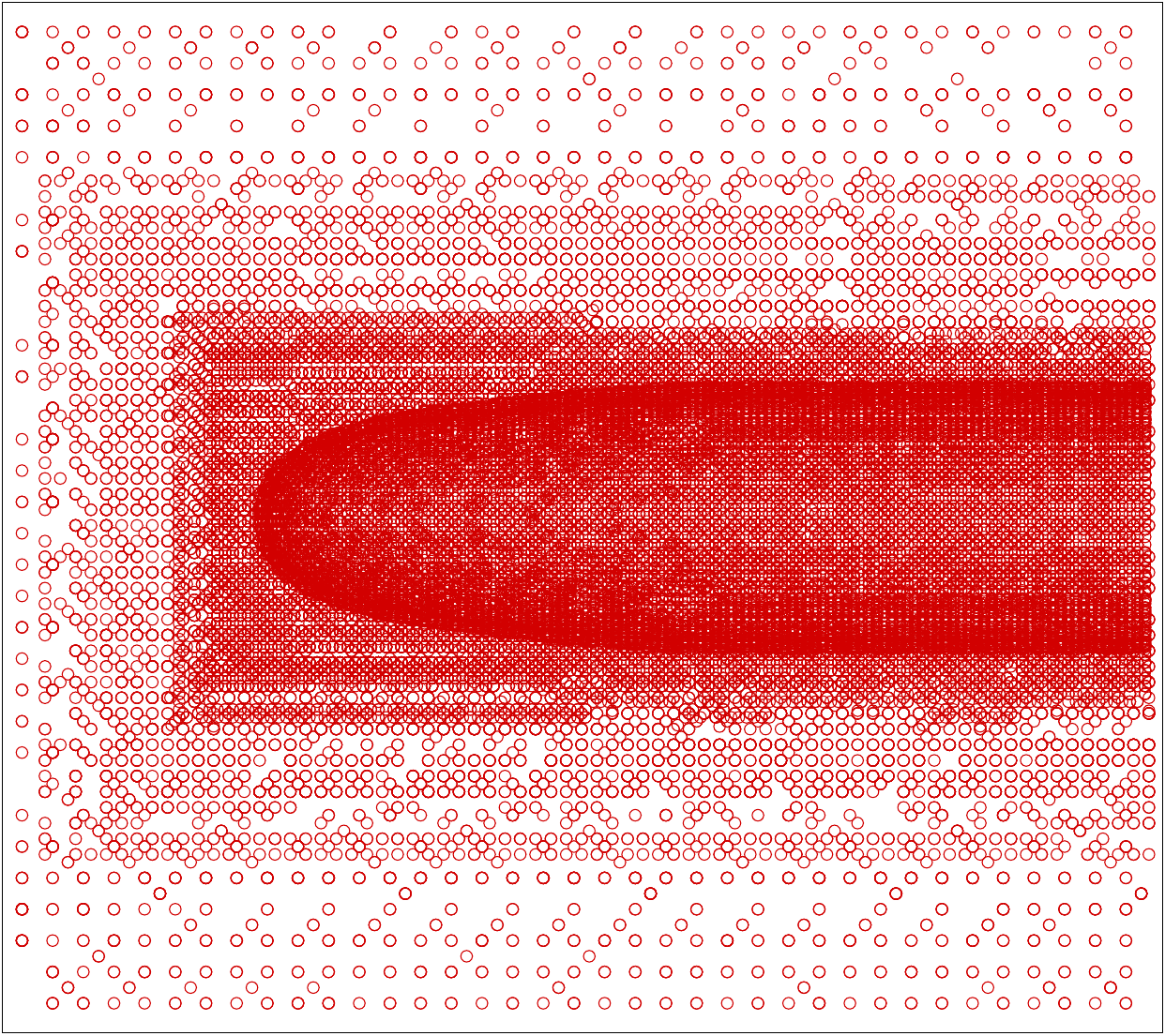}
    \caption{Sample interval=4} 
    \label{fig:phase_a}
\end{subfigure}
\quad  
\quad  
\quad
\quad
\begin{subfigure}[b]{0.4\textwidth}
    \centering
    \includegraphics[width=0.95\textwidth]{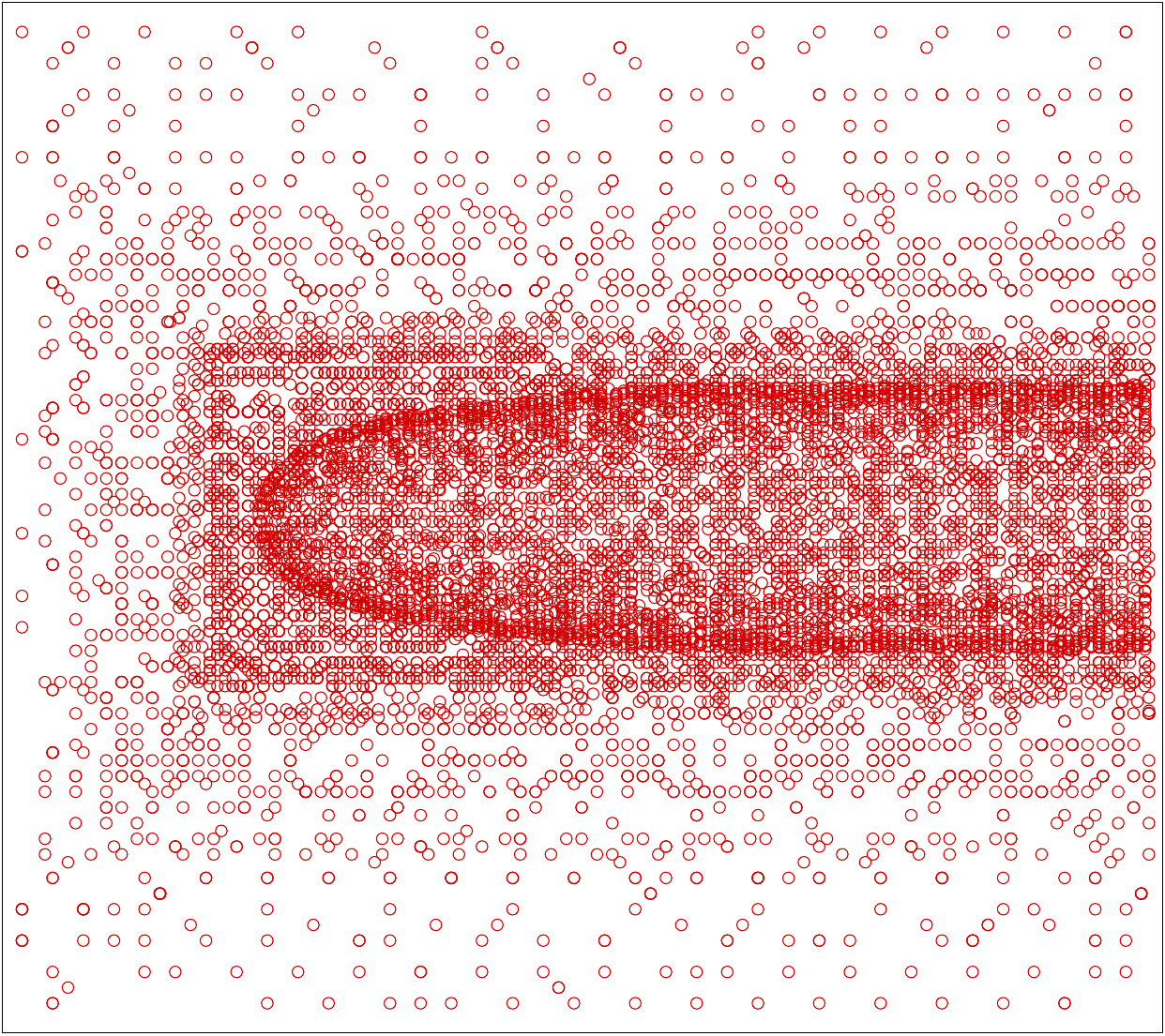}
    \caption{Sample interval=16}
    \label{fig:phase_b}
\end{subfigure}

\vspace{0.5em}  

\begin{subfigure}[b]{0.4\textwidth}
    \centering
    \includegraphics[width=0.95\textwidth]{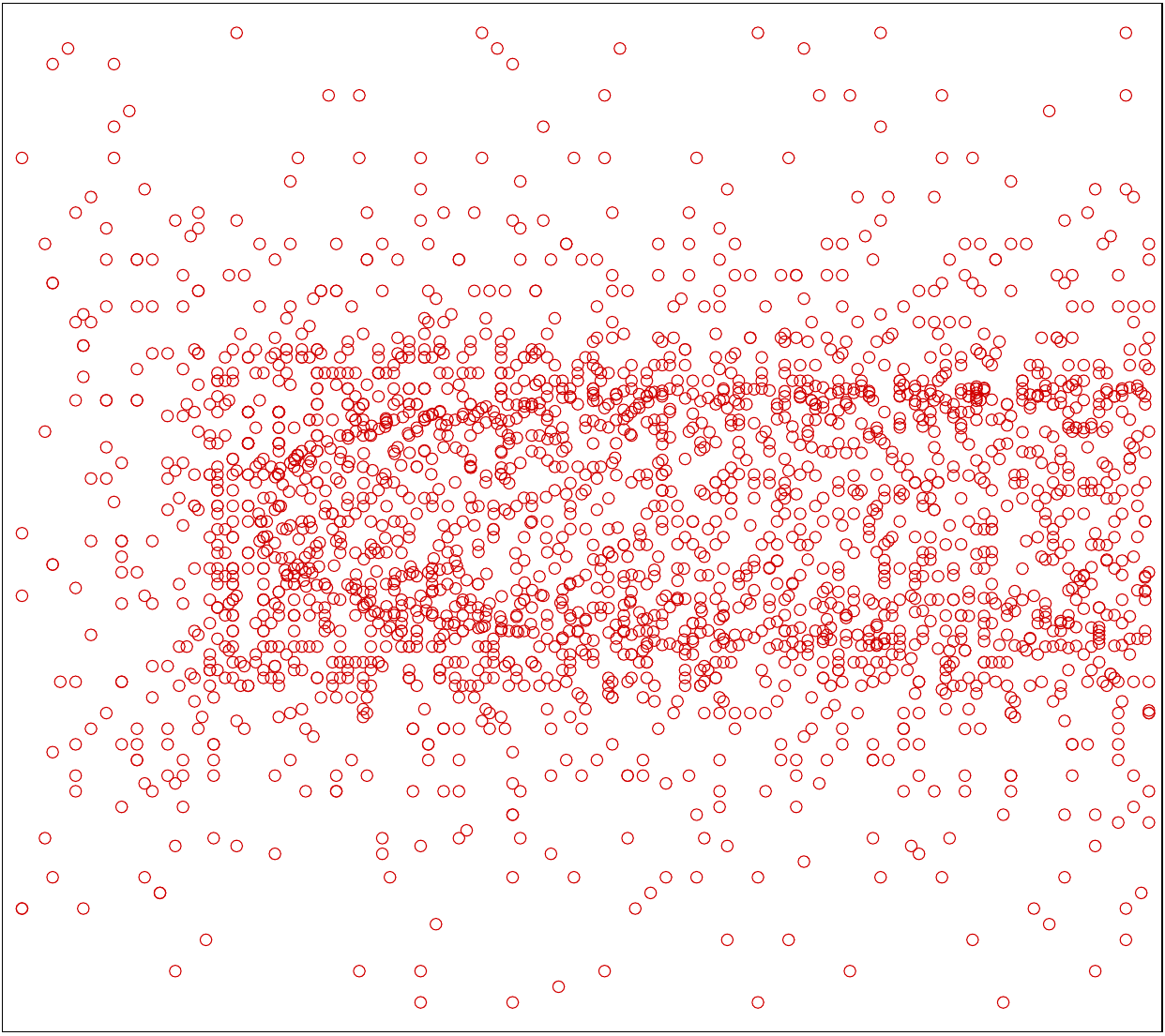}
    \caption{Sample interval=64}
    \label{fig:phase_d}
\end{subfigure}
\quad
\quad  
\quad
\quad
\begin{subfigure}[b]{0.4\textwidth}
    \centering
    \includegraphics[width=0.95\textwidth]{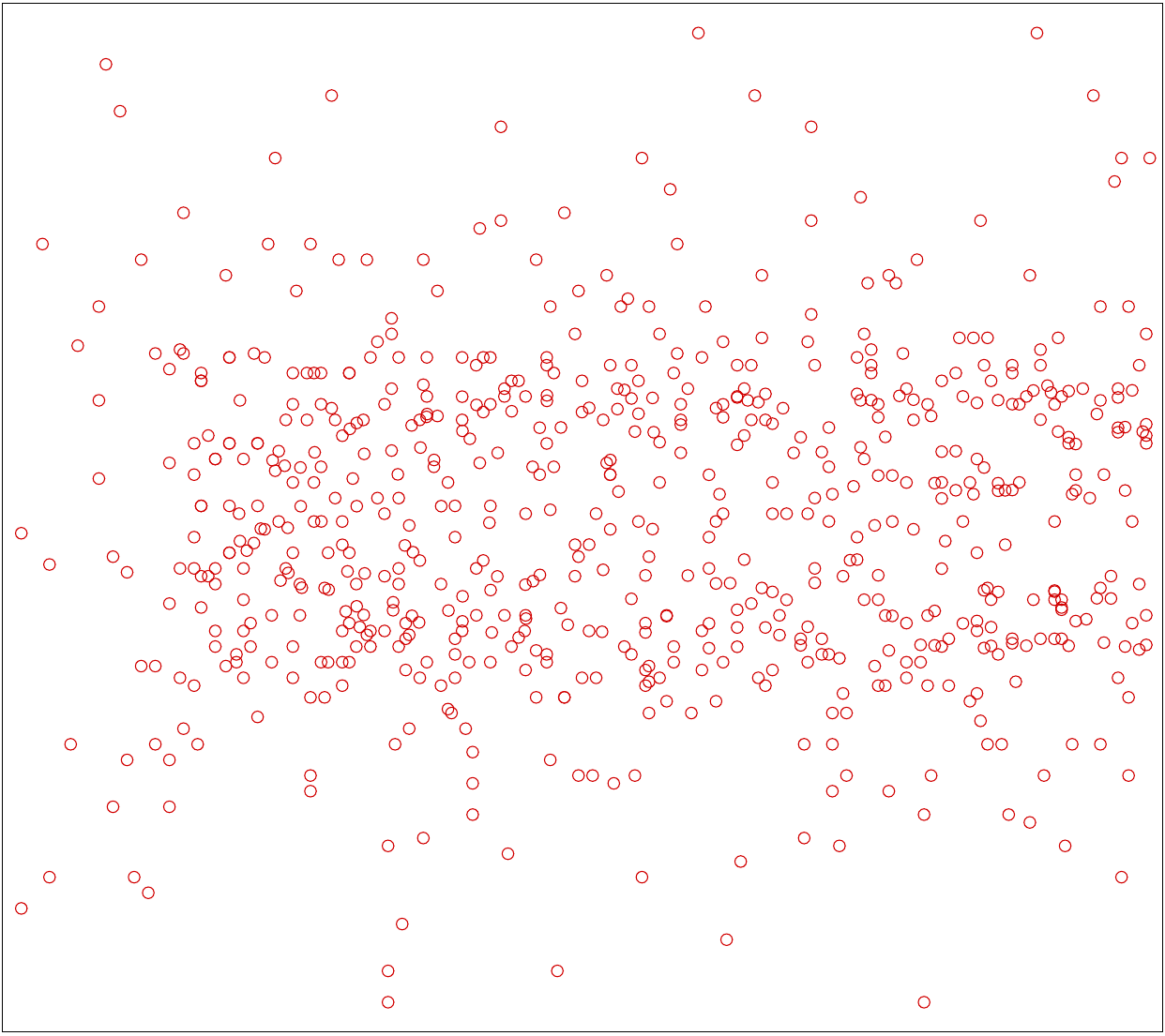}
    \caption{Sample interval=256}
    \label{fig:phase_e}
\end{subfigure}

\caption{Schematic diagram of the distribution of mesh cell  centroid near the bow of the suboff model at four sample intervals.}
\label{Fig:sample}
\end{figure*}

\begin{figure*}[!t]
\centering
\setlength{\tabcolsep}{0pt}
\renewcommand{\figurename}{Fig.}

\begin{subfigure}[b]{0.475\textwidth}  
    \centering
    \includegraphics[width=0.975\textwidth]{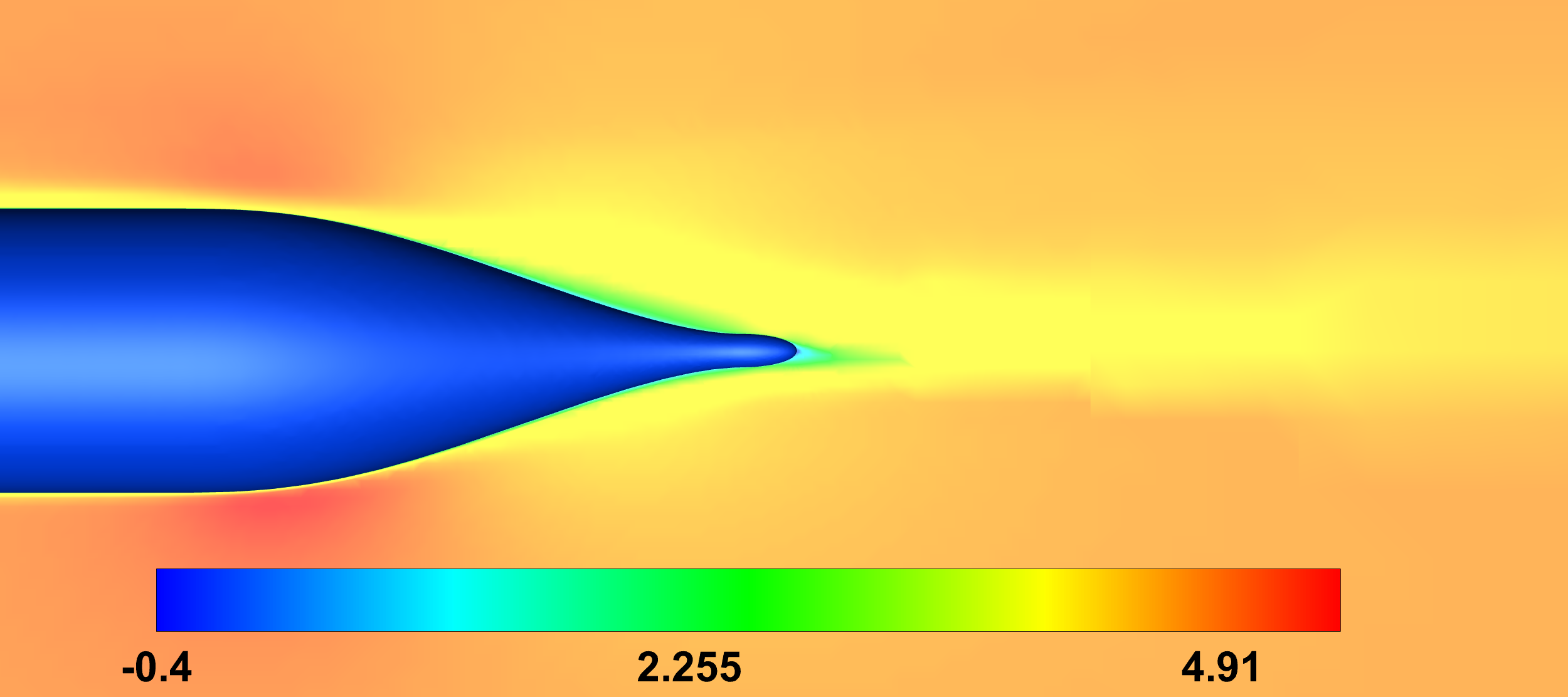}
    \caption{ CFD} 
    \label{subfig:case1_1}
\end{subfigure}
\hfill  
\begin{subfigure}[b]{0.475\textwidth}
    \centering
    \includegraphics[width=0.975\textwidth]{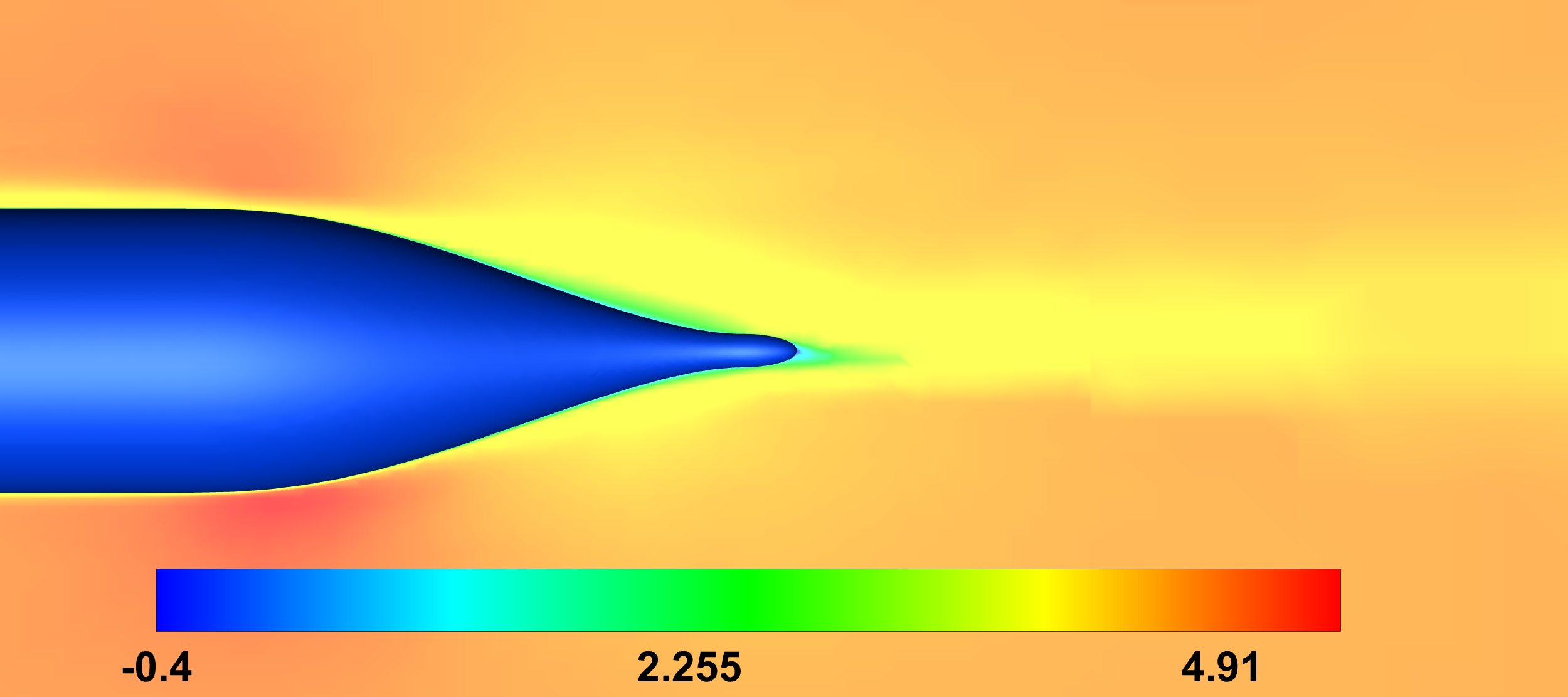}
    \caption{ NN-predicted (1)}  
    \label{subfig:case1_2}
\end{subfigure}
\vspace{0.2cm}  
\begin{subfigure}[b]{0.475\textwidth}
    \centering
    \includegraphics[width=0.975\textwidth]{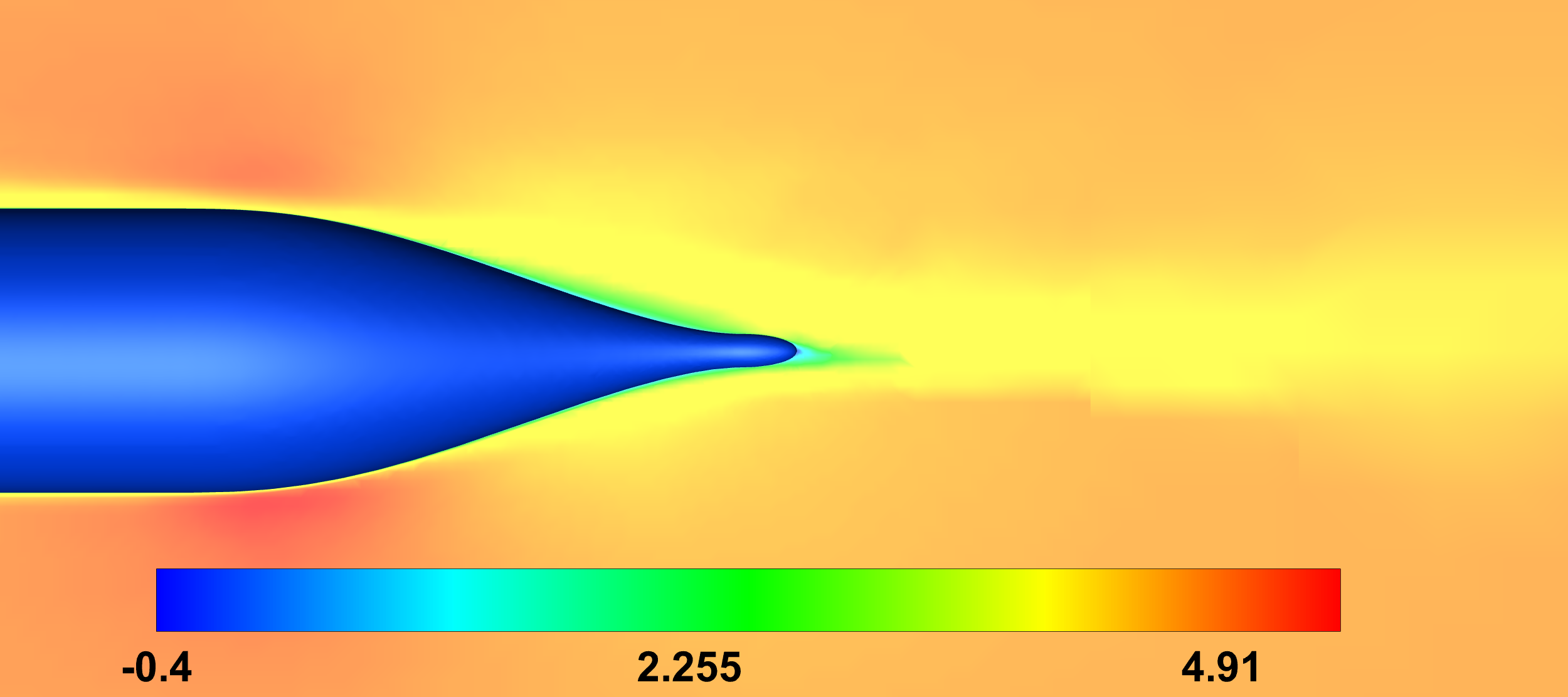}
    \caption{NN-predicted (4)} 
    \label{subfig:case2_1}
\end{subfigure}
\hfill
\begin{subfigure}[b]{0.475\textwidth}
    \centering
    \includegraphics[width=0.975\textwidth]{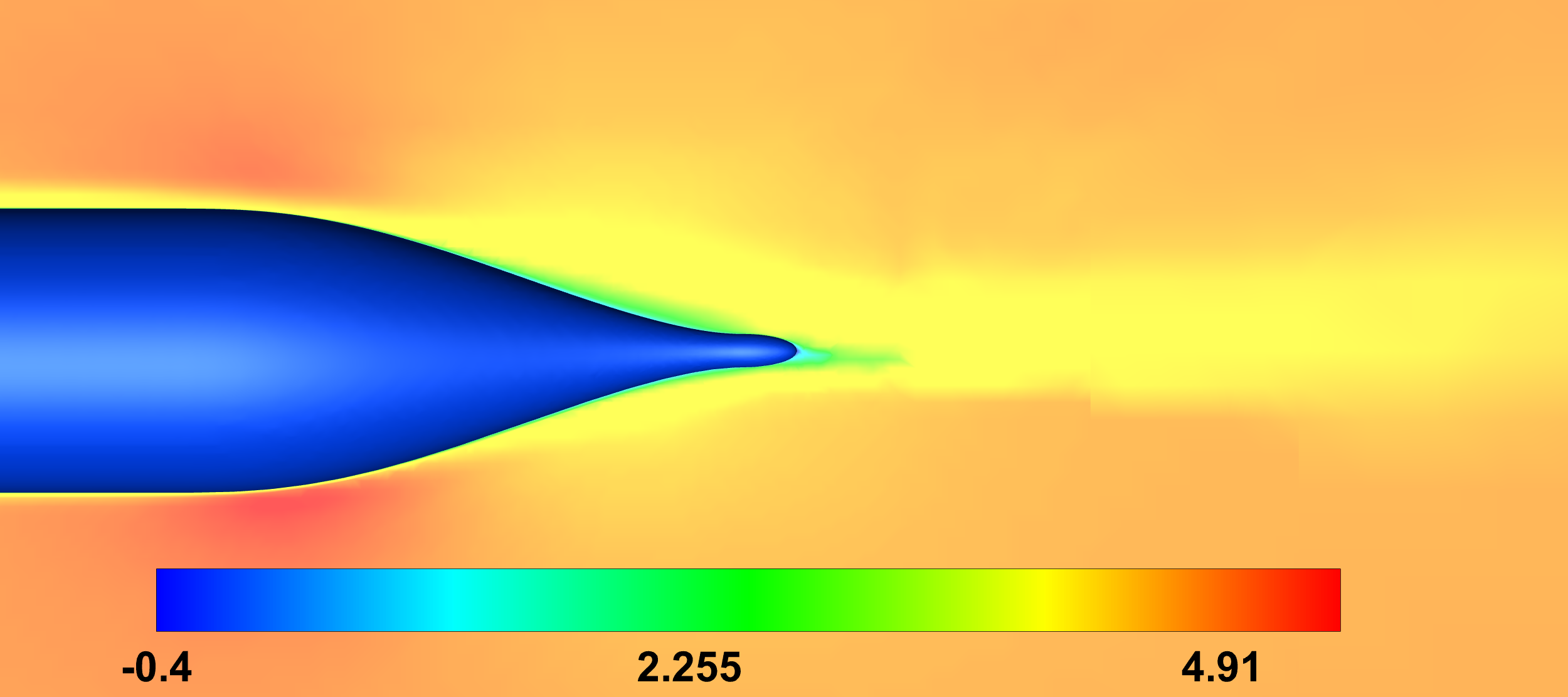}
    \caption{ NN-predicted (16)} 
    \label{subfig:case2_2}
\end{subfigure}
\vspace{0.2cm}

\begin{subfigure}[b]{0.475\textwidth}
    \centering
    \includegraphics[width=0.975\textwidth]{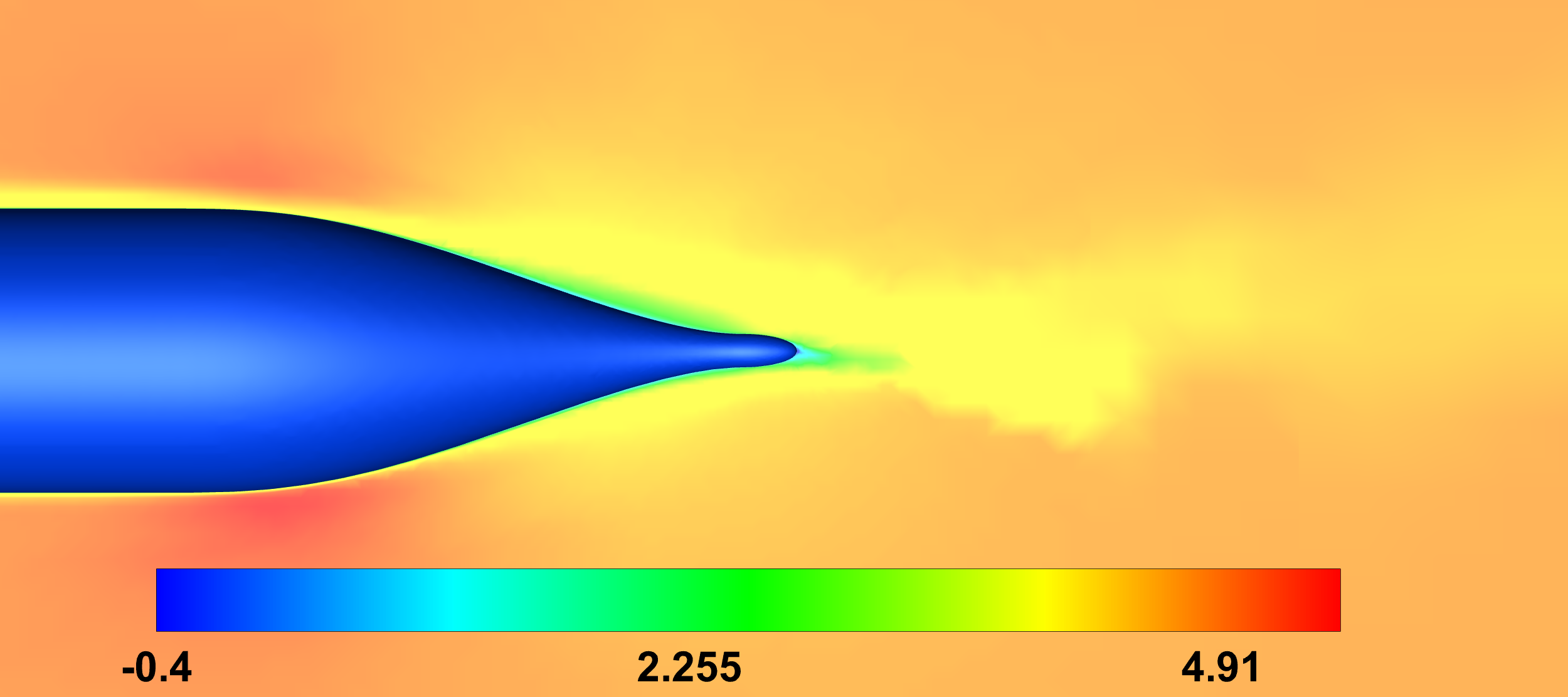}
    \caption{ NN-predicted (64)}  
    \label{subfig:case3_1}
\end{subfigure}
\hfill
\begin{subfigure}[b]{0.475\textwidth}
    \centering
    \includegraphics[width=0.975\textwidth]{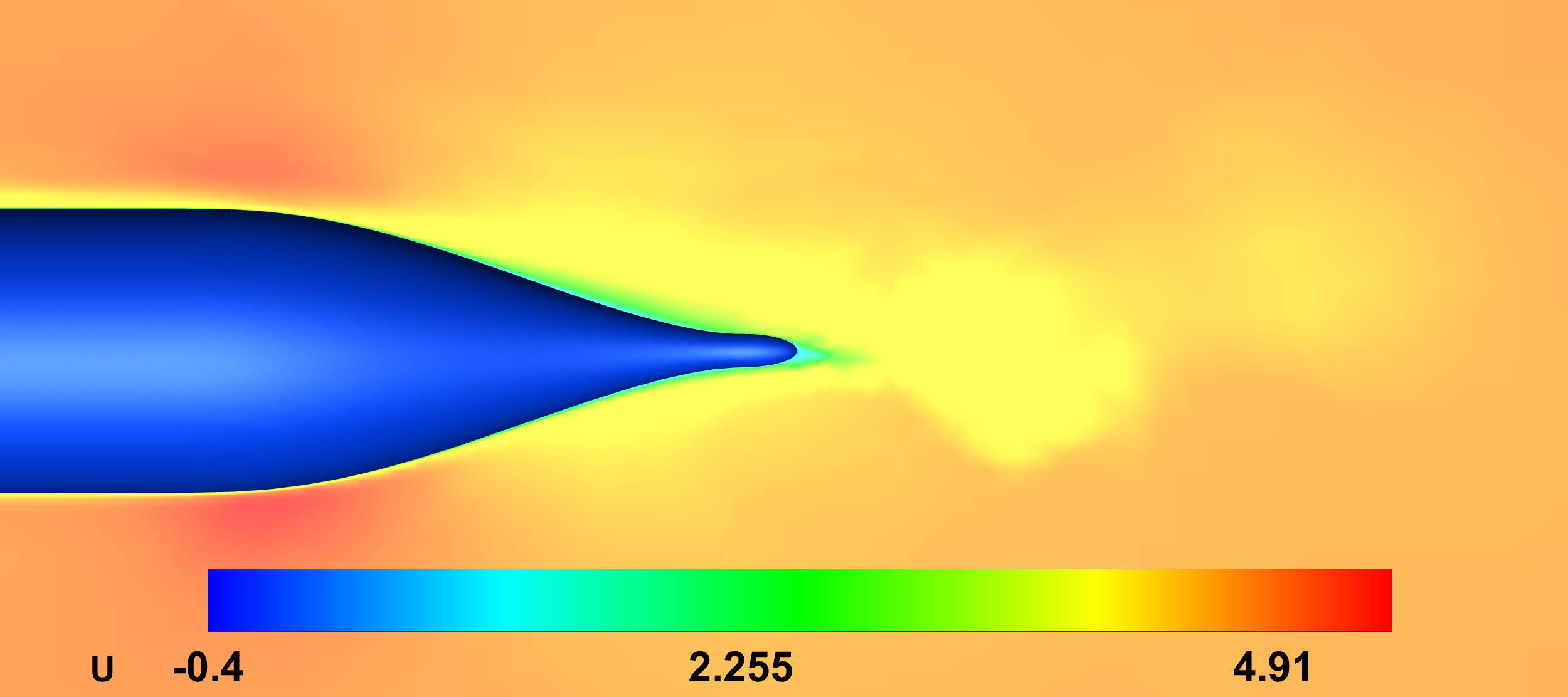}
    \caption{NN-predicted (256)}  
    \label{subfig:case3_2}
\end{subfigure}

\caption{Comparison of results of the $x$-direction velocity field $U$ at the stern of the suboff model for test case $T_{5}$ ($5.0$m/s$,\,6^{\circ}$). (a) CFD results. (b) NN-predicted results without sampling (sample interval=1), (c)-(f) sample interval =4,16,64,256, respectively.}
\label{Fig:sample_U}
\end{figure*}

Based on this, the prediction results of the neural network are employed as the initial flow field for CFD calculations, and a comparison is conducted with the case where a uniform initial field is used as the starting point for CFD solution, so as to quantify the acceleration effect of the former on the CFD solution process. In addition, to eliminate a potential question, i.e. whether CFD solution acceleration can also be achieved by directly adopting the flow field of similar operating conditions in the training set as the initial condition, two operating conditions in the training set that are closest to the test condition are additionally selected (hereafter referred to as CTS1 and CTS2), and their corresponding flow fields are also used as initial conditions to perform CFD calculations under the test condition. Besides, when the inlet velocity of the closest training set is inconsistent with that of the test condition, the dimensionless normalization method described in Section.~\ref{method} is adopted for scaling. The results demonstrate that the core of CFD solution acceleration lies in utilizing the neural network to learn general flow laws for accurate flow field prediction, rather than merely relying on the flow field data of similar operating conditions in the training set.

Specifically,  Figs.~\ref{Fig:acc_abc} and \ref{Fig:acc_de} illustrate the  convergence characteristics of flux residuals with iteration steps under different test cases. Obviously, across all test conditions, the decline rate of flux residuals is consistently faster when the flow field predicted by the neural network is adopted as the initial condition, compared with that when the uniform field is used as the initial condition.

Further investigation reveals that even the flow field of the most closest training condition  to the test condition is selected as the initial condition, its convergence rate is not superior to that of the uniform field initial condition. In particular, when the nonlinear characteristics of the test condition are enhanced (i.e., the inlet flow velocity increases), the closest training case initial condition often requires more iteration steps to reach the same residual convergence threshold.

In fact, when the flow field of an approximate condition is used as the initial condition, a phenomenon called pseudo-convergence tends to occur in the early stage of iteration: the flux residual drops rapidly to a magnitude close to $10^{-5}$, followed by a quick rebound, and then enters a plateau phase with minimal decline for hundreds of iteration steps. This is because even if the operating parameters are similar, the flow field distribution near the wall still exhibits significant differences, which requires the CFD solver to consume a large number of iteration steps for correction.

Fortunately, the neural network can generate flow field results with higher matching degree and accuracy corresponding to the target conditions, which implies a faster convergence speed and stronger robustness for the CFD solution process. As shown in the first column of Figure.~\ref{Fig:acc_abc} and~\ref{Fig:acc_de},  the flow field predicted by the neural network completely avoids the pseudo-convergence problem. Compared with the other three initialization strategies, the flux residual presents a faster and smoother downward trend. This result fully demonstrates the unique advantages of the DDFI strategy proposed in this paper in improving CFD solution performance.

The second column of Figure.~\ref{Fig:acc_abc} and ~\ref{Fig:acc_de}  quantitatively presents the acceleration effect of the DDFI strategy under three residual thresholds(). At the two relatively loose flux residual thresholds of $5\times10^{-6}$ and $1\times10^{-5}$, the DDFI strategy can achieve a maximum speedup ratio of $3.5$. Even at the two strictest flux residual thresholds of $5\times10^{-8}$ and $1\times10^{-7}$, DDFI can still realize a speedup ratio close to $2$  under the vast majority of test conditions. It should be noted that, the enhanced nonlinearity of the flow field induced by the increase in inlet flow velocity will lead to more oscillations during the convergence of flux residuals. The decline in acceleration effect observed in Fig.~\ref{Fig:acc_de} (d) is attributed to this phenomenon, which is also a common technical challenge faced in all CFD numerical solution processes.

\section{Discussion}
\subsection{Influence of Training Set Simplification}
\begin{figure*}[htbp]
\centering
    \begin{subfigure}[b]{0.9\textwidth}
    \centering
    \includegraphics[width=0.48\textwidth]{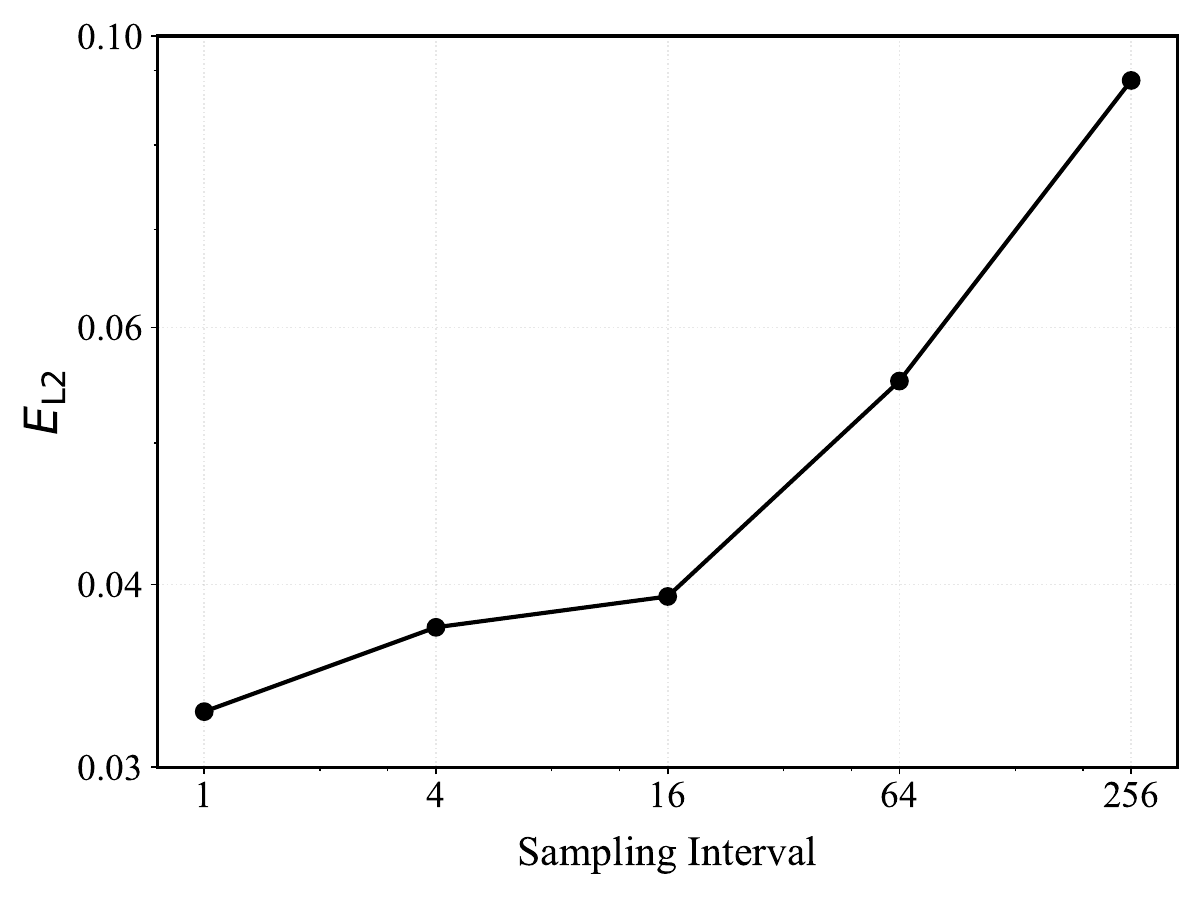}
    \end{subfigure}
\caption{The relative L2 error $E_{L2}$ (defined in Eq.~\ref{E_L2_1}) between CFD results and NN-predicted results under different sampling intervals for the test cases $T_{5}$.}
\label{Fig:E_L2_5_sample} 
\end{figure*}

\begin{figure*}[!t]
\centering
\setlength{\tabcolsep}{0pt}
\renewcommand{\figurename}{Fig.}
\begin{subfigure}[b]{0.475\textwidth}
    \centering
    \includegraphics[width=0.975\textwidth]{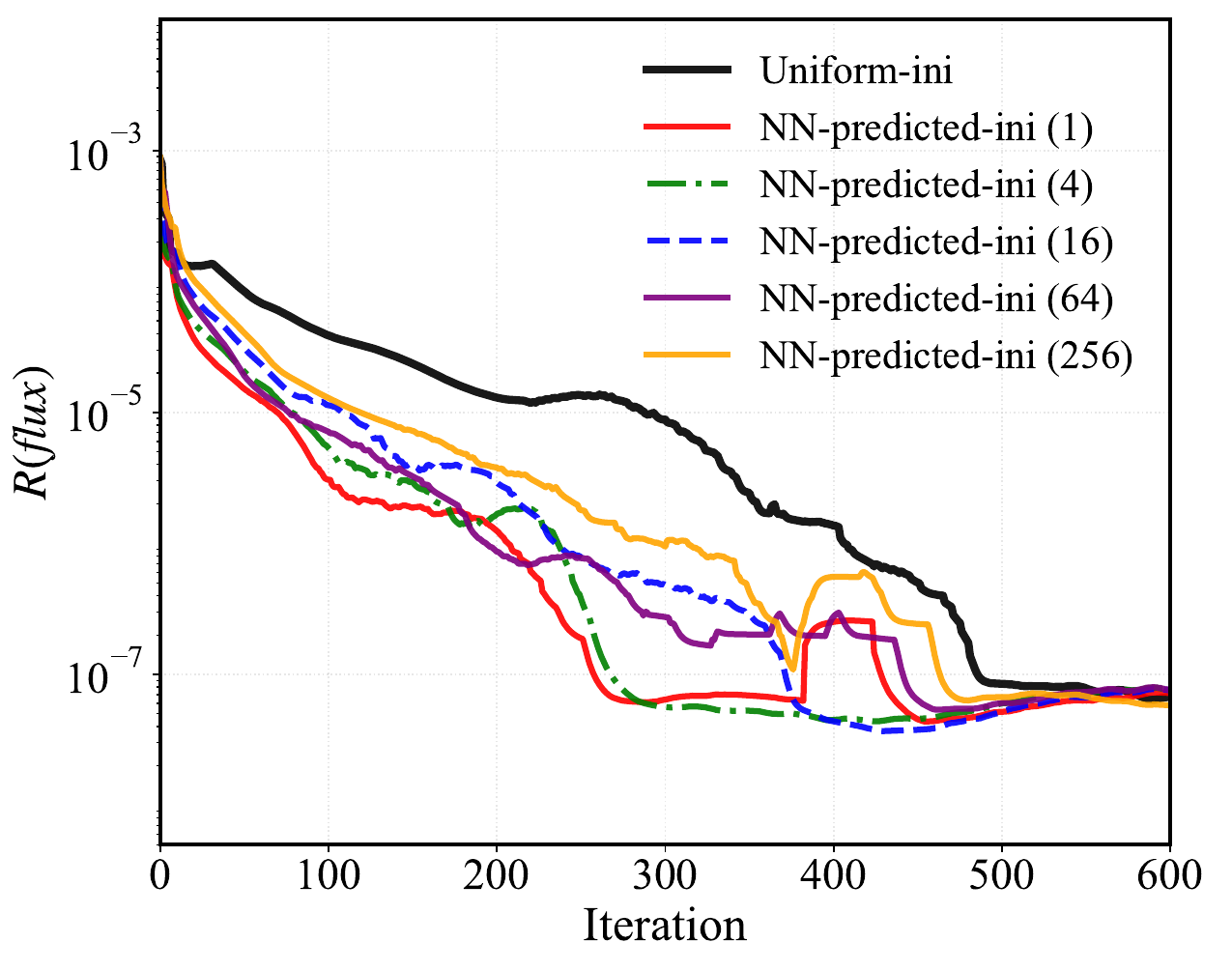}
\end{subfigure}
\quad
\begin{subfigure}[b]{0.475\textwidth}
    \centering
    \includegraphics[width=0.975\textwidth]{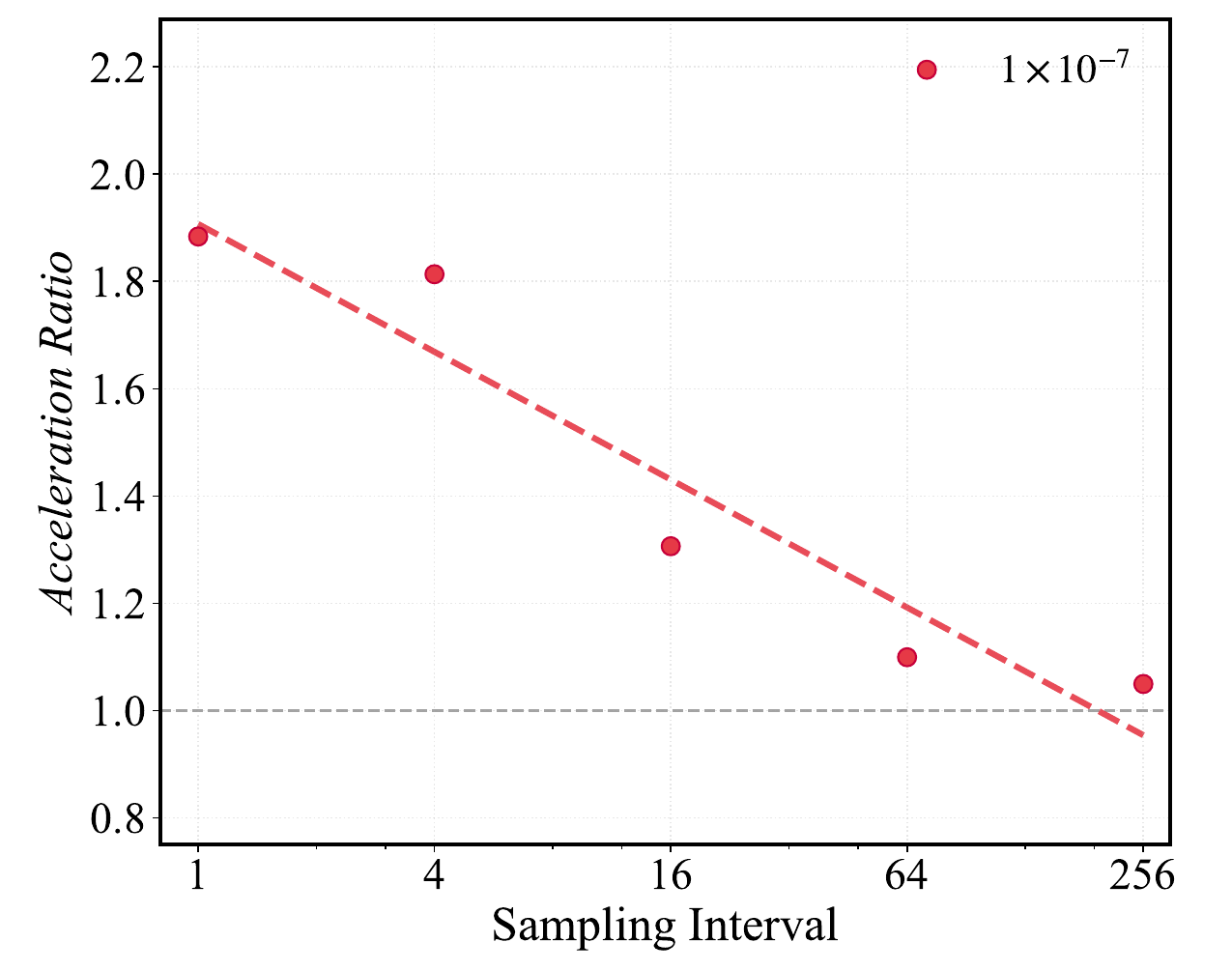}
\end{subfigure}
\\
\caption{Left: comparison of flux residual convergence curves with Uniform initialization and NN-Predicted results at different sampling intervals as initial conditions; Right: Acceleration ratio of NN-Predicted results at different sampling intervals compared with Uniform initialization when the convergence threshold is $R(flux)=1\times10^{-7}$ (based on a 40-step moving average). The test condition is $T_{5}$.
}
\label{Fig:V5_sampling}
\end{figure*}

For the aforementioned DDFI strategy, a core challenge lies in the fact that high-precision CFD data significantly expands the scale of the constructed neural network training dataset. Not only does this greatly increase the model training time, but it also limits the feasibility of incorporating more operating conditions into the training set. To explore the scalability of the DDFI strategy in multi-operating condition scenarios, this section focuses on investigating the technical feasibility of reducing the dataset size through sampling methods. Without loss of generality, this study adopts the most basic sampling paradigm, i.e. the mesh-number-based equidistant sampling strategy. Fig.~\ref{Fig:sample} presents schematic diagrams of the distribution of mesh cell centroid near the bow of the suboff at sampling intervals of 4, 16, 64, and 256. During the training process, the batch size is reduced by the same multiple as the sampling interval, while other training strategies remain consistent with those adopted for the full training set. It is noted that an early stopping strategy is employed to mitigate  overfitting. Accordingly, we constructed four subsets of different sizes from the original training set to investigate whether the DDFI strategy remains effective under insufficient training data conditions. Table.~\ref{tab:sample} presents the number of samples contained in each subset.

\begin{table}[h!t]
    \centering
    \small
    \caption{Number of input samples at different sampling intervals}
    \label{tab:sampling_dataset_params}
    \begin{tabular*}{\linewidth}{@{\extracolsep{\fill}} c c c c@{}}
        \toprule
        \multirow{2}{*}{\bfseries Sampling interval} & \bfseries Cells & \bfseries Faces & \bfseries Total \\
        & \bfseries  & \bfseries  & \bfseries  \\  
        \midrule
        1    & 5,341,428  & 182,520 & 5,523,948\\
        4    & 1,335,357  & 45,630   & 1,380,987\\
        16   & 333,839   & 11,408  & 345,247 \\
        64   & 83,460  & 2,852  & 86,312  \\
        256  & 20,866   & 712     & 21,578\\
        \bottomrule
    \end{tabular*}
    \label{tab:sample}
\end{table}

Fig.~\ref{Fig:sample_U} presents the pressure field prediction results of the neural network for the bow of the suboff model under test case $T_{5}$ ($\left|\mathbf{V}^{*} \right|=5.0 $m/s$, \alpha=6^{\circ}$). It can be observed that when the sampling interval is 4, the pressure field predicted by the neural network is visually identical to both the converged CFD results and the non-sampled results. As the sampling interval increases, the prediction accuracy gradually decrease. When the sampling interval reaches 256, obvious discrepancies emerge in the flow field prediction of the suboff bow region, with the range of the red high-pressure zone being significantly smaller than that of the CFD numerical solution. Fig.~\ref{Fig:E_L2_5_sample} presents the relative L2 errors under different sampling intervals according to Eq.(\ref{E_L2_1}). Besides, even under a considerably large sampling interval, the prediction results of the neural network do not deviate significantly from the CFD solution, which indicates the application potential of the DDFI strategy under sparse data conditions.

To further quantify the impact of sampling on the iterative acceleration effect of CFD flow fields, this section employs the NN-predicted flow fields trained with datasets of different sampling intervals as the initial conditions for CFD iterative calculations.  Fig.~\ref{Fig:V5_sampling}  presents the corresponding residual convergence curves and the speedup ratio (iteration number) relative to the uniform flow field initial condition. It can be observed that acceleration effect decreases as the interpolation interval increases. However, the convergence curve corresponding to the uniform flow field initial condition always lies above all other curves, indicating that under the framework of the DDFI strategy, the predicted flow fields corresponding to almost all the aforementioned sampling intervals can effectively improve the iterative convergence efficiency of CFD. Specifically, when the sampling interval is 4, the acceleration effect is nearly consistent with that of the non-sampled case. When the sampling interval is 256 (i.e., using only $0.39\%$ of the complete dataset), the strategy can still ensure that its convergence efficiency is consistently better than the uniform field initial condition, and achieves the strictest convergence threshold in nearly the same number of iterations as  the latter.

\begin{figure*}[!t]
\centering

\setlength{\tabcolsep}{0pt}
\renewcommand{\figurename}{Fig.}
\begin{subfigure}[b]{0.475\textwidth}  
    \centering
    \includegraphics[width=0.975\textwidth]{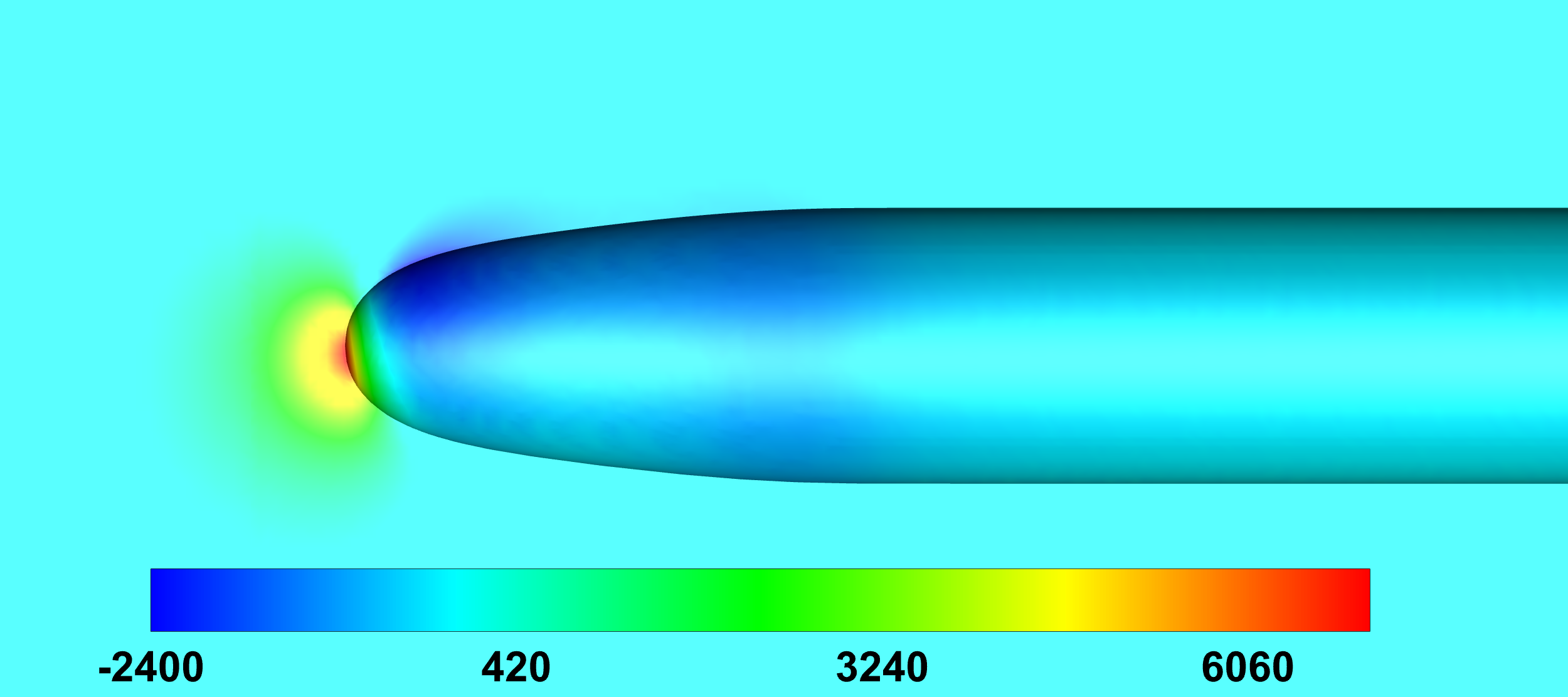}
    \caption{ CFD} 
    \label{subfig:case1_1}
\end{subfigure}
\hfill  
\begin{subfigure}[b]{0.475\textwidth}
    \centering
    \includegraphics[width=0.975\textwidth]{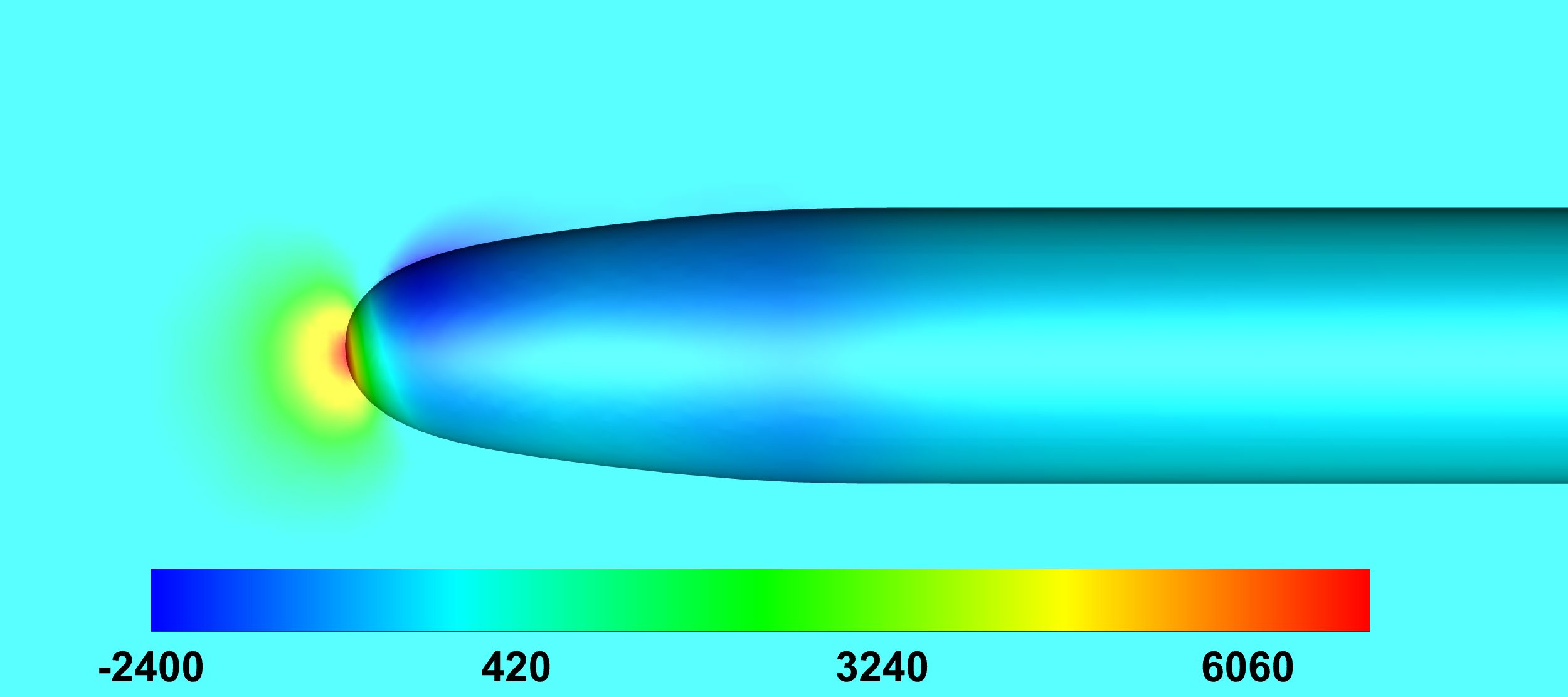}
    \caption{NN-predicted (1)}  
    \label{subfig:case1_2}
\end{subfigure}
\vspace{0.2cm} 

\begin{subfigure}[b]{0.475\textwidth}
    \centering
    \includegraphics[width=0.975\textwidth]{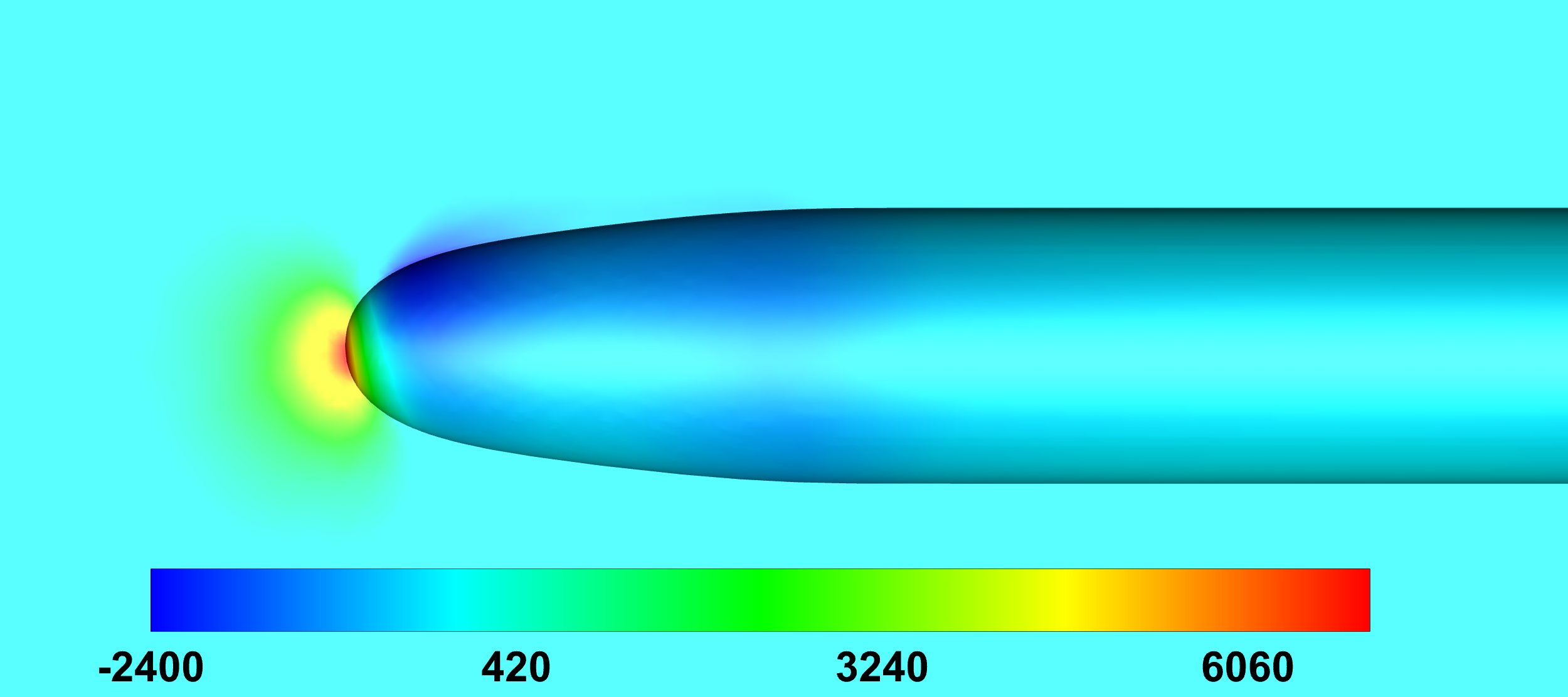}
    \caption{ NN-predicted (4)} 
    \label{subfig:case2_1}
\end{subfigure}
\hfill
\begin{subfigure}[b]{0.475\textwidth}
    \centering
    \includegraphics[width=0.975\textwidth]{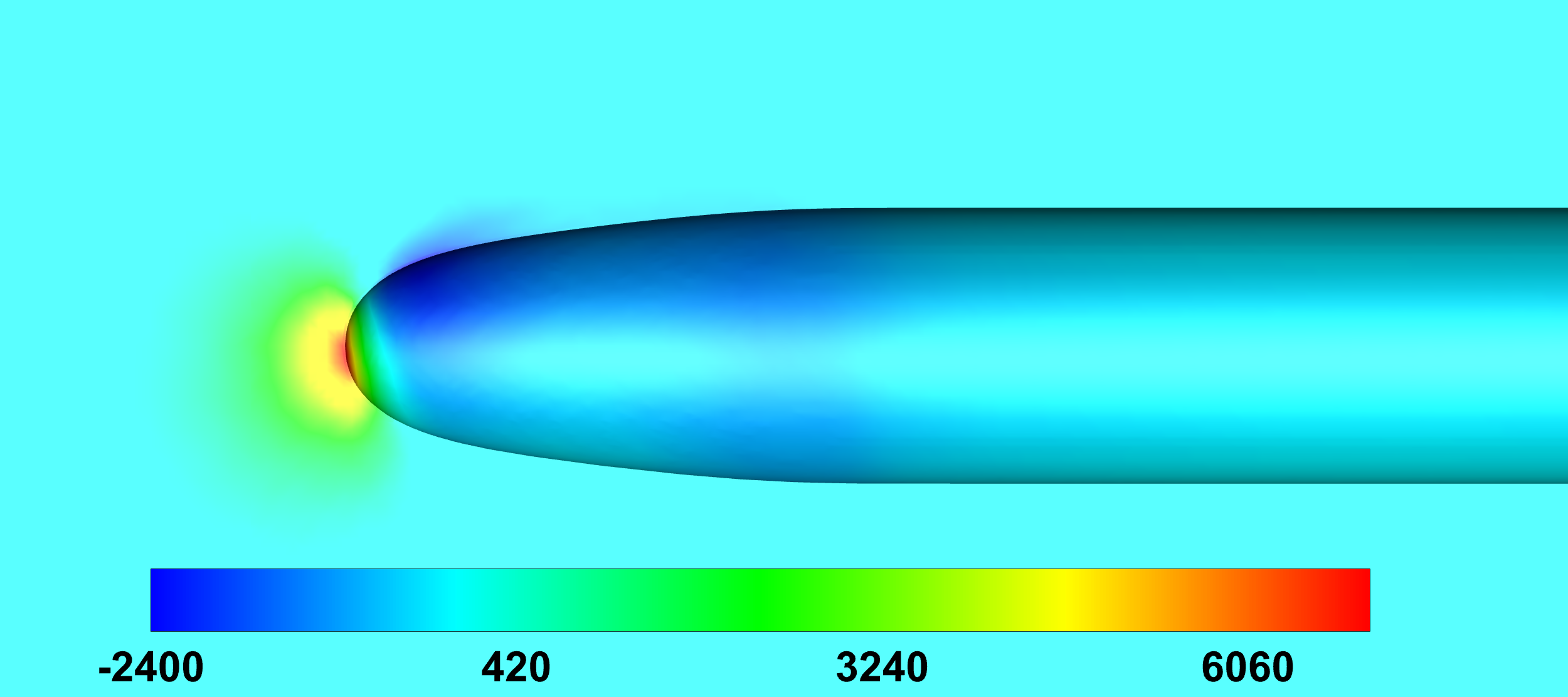}
    \caption{NN-predicted (16)} 
    \label{subfig:case2_2}
\end{subfigure}
\vspace{0.2cm}

\begin{subfigure}[b]{0.475\textwidth}
    \centering
    \includegraphics[width=0.975\textwidth]{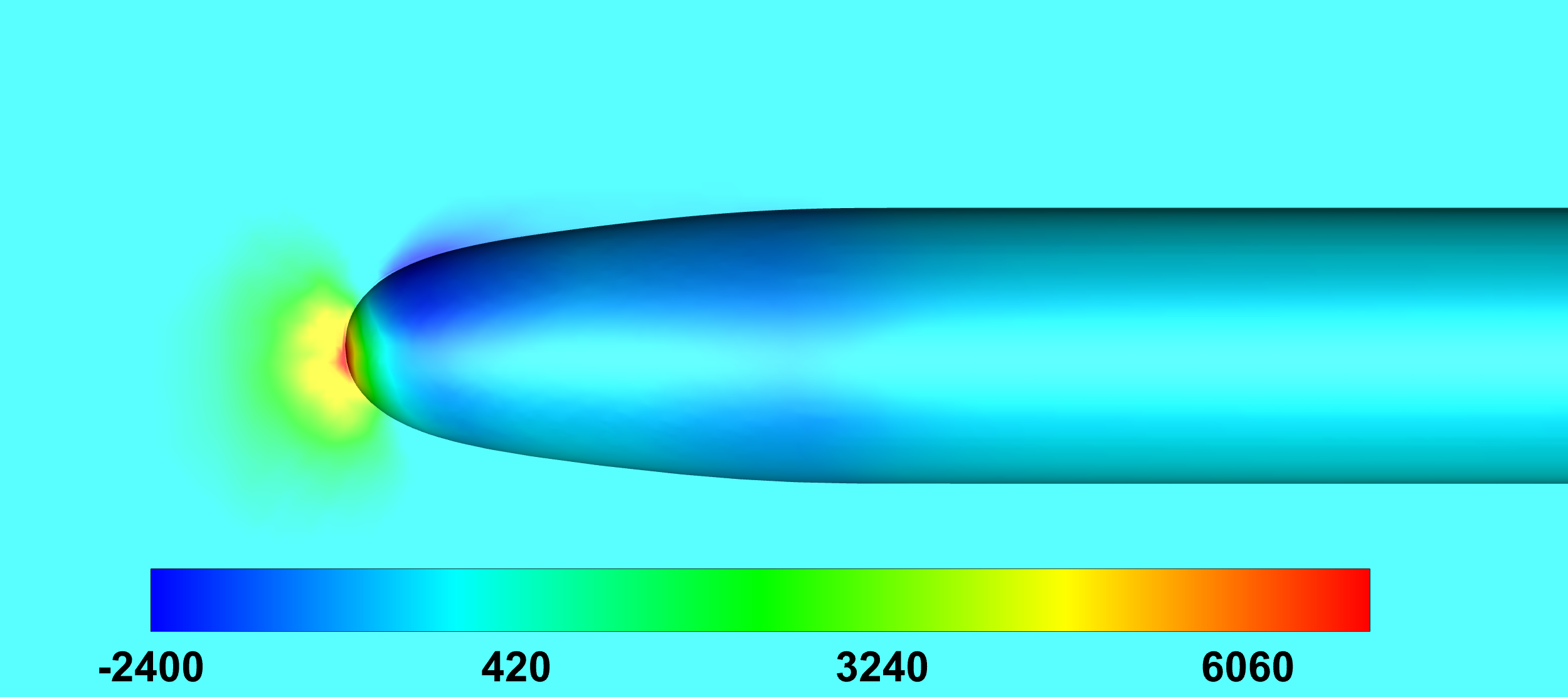}
    \caption{ NN-predicted (64)}  
    \label{subfig:case3_1}
\end{subfigure}
\hfill
\begin{subfigure}[b]{0.475\textwidth}
    \centering
    \includegraphics[width=0.975\textwidth]{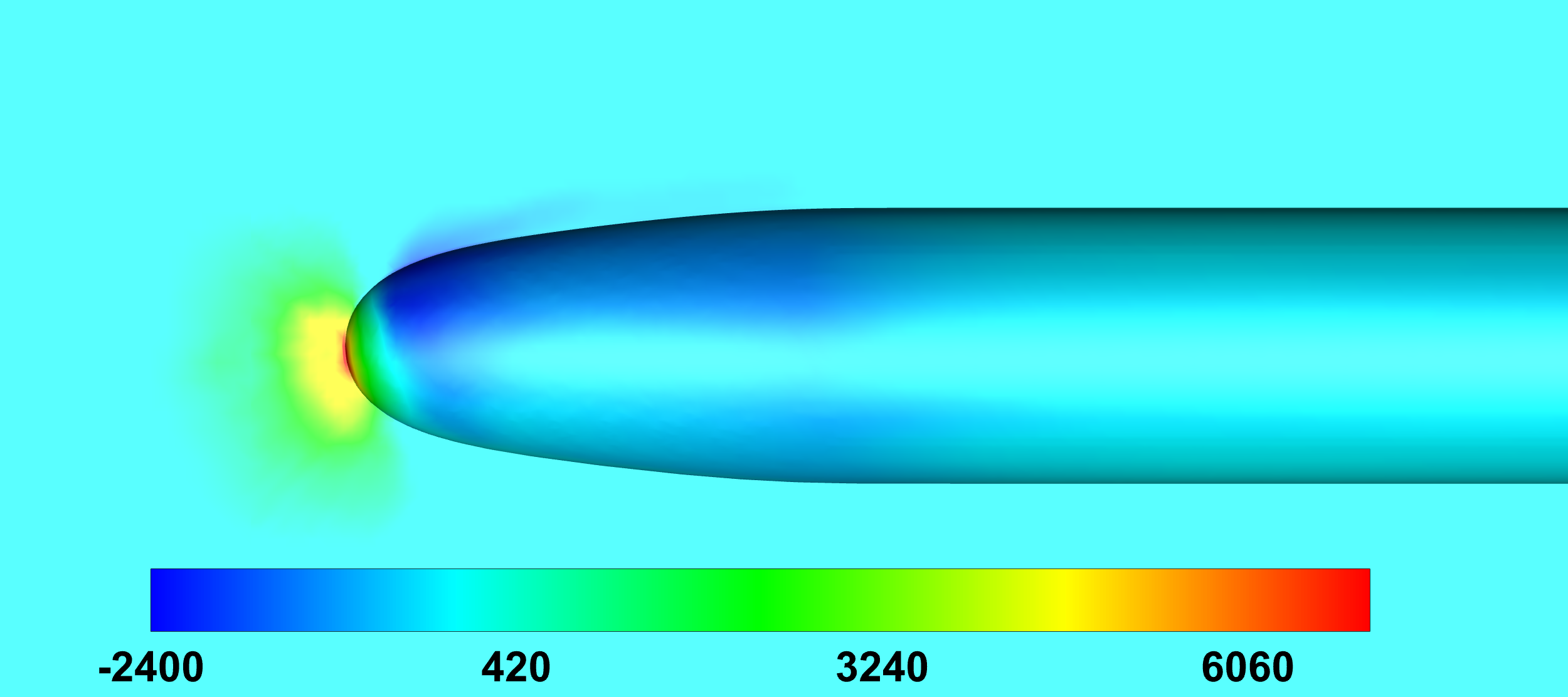}
    \caption{NN-predicted (256)} 
    \label{subfig:case3_2}
\end{subfigure}

\caption{Comparison of results of the pressure field $P$ at the bow of the suboff model for test case $T_{3}$ ($3.765$m/s$,\,6^{\circ}$). (a) CFD results. (b) NN-predicted results without sampling (sample interval=1), (c)-(f) sample interval =4,16,64,256, respectively.}
\label{Fig:sample_pres}
\end{figure*}

\begin{figure*}[!h]
\centering
    \begin{subfigure}[b]{0.9\textwidth}
    \centering
    \includegraphics[width=0.492\textwidth]{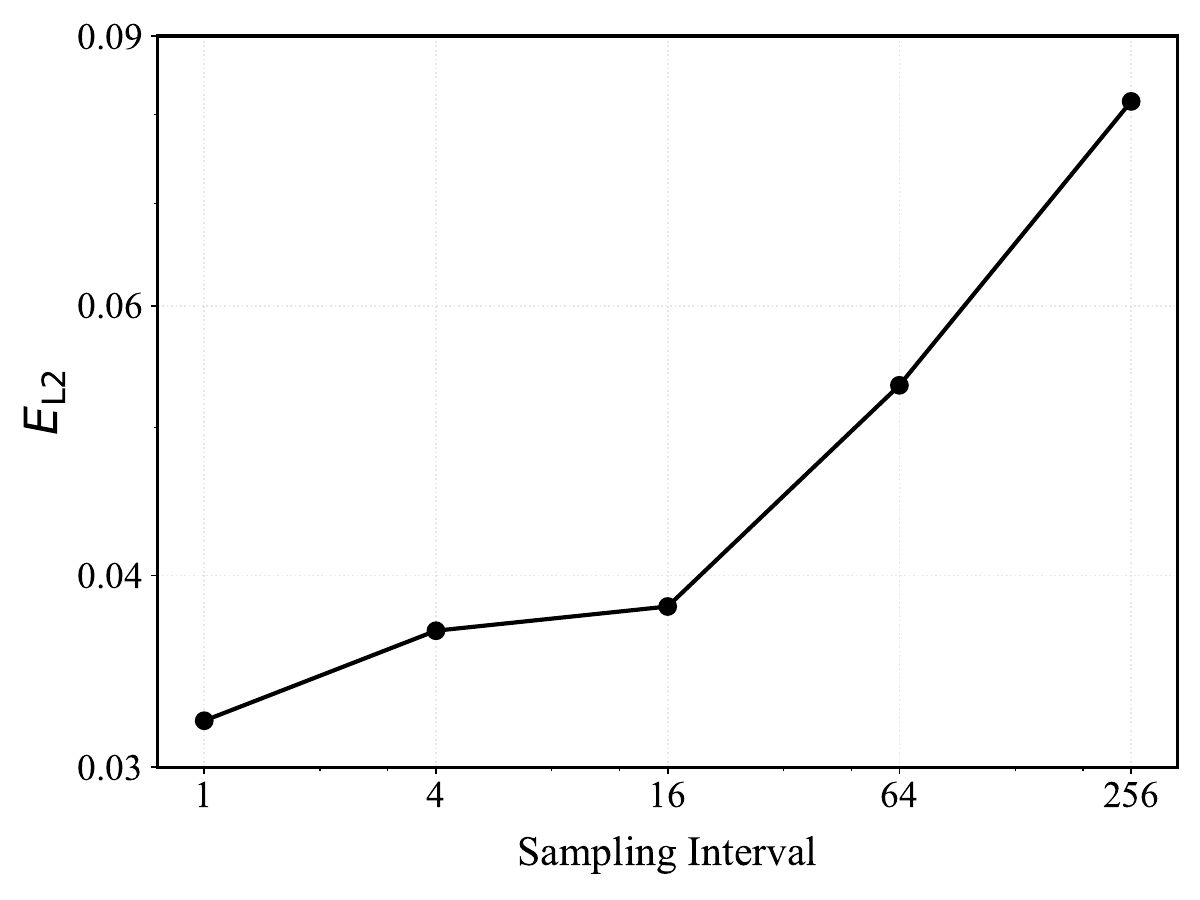}
    \end{subfigure}
\caption{The relative L2 error $E_{L2}$ (defined in Eq.~\ref{E_L2_1}) between CFD results and NN-predicted results under different sampling intervals for the test cases $T_{3}$.}
\label{E_L2_3.765_sample} 
\end{figure*}

\begin{figure*}[htbp]
\centering
\setlength{\tabcolsep}{0pt}
\renewcommand{\figurename}{Fig.}
\begin{subfigure}[b]{0.475\textwidth}
    \centering
    \includegraphics[width=0.975\textwidth]{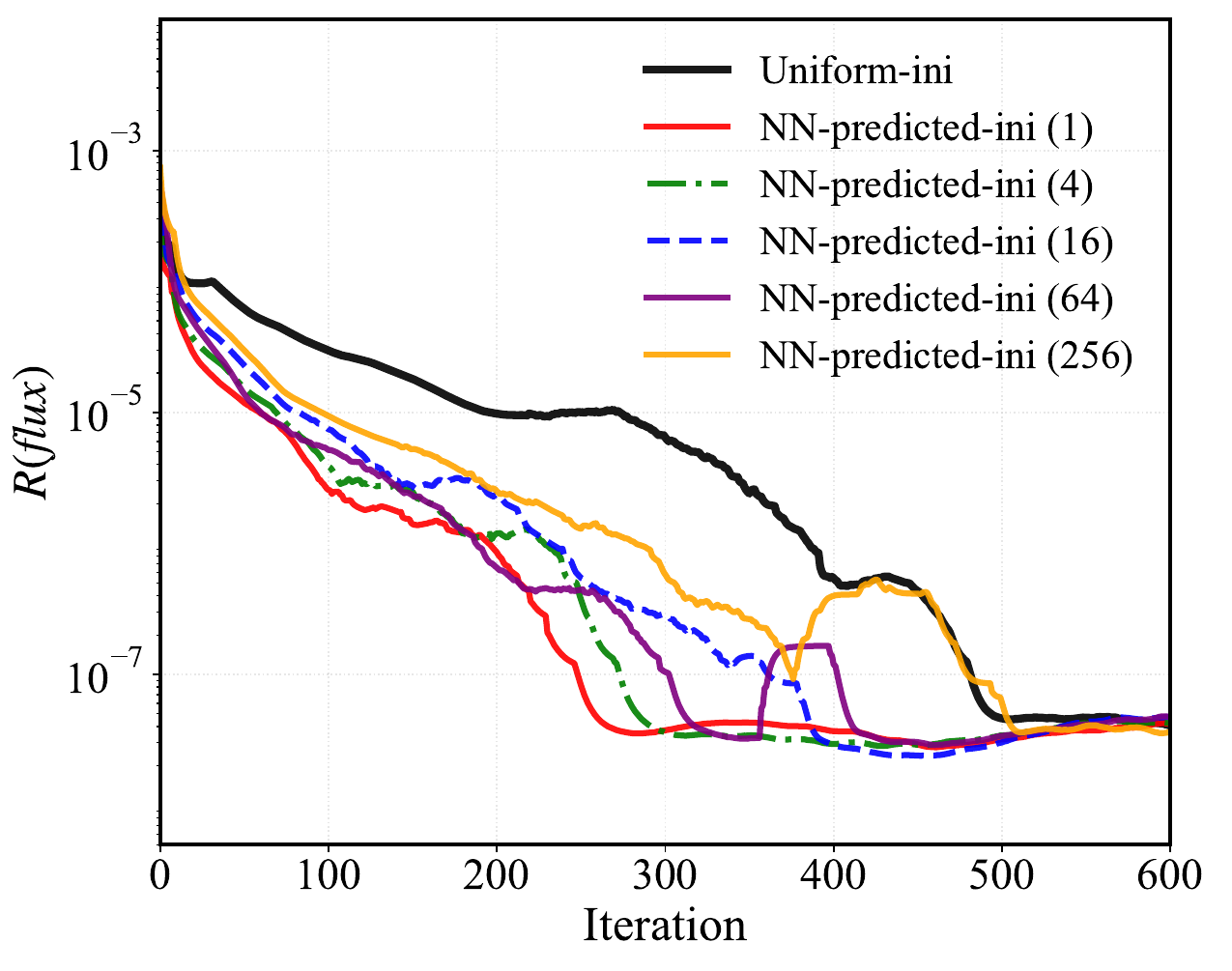}
\end{subfigure}
\quad
\begin{subfigure}[b]{0.475\textwidth}
    \centering
    \includegraphics[width=0.975\textwidth]{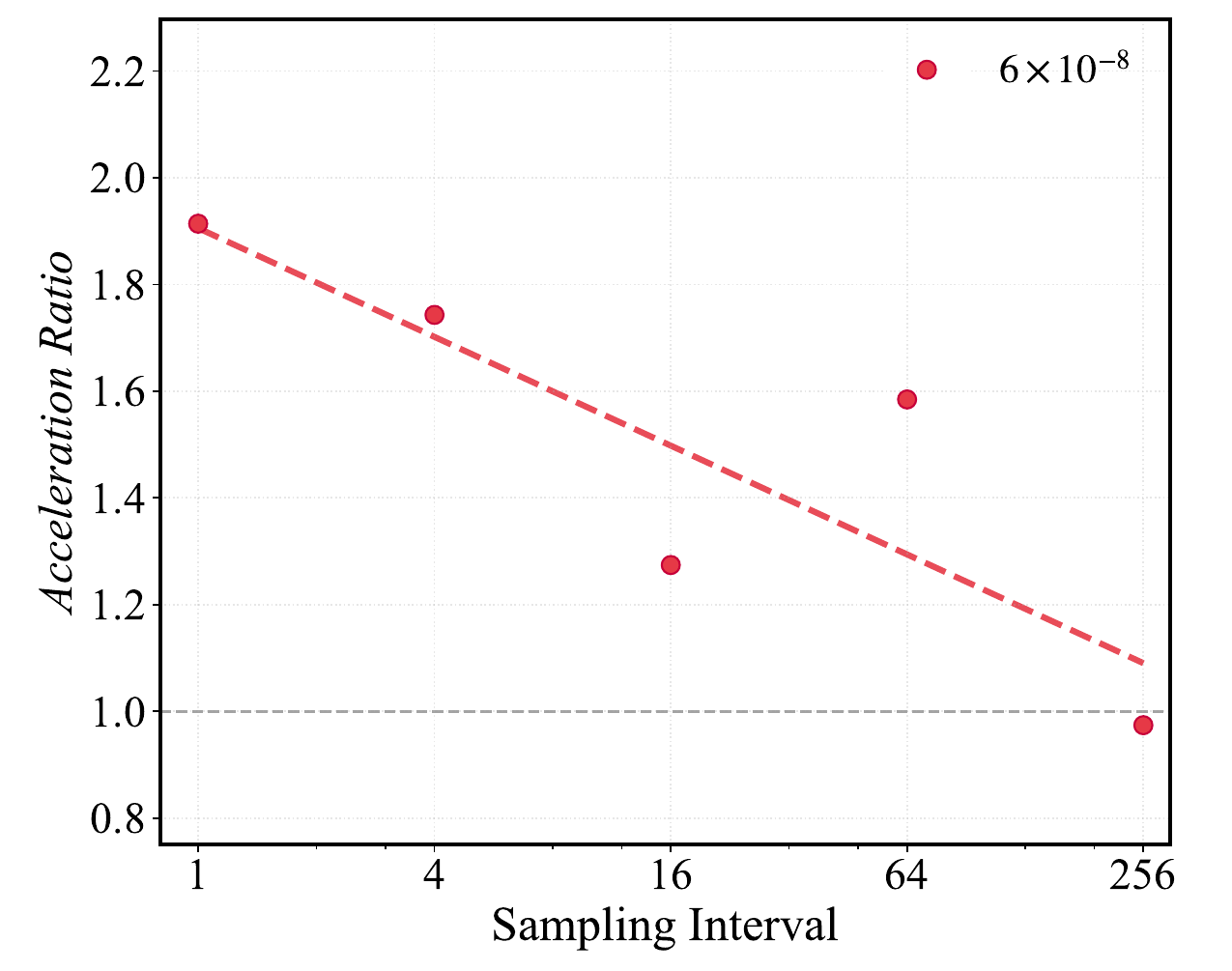}
\end{subfigure}
\\
\caption{Left: comparison of flux residual convergence curves with Uniform initialization and NN-Predicted results at different sampling intervals as initial conditions; Right: Acceleration ratio of NN-Predicted results at different sampling intervals compared with Uniform initialization when the convergence threshold is $R(flux)=6\times10^{-8}$ (based on a 40-step moving average). The test condition is $T_{3}$.}
\label{Fig:V3.765_sampling}
\end{figure*}

Fig.~\ref{Fig:sample_pres} and Fig.~\ref{E_L2_3.765_sample} present the  results of another test case $T_{3}$ ($\left|\mathbf{V}^{*} \right|=3.765$m/s, $\alpha=6^{\circ}$). Similarly, the prediction accuracy decreases as the sampling interval increases, the flow field distribution remains generally consistent with the CFD numerical simulation results. Fig.~\ref{Fig:V3.765_sampling} illustrates the residual convergence characteristics under this operating condition,  as well as the speedup ratio. The sampled dataset can still generate flow field initial conditions that accelerate the iterative solutions of CFD solver. Although the increase in sampling interval impairs the speedup effect, the acceleration function of the DDFI strategy remains effective even when the sampling interval is considerably large.

\begin{figure*}[!t]
\centering
\setlength{\tabcolsep}{0pt}
\renewcommand{\figurename}{Fig.}
\begin{subfigure}[b]{0.475\textwidth}
    \centering
    \includegraphics[width=0.975\textwidth]{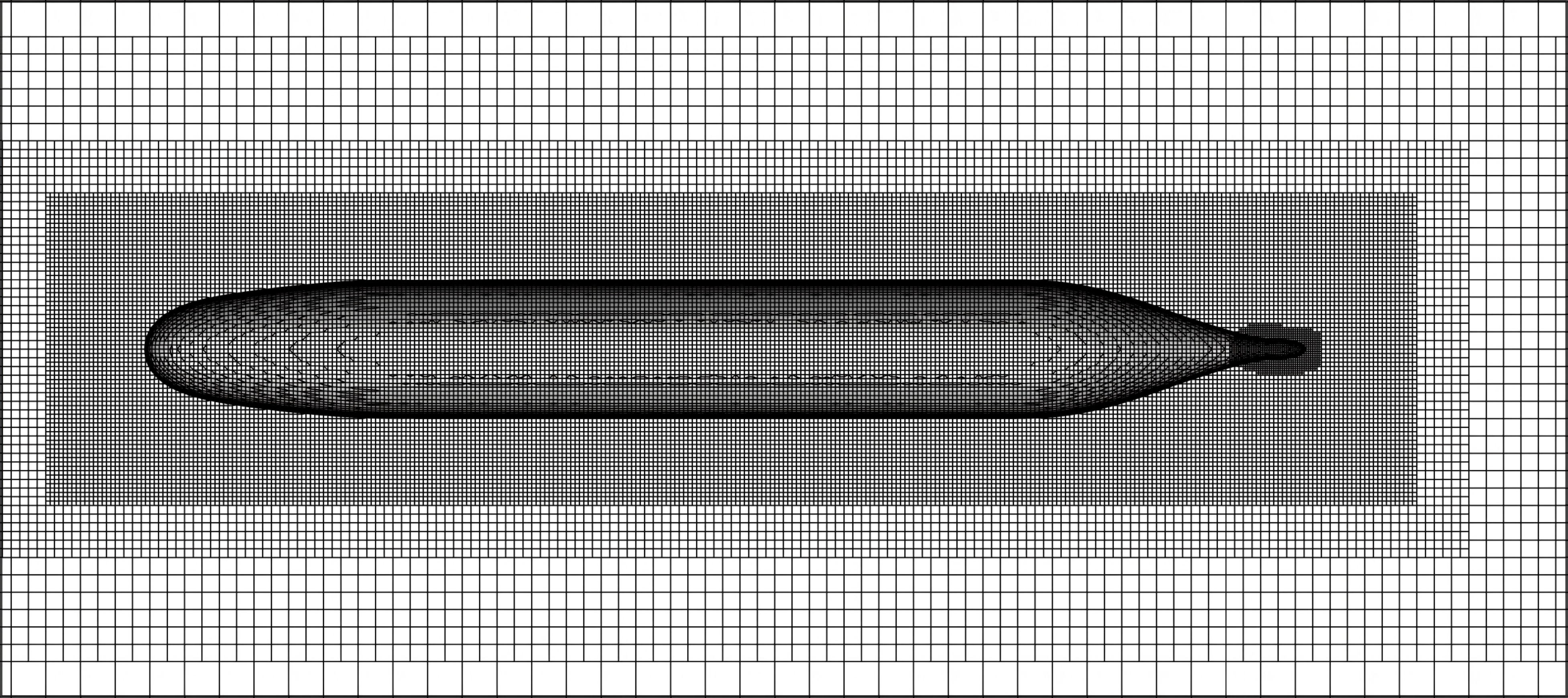}
    \caption{ Mesh$_{f1}$}  
\end{subfigure}
\quad
\begin{subfigure}[b]{0.475\textwidth}
    \centering
    \includegraphics[width=0.975\textwidth]{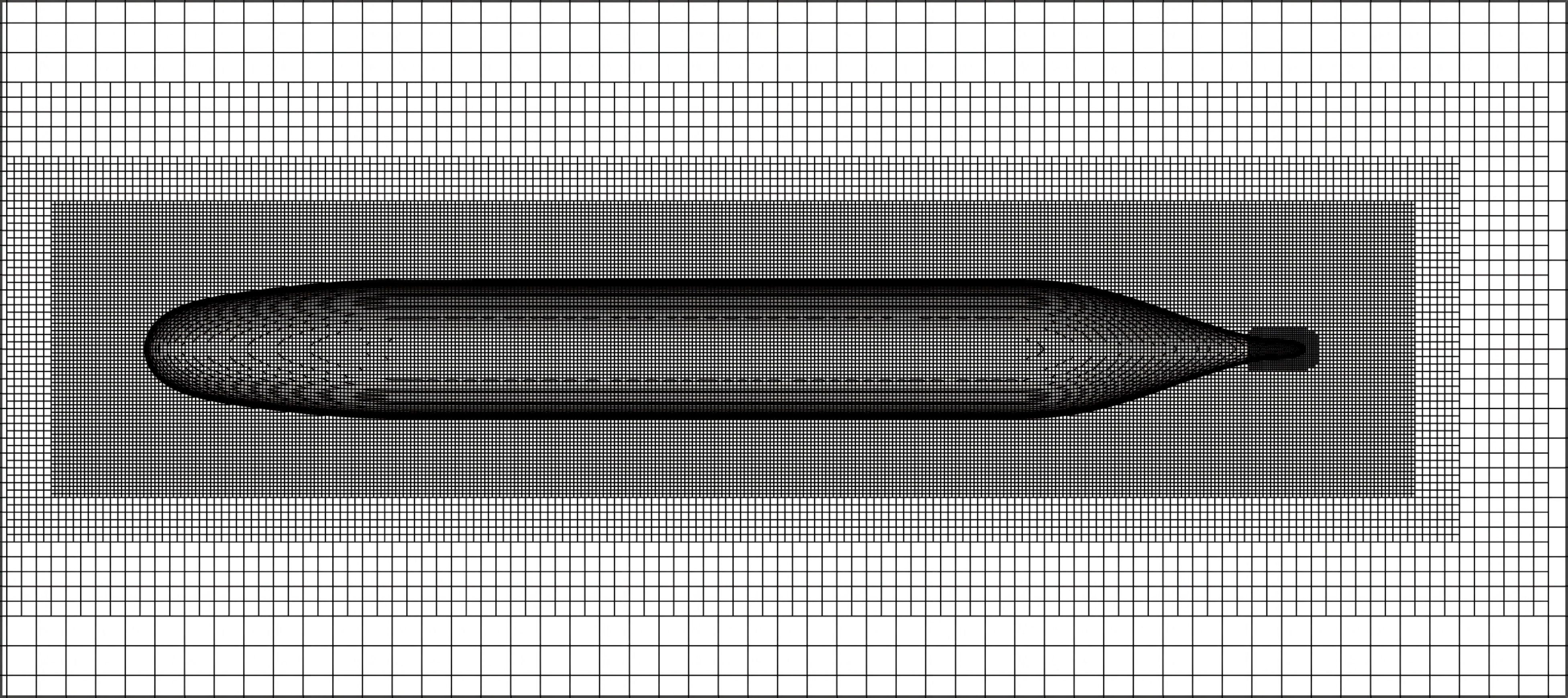}
    \caption{ Mesh$_{f2}$}  
\end{subfigure}
\\
\caption{\label{Fig:finemesh} 
Mesh distribution in the near-wall region of two refined meshes. (a) $7.9\times10^{5}$ cells.(b) $1.15\times10^{6}$ cells.} 
\end{figure*}
In summary, these two cases mentioned above draw an encouraging conclusion: reducing the size of the training set does not exert an essential impact on the acceleration performance of the DDFI strategy. For the cases considered in this study, when the sampling interval is set to 4, its influence on the acceleration effect of CFD solution is negligible. Under extreme conditions, a sampling interval of 64 or even 256 can be adopted. Such large sampling intervals enable the training set to cover more operating conditions, thereby enhancing the generalization capability of the model, which is of great significance for the engineering application of the DDFI strategy.

\subsection{Generalization Ability for Cross-Mesh Outputs}

Another core issue lies in the core-mesh generalization capability of the  neural network model. Nowadays, high-resolution meshes are often required in industrial-grade application scenarios to accurately capture the fine evolution characteristics of flow fields. Obviously, once the mesh topology is altered, it becomes necessary to reconstruct the dataset and retrain the neural network, and the resulting computational resource consumption is unacceptable.  We expect the constructed neural network model to possess a certain degree of cross-mesh generalization capability , i.e., without retraining, it can output the predicted flow field at mesh scales not corresponding to the training set simply by inputting new mesh topology information.

Based on this, this section employs two refined suboff meshes, namely Mesh$_{f1}$ and Mesh$_{f2}$, which consist of $7.9\times10^{5}$ and $1.15\times10^{6}$ cells, respectively, as shown in the Fig.~\ref{Fig:finemesh}. Without loss of general, the test case $T_{5}$ ($\left|\mathbf{V}^{*} \right|=5.0 $m/s, $\alpha=6^{\circ}$) is considered, and  we input the geometric information of these two meshes into the trained model to obtain the NN-predicted initial conditions.

\begin{figure*}[htbp]
\centering
\setlength{\tabcolsep}{0pt}
\renewcommand{\figurename}{Fig.}
\begin{subfigure}[b]{0.475\textwidth}
    \centering
    \includegraphics[width=0.975\textwidth]{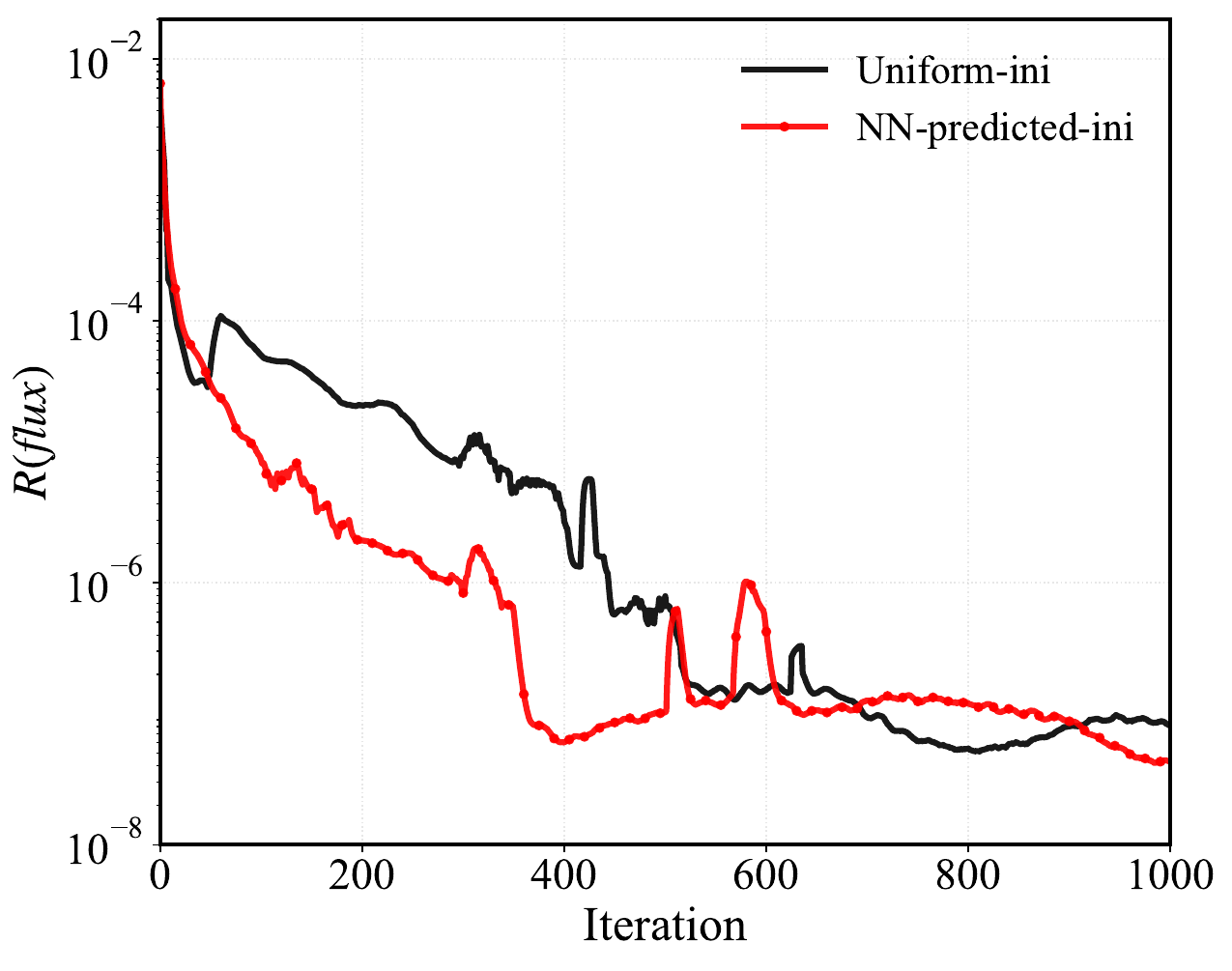}
\end{subfigure}
\quad
\begin{subfigure}[b]{0.475\textwidth}
    \centering
    \includegraphics[width=0.975\textwidth]{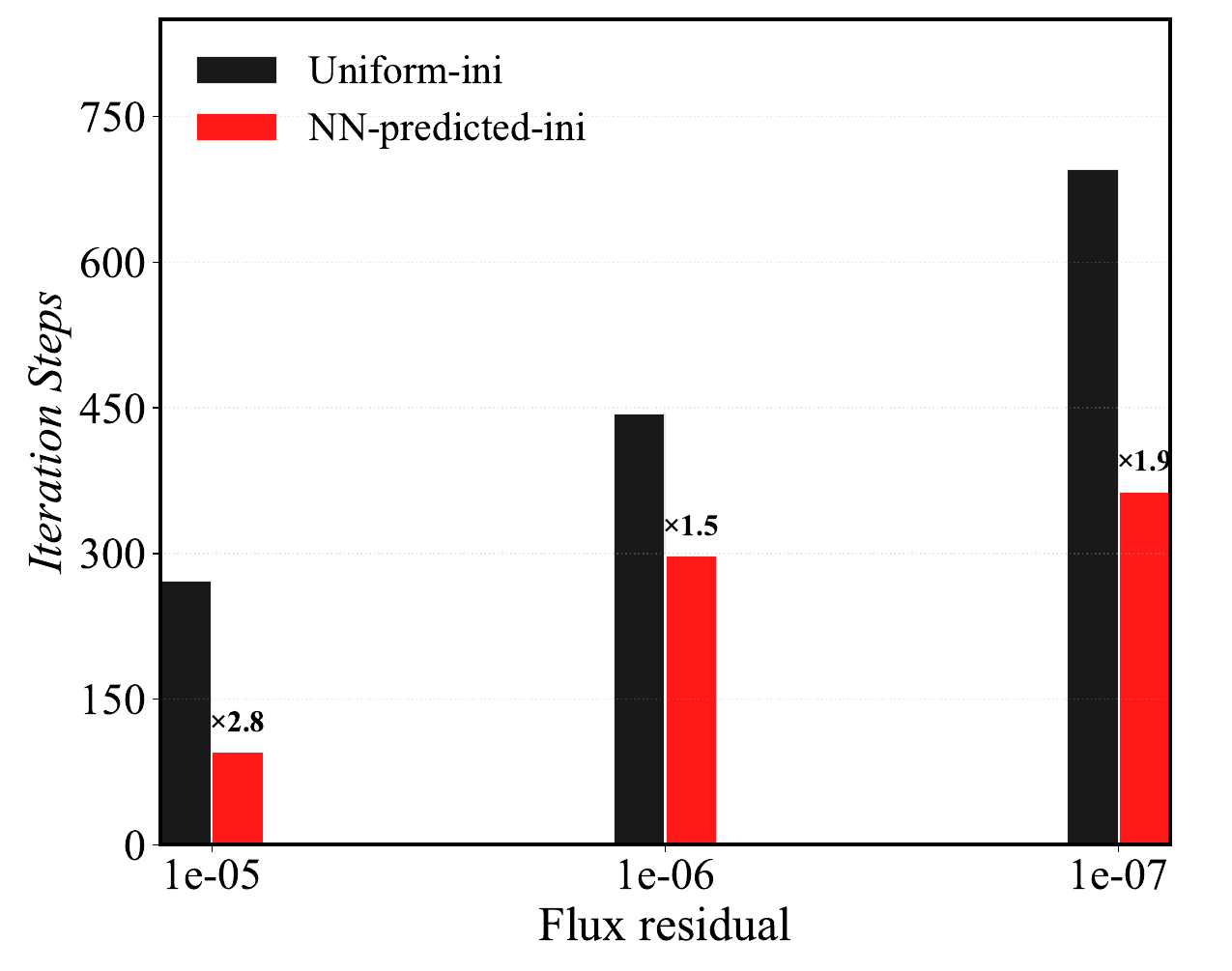}
\end{subfigure}
\\
\caption{
Comparison of flux residual convergence curves (left) and computation time required at threshold of $1\times10^{-5}$, $1\times10^{-6}$ and $1\times10^{-7}$  (right) respectively under uniform and NN-predicted initialization for Mesh$_{f1}$ (based on a 10-step moving average). The test condition is $T_{5}$.}
\label{Fig:V5_790sampling}
\end{figure*}

\begin{figure*}[htbp]
\centering
\setlength{\tabcolsep}{0pt}
\renewcommand{\figurename}{Fig.}
\begin{subfigure}[b]{0.475\textwidth}
    \centering
    \includegraphics[width=0.975\textwidth]{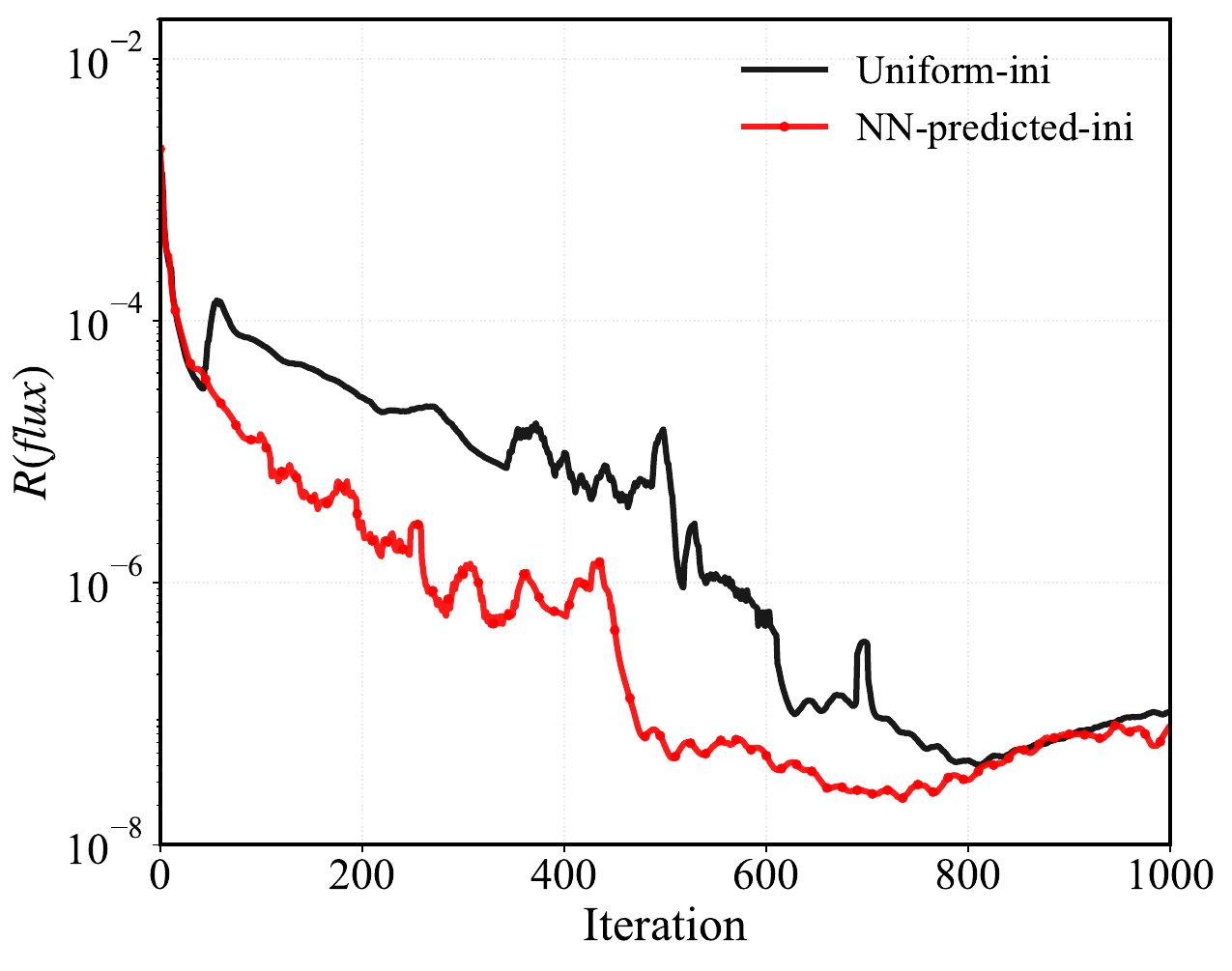}
\end{subfigure}
\quad
\begin{subfigure}[b]{0.475\textwidth}
    \centering
    \includegraphics[width=0.975\textwidth]{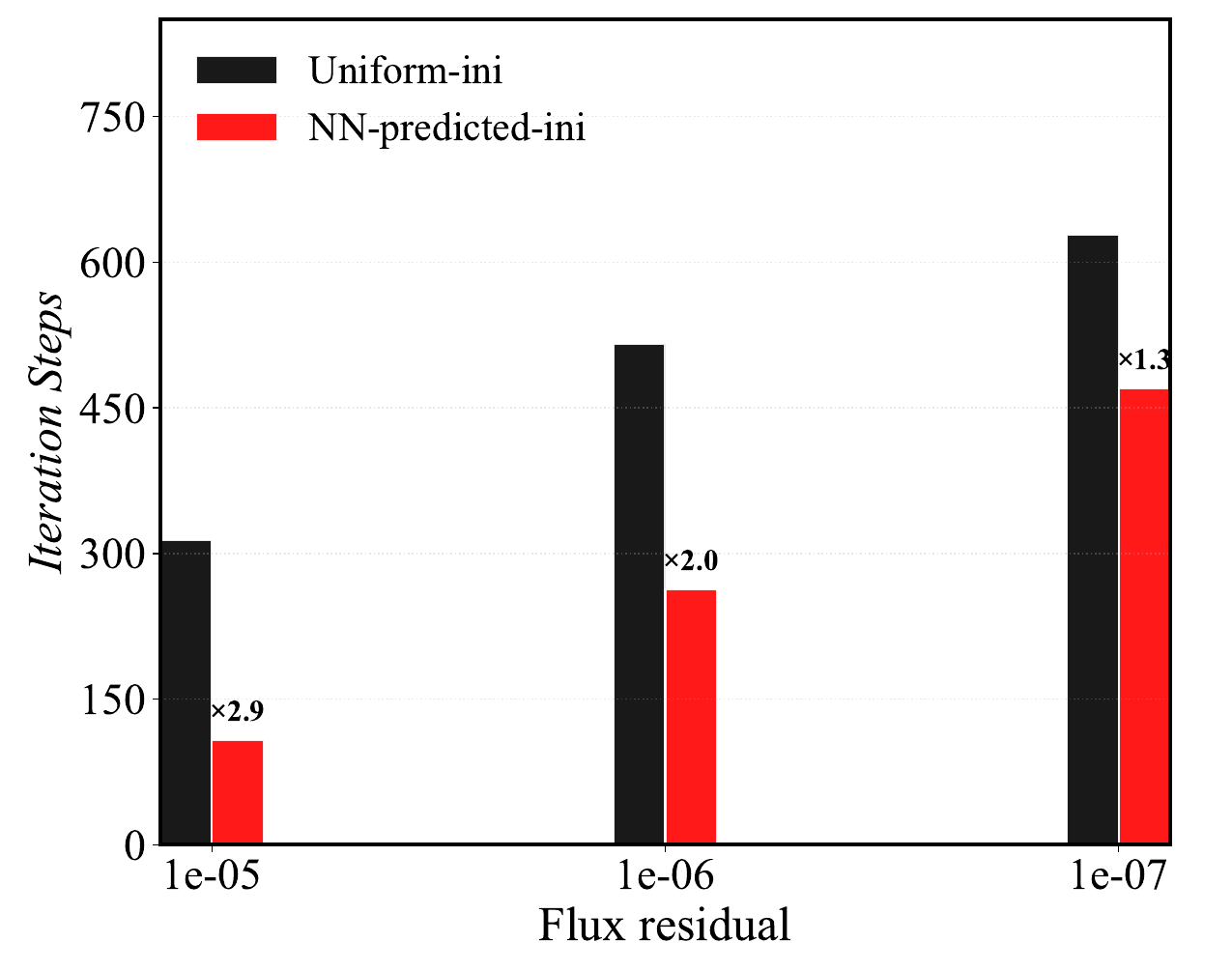}
\end{subfigure}
\\
\caption{
Comparison of flux residual convergence curves (left) and computation time required at threshold of $1\times10^{-5}$, $1\times10^{-6}$ and $1\times10^{-7}$  (right) respectively under uniform and NN-predicted initialization for Mesh$_{f2}$ (based on a 10-step moving average). The test condition is $T_{5}$.}
\label{Fig:V5_1145sampling}
\end{figure*}

Fig.~\ref{Fig:V5_790sampling} and Fig.~\ref{Fig:V5_1145sampling} demonstrate the acceleration effect of NN-predicted initial conditions on the iterative solution of flow fields, when numerical calculations are performed based on Mesh$_{f1}$ and Mesh$_{f2}$. It can be seen from the flux residual convergence curves that the residual decline rate under the NN-predicted initial conditions is consistently faster than that under the uniform flow field initial conditions. In particular, when the iteration step number $\approx100$, a residual rebound is observed in the uniform flow field, a phenomenon that signals the requirement for re-correcting  the flow field configuration. By contrast, the residual of the NN-predicted initial conditions maintains a continuous downward trend, an advantage derived from its flow field distribution being more consistent with the current operating conditions.

It should be noted that for strongly nonlinear operating conditions such as suboff ascent/descent motions, numerical errors induced by meshes with different resolutions will lead to discrepancies in the detailed CFD flow field characteristics even under identical operating conditions, which further increases the difficulty of cross-mesh prediction. Nevertheless, for the test scenarios in this section where the mesh resolution is increased by $1.8$ times and $2.5$ times compared with the training set, the neural network proposed in this paper can still generate initial conditions with satisfactory accuracy and significantly improve the convergence efficiency of CFD iterations, fully demonstrating the application potential of the DDFI strategy in cross-mesh flow field acceleration tasks.

\section{Conclusion}
This paper proposes a DDFI strategy for the fast prediction of flow fields induced by the vertical-plane oblique motion of underwater vehicles under steady-state conditions. Balancing computational efficiency and result reliability, this strategy abandons the conventional practice of using a uniform field as the initial condition in the initialization phase of the CFD solver. Instead, it utilizes a DNN model that takes mesh
geometric, operating conditions, and mesh-operation condition hybrid vectors as inputs to generate a refined initial flow field, thereby significantly reducing the convergence time of CFD iterations.

The suboff model with ascending and descending motions is selected as the research object for validation. The results demonstrate that, under such nonlinear operating conditions, the constructed neural network can achieve adequate-precision flow field prediction across the entire three-dimensional computational domain. Even for navigation angles and velocities not covered in the training dataset, the predicted flow field distributions remain in good agreement with the CFD simulation results. With the aforementioned optimization of initial flow fields, the DDFI strategy can significantly accelerate the iterative convergence speed of the solver: under loose convergence thresholds, the maximum speedup ratio reaches 3.5; even under strict convergence thresholds, the average speedup ratio still achieves 1.8. In addition, compared with traditional strategies, the DDFI strategy exhibits superior computational stability, where the residual converges steadily and monotonically without significant rebound. It is worth emphasizing that this strategy retains physical equations as the core guarantee for result correctness, a feature that is indispensable for practical engineering applications.

On this basis, we further explores two core issues concerning the engineering application of the DDFI strategy. The results indicate that reasonable sampling of the training dataset does not fundamentally compromise the effectiveness of the strategy. Therefore, we can expand the coverage of training conditions without increasing training costs by increasing the sampling interval. Besides, we demonstrates the cross-mesh generalization capability inherent in this strategy, which also acts as a core enabler for reducing the costs of dataset construction and model training.

In summary, the core objective of this paper is to provide a practical  and well-balanced new approach for the integration of data-driven machine learning and CFD solvers, ensuring the accuracy and reliability of simulation results while fully leveraging the efficiency of machine learning. Nevertheless,  enhancing generalization ability remains a core and enduring topic for data-driven machine learning methods. In future work, we plan to extend this strategy to more operating conditions and numerical cases, including two specific scenarios: 1). operating conditions with larger intervals of angles/velocities and broader coverage; 2). further predicting the flow field distribution characteristics within the entire computational domain after modifying the shape of the submersible. We anticipate that these efforts will contribute to the advancement of  machine learning-enhanced CFD technologies and help solve the key challenges in practical applications.
\section*{Acknowledgments}
This work is partly supported by funds from the Independent Research Project of the State Key Laboratory of Ocean Engineering, and the Shanghai Jiao Tong University New Faculty Start-up Program.










\printcredits

\bibliographystyle{cas-model2-names}

\bibliography{cas-refs}



\end{document}